\documentstyle[aps,preprint,prb]{revtex}
\tightenlines 
\include{epsf}
\begin{document}

\title{ Correlations Between Reconstructed $EUR$ Exchange Rates vs. 
$CHF$, $DKK$,
$GBP$, $JPY$ and $USD$}

\author{ \large \bf M. Ausloos$^1$ and K. Ivanova$^2$} \address{ $^1$ GRASP,
University of Li\`ege, Sart Tilman B5, B-4000 Li\`ege, Belgium \\ 
$^2$ Department
of Meteorology, Pennsylvania State University, \\ University Park, PA 
16802, USA}


\maketitle

\begin{abstract}

On Jan. 1, 1999 the European Union introduced a common currency Euro ($EUR$), to
become the legal currency in all eleven countries which form the 
$EUR$. In order
to test the $EUR$ behavior and understand various features, the $EUR$ exchange
rate is artificially extrapolated back to 1993 by a linear superposition of the
exchange rates of the 11 currencies composing $EUR$ with respect to several
currencies not belonging to the $EUR$, i.e. Swiss Franc
($CHF$), Danish Kroner ($DKK$), British Pound ($GBP$), Japanese Yen ($JPY$) and
U.S. Dollar ($USD$) of interest for reasons given in the text. The distribution of fluctuations of
the exchange rates is shown to be Gaussian for the central part of the
distribution, and having fat tails for the large size fluctuations. Within the
{\it Detrended Fluctuation Analysis} ($DFA$) statistical method we 
have obtained
the power law behavior describing the root-mean-square deviation of 
the exchange
rate fluctuations as a function of time. For the period between Jan. 1995 and
Jan. 1999 we have compared the time-dependent exponent of these exchange rate
fluctuations for $EUR$ and that of the 11 currencies which form the $EUR$. The
German Mark ($DEM$) and the French Franc ($FRF$) have been the currencies
primarily leading the fluctuations of the exchange rates, while Italian Lira 
($ITL$) and ($PTE$) Portuguese
Escudo are the less relevant currencies from this point of view. Technical
considerations for the $EUR$ implementation are given as conclusions. The cases
of exchange rates with $DKK$ appear quite different from the other four major
currencies.\end{abstract}

\vskip 2cm


\newpage

\section{Historical Introduction}

The theory of FExCR (foreign exchange currency rates) does not really 
exist or if
any has been proposed it is controversial.\cite{hartmann} The idea 
has been that
FExCR fluctuations occur due to the differences in Gross National Products, -
measured by {\it national moneys}. It has been proposed that FExCR 
fluctuations are
varying as a function of the trade balance (deficit or not) between 
countries, -
measured by {\it international moneys}.\cite{hartmann} Fluctuations 
in FExCR surely
hamper the trades because one is never sure of the FExCR variation between the
order, sale, delivery and invoice time, - not counting transaction fees.
Therefore, it is extremely
interesting to predict the evolution of FExCR and to discover similarities
between their behaviors. One technical way that 11 nations in Europe have found
to reduce FExCR variations was to define a new currency $EUR$, based on fixed
parities between these eleven European countries. How the rates were 
obtained and
agreed upon remain a political mystery known to a few. We have 
already looked at
the evolution of 
FExCR,\cite{nvma,nvma2,kiandma2,maladek,maki3,kimalg} and of the
$EUR$ in particular, even inventing a False $EUR$,\cite{maki3,kimalg} 
backward in
time in order to observe whether previous not-so-artificial fluctuations were
conserved, and whether in some sense the invention of $EUR$ made sense from a
statistical physics point of view ! We have used the detrended fluctuation
analysis technique (and previously a multifractal
approach\cite{nvma2,kiandma2,Kada}) to do so, a method which when applied to
study the fluctuation correlations in a signal on a finite time 
interval in which
some scaling law holds gives some information on the future evolution of the
signal.

Hartmann\cite{hartmann} has examined the competition between $USD$, $JPY$, and
$EUR$. Hereby we are also trying to answer the question whether the apparently
strongest currency, i.e. the $DEM$ was/is the main basis for comparing the 11
currencies forming $EUR$ in the past. We want to search whether there is a
mathematical evidence for {\it said to be obvious or well known historical}
correlations between national currencies. In order to do so we construct false
exchange rates with respect to currencies like $CHF$, $DKK$, $GBP$, $JPY$, and
$USD$. These currencies have been selected for various reasons. The $CHF$ is a
European currency from a country NOT belonging to the European Union, 
but usually
a reference currency. The $DKK$ and $GBP$ are currencies for a 
country belonging
to the European Union but not in the $EUR$ system, with different financial
and/or political views on whether or not to join $EUR$, and from 
countries having
different sizes, economy, historical relevance, $etc.$. The UK Government has
decided that the UK economy is not ready for joining $EUR$ on 1 January 1999.
However, the UK Government supports the principle of joining the 
single currency,
if that is in the national economic interest. Denmark had a referendum and
raised not necessarily $ad$ $hoc$ arguments about $DKK$ to becoming
part of the $EUR$, $etc.$ The $JPY$ and $USD$ are both well known major
currencies outside Europe.

In the following we present a rather complete study, reporting many graphs,
keeping as much as possible the same scales on axes for comparisons for several
FExCR. We give the exchange rates, and their fluctuation distribution of the
$EUR$ eleven currencies and $EUR$ itself with respect to the five selected main
non-$EUR$ currencies over the time interval Jan.01, 1993 to June 30, 2000. We
perform the usual DFA and obtain the mean $\alpha$ exponent characterizing the
power law over the longest possible scaling range pertaining to this study. We
perform next the local DFA study and present in graphical form the resulting
so-called correlation matrix eliminating the time as a series parameter. We
summarize the data through the mean, the variance and the median of the local
$\alpha$ exponents. We conclude with some financial policy statement and
historical considerations arising from our observations.

\section{Experimental Data}

The conversion rates of the $EUR$ participating countries were fixed 
by political
agreement based on the bilateral market rates of December 31, 1998. 
\cite{maki3}
Using these rates,\cite{quoteEUR} one Euro ($EUR$) can be represented as a
weighted sum of the eleven currencies $C_i$, $i= 1, 10$:

\begin{equation} 1 EUR = \sum_{i=1}^{10} (\delta_{i,2} + 1) \frac{\gamma_i}{11}
\, C_i \end{equation}

\noindent where $\gamma_i$ are the conversion rates and $C_i$ denote the
respective currencies, i.e. Austrian Schilling ($ATS$, $i$=1), Belgian Franc
($BEF$, $i$=2), German Mark ($DEM$, $i$=3), Spanish Peseta ($ESP$, $i$=4),
Finnish Markka ($FIM$, $i$=5), French Franc ($FRF$, $i$=6), Irish Pound ($IEP$,
$i$=7), Italian Lira ($ITL$, $i$=8), Dutch Guilder ($NLG$, $i$=9), Portuguese
Escudo ($PTE$, $i$=10). In view of the financial identity of the 
Luxemburg Franc
($LUF$), with the Belgian Franc ($BEF$), the latter is weighted by a factor of
two, whence the $\delta$ Kronecker symbol in the above equation. In order to
study correlations in the $EUR$ exchange rates, the $EUR$ existence can be
artificially extended backward, i.e., before Jan. 01, 1999 and thereby defining
an artificial $EUR$ before its birth.\cite{maki3}

The behavior of $EUR$ with respect to currencies outside the European Union is
still of interest in view of the recent fall, in particular with respect to
$USD$, then resurgence of the $EUR$ : considerations with respect to 
Swiss Franc
($CHF$), Danish Kroner ($DKK$), British Pound ($GBP$), Japanese Yen ($JPY$) and
U.S. Dollar ($USD$) are presented below. In the following we call 
$B_j$, $j$ = 1,
5 these 5 currencies outside the European Union, keeping $j$ as an index; the
order is  alphabetic in currency acronym. We have constructed a
data series of $EUR$ exchange rates with respect to these currencies following
the linear superposition rule:

\begin{equation} 1 EUR/B_j = \sum_{i=1}^{10} (\delta_{i,2} + 1)
\frac{\gamma_i}{11} \, (C_i/B_j) \end{equation}

Since the number of data points of the exchange rates for the period starting
Jan. 1, 1993 and ending Dec. 31, 1998 is different for the eleven 
currencies, due
to different national and bank holidays a linear interpolation has 
been used for
the days when the banks are closed and official exchange rates are 
not defined in
some countries. The number $N$ of data points as equalized is $N = 1902$,
spanning the time interval from January 1, 1993 till June 30, 2000.

The latest date is so chosen for the present report and studies in order to
remain coherent and avoid a possible spurious effect in later considerations
arising from the Greek Drachma ($GRD$) was introduced as a 
supplementary currency
in $EUR$ on June 19, 2000.\footnote {The greek drachma exchange rate has been
locked on June 19, 2000 and the greek drachma becomes the 12th $EUR$ equivalent
currency starting on Jan. 01, 2001.}

The normalized $EUR$ exchange rates so reconstructed with respect to the $CHF$,
$DKK$, $GBP$, $JPY$ and $USD$ are plotted in Fig. 1(a-e). Indicative 
values, for
normalization purposes of the exchange rates on Oct. 2, 1996 are given in Table
1.

While the $EUR/CHF$ and $EUR/DKK$ exchange rate (ExR) are pretty 
stable and thus
not much disturbed by the transition to the real $EUR$, the other currency ExR,
in particular with respect to $USD$, $JPY$ and $GBP$ have been much 
sensitive to
the transition, with a noticeable decay of the $EUR$ value after Jan. 01, 1999.
With respect to $CHF$ several currencies seem to be pretty stable, as $ ATS$,
$BEF$, $DEM$, $FRF$, and $NLG$, while a neighboring country currency $ITL$
presents a marked dip in 1995. Interestingly, with respect to $DKK$, $FIM$ has a
better value than all the other currencies. See again the 1995 dip of 
$ITL$, and
end of 1993 bumps in $ATS$, $BEF$, $DEM$ and $NLG$. The same $grouping$ of
currencies is observed for the ExR with respect to $GBP$, the rate evolution
being rougher. For the ExR with respect to $JPY$ and $USD$ a wilder variation
exists, with the rate being greater or smaller than unity, though with some
general consistency in the evolution. The $ESP$, $FIM$ , and $ITL$ 
seem to be the
currencies following weakly the majority (or mean for $EUR$) evolution.

\section{Distribution of the fluctuations}

The distributions of the exchange rate fluctuations for $EUR/CHF$, $EUR/DKK$,
$EUR/GBP$, $EUR/JPY$, and $EUR/USD$ are shown in Fig. 2(a-e) for the time
interval of interest. The central part of each distribution, i.e. the small
fluctuations, is close to a Gaussian, while the tails of the 
distributions, i.e.
the large fluctuations, strictly depart from the normal distribution. These
so-called fat tails are found to follow a power-law
distribution\cite{MantegnaStanleybook} with a slope equal to $ca.$ 2.9 for
$EUR/CHF$, of order of 4.0 for $EUR/GBP$, 3.2 for the negative and 4.0 for the
positive tail of $EUR/JPY$, and 3.2 for the negative and about 4.5 for the
positive tail in $EUR/USD$ ExR, as found in the {\it elastic energy price 
resistance - linear potential momentum trading} model of Castiglione et
al.\cite{castiglione} Notice the asymmetry in some 
distributions, like
the positive tail for $EUR/JPY$ fluctuations. The distribution of 
fluctuations in
exchange rates for each 11 currency of interest has also been studied 
but is not
shown here for space saving. The case of the exchange rate fluctuations of the
$DEM$ with respect to $DKK$, $CHF$, $JPY$, and $USD$ has already been 
illustrated
in Ref.\onlinecite{tokyokima} in which the reader can observe a typical
distribution as found for most of the European currencies. It is 
known that such
tails usually have a slope markedly different from $-2$. \cite{gopi}

\section{Intracorrelations between fluctuations in a specific exchange rate}
\subsection{DFA results}

The $DFA$ technique\cite{DFA} has been often described and is not 
recalled here.
It leads to investigate whether the root-mean-square deviations of the
fluctuations of an investigated signal $y(n)$ have a scaling behavior, 
e.g. if the
so-called $DFA$ function $<F^2(\tau)>$ scales with time as

\begin{equation} \langle {1 \over \tau } {\sum_{n=k\tau+1}^{(k+1)\tau}
{\left[y(n)- z(n)\right]}^2} \rangle \sim \tau^{2\alpha} \end{equation}

\noindent
where $z(n)$ is hereby a linear function fitting at best the data in the
$\tau$-wide interval which is considered. A value $\alpha = 0.5$ 
corresponds to a
signal mimicking a Brownian motion. A thorough discussion of various features
found in $DFA$ results can be found in 
Ref.\onlinecite{ivanovchinese}. Some care
must be taken concerning crossover behaviors and subliminal noise features.

A log-log display of the $DFA$ function leading to a measure of 
$\alpha$ for the
11 exchange rates of interest is found in Fig. 3 (a-e). The $\alpha$-exponent
values are summarized with their scaling range in Table 2.

Let it be observed that the time scale invariance for $EUR/CHF$, $EUR/JPY$, and
$EUR/USD$ holds from 5 days (one week) to about 250 days ($ca.$ 52 weeks or one
banking year) showing Brownian-like type of correlations. Two different scaling
ranges are found for the $ATS/DKK$, $DEM/DKK$, $FIM/DKK$, $ITL/DKK$, $NLG/DKK$,
and $EUR/DKK$; one, from five to 60 days (or twelve weeks) with a non-Brownian
$\alpha<0.5$ and thus $anti-persistent$ type of behavior, and 
another, after that
for up to one year with mostly Brownian-like fluctuations within the standard
error bars (see Table 2).

Notice that $ITL$ has usually the largest $\alpha$, even markedly greater than
0.5. An anomalous value (with respect to those quoted) is that of $ATS$ with
respect to $CHF$, which is interesting to notice in view of the geographical
proximity of the corresponding countries, thus economies. With respect to $JPY$
and $USD$ the correlations are similar to those in Brownian motion within the
standard error bars. The case of ExR with respect to $DKK$ is {\it 
clearly different}
from others, with a marked double set of correlations, one $anti-persistent$ at
short time lag, and the other with a quite $persistent$ one at longer time lag.
In all cases, notice that the value of $\alpha$ for the $EUR$ ExR 
falls close to
the Brownian motion value, - a result indicating the good sense of creating and
using the $EUR$.

\subsection{Local DFA studies}

As done elsewhere,\cite{nvma} in order to probe the existence of {\it locally
correlated} and {\it decorrelated} sequences, we have constructed an 
observation
box, i.e. a 514 days (two years) wide window probe placed at the 
beginning of the
data, calculated $\alpha$ for the data in that box, moved this box by one day
toward the right along the signal sequence, calculated $\alpha$ in that box,
a.s.o. up to the $N$-th day of the available data. A local, time dependent
$\alpha$ exponent is thus found.

The time dependent $\alpha$-exponent for $EUR$ and each currency 
(which forms the
$EUR$) exchange rate toward $DKK$, $CHF$, $GBP$, $JPY$, and $USD$ are shown in
Fig. 4(a-e).

For all currencies, $\alpha$ is determined from the best fit over the central
box/interval, which is maintained to be constant, for calculation ease, between
11 and 67 days (2 and 13 weeks), whence skipping the first few points from the
DFA-function, known often to be deficient and also avoiding the points from the
large time intervals that come with not sufficient statistics. While the
differences in the $\alpha$-behavior after Jan. 1, 1999 are almost
undistinguishable, the time dependent $\alpha$-exponents before that 
day exhibit
quite different correlated fluctuations depending on the currency.\footnote
{Different fluctuations still exist, though with a narrower distribution, even
after Jan. 01, 1999 because of the memory effect in calculating $\alpha$ in the
DFA.}

We stress that the evolution of fluctuations in the currency ExR are markedly
seen. The behavior of the European currencies with respect to $e.g.$ $CHF$ is
widely different from time to time and uncorrelated with each other. For the
$CHF$ ExR notice how different the $EUR$ is, say with respect to $ATS$, $BEF$,
$FIM$ and $IEP$, which tend toward a $EUR$ fluctuation behavior as 
time goes by.
Interestingly the $\alpha$-exponents for $ITL$ and $PTE$ are very close to the
$EUR$-$\alpha$-exponent behavior over the whole period, indicating 
the dependence
of such currencies with respect to the nine others. The same is true 
for the ExR
with respect to $GBP$, $ESP$, $ITL$ and $PTE$ forming a different 
group from the
others. Again, notice the special behavior of a country currency quite tied to
the UK, i.e. the behavior of $IEP$$-\alpha$-exponent is markedly different from
others. Observe the bumps in the $\alpha$-exponent curve in $ESP$ and $FRF$ for
example for the $JPY$ exchange rate in the 1995 year, similar to the 
one seen in
the $EUR$. Observe the very similar behavior of all currencies with respect to
major currencies like the $JPY$ and the $USD$. Finally, the 
$\alpha$-exponents of
ExR with respect to the $DKK$ are {\it rougher and rather different} 
from one currency
to the other, with sometimes huge fluctuations.

\subsection{Time evolution of elementary statistical parameters}

In Figs. 5 (a-d) the time evolution of the mean, median and standard 
deviation of
the $\alpha$ exponents for the currencies forming the $EUR$, are 
compared to that
of the $EUR$ and $DEM$ for $CHF$, $GBP$, $JPY$ and $USD$ as reference 
currencies.
The ratio mean/median of the $\alpha$ exponents for the four cases is next
considered.

Several remarks follow. It is clear that the leading currencies from 
the point of
view of the exchange rate fluctuations were $DEM$ and $FRF$, with 
$PTE$ far away
from the main stream. Larger fluctuations in the mean/median ratios 
are observed
for $CHF$ as reference currency in comparison to the other currencies. The
largest fluctuations predominantly occur for the years before 1997. 
Fig. 6 allows
to sort out the behavior of the fluctuations and currencies with respect to
$DKK$. Notice that $\alpha_{mean}$ and $\alpha_{median}$ markedly evolve away
from 0.5. The variation of the ratio ($\alpha_{mean}/\alpha_{median}$) in
particular after Jan. 01, 1999 is due to the fact that both $\alpha_{mean}$ and
$\alpha_{median}$ become small, and therefore even small discrepancies between
them lead to large fluctuations of their ratio. After Jan. 01, 1999
$DKK$-$\alpha$-values are close to that of noise-like correlations of 
fluctuations.

\section{Intercorrelations between fluctuations in exchange rates}

A graphical correlation matrix of the time-dependent $\alpha$ exponent has been
constructed for the various exchange rates of interest. In Figs. 7 
(a-e), we plot
$\alpha_{C_i/B_j}$ {\it vs.} $\alpha_{EUR/B_j}$ for all $i$ and $j$ 
values. This
so-called correlation matrix is displayed for the time interval hereby
considered, i.e. from Jan. 01, 1995 till Dec. 31, 1998. The interval 
is so chosen
because the latter has to be reduced on one hand at the lower end due 
to the size
of the testing window box, and at the upper end by the fact that after Jan. 01,
1999 the 11 currencies are not independent any more since their 
conversion rates
are fixed within the $EUR$.

As described elsewhere, e.g. in Ref.\onlinecite{kimalg}, such a correlation
diagram can be divided into its main sectors through a horizontal, a 
vertical and
diagonal lines crossing at (0.5,0.5). If the correlation is strong the cloud of
points should fall along the slope $= + 1$ line. It appears that the strongest
correlation is that between $DEM/CHF$ and $EUR/CHF$, while $ATS$, $IEP$, and
$PTE$ show weak correlations with $EUR$ for $CHF$ ExR as a reference currency
(Fig. 7a).

The rest of the referencing currencies, i.e. $GBP$, $JPY$ and $USD$, view $DEM$
and $FRF$, and also $BEF$, and $NLG$ as the leading correlated currencies in
defining the $EUR$ behavior, while the $ESP$, $ITL$ and $IEP$ seem to 
be outside
this leading process. $FIM$ and $PTE$ being rather peculiar with respect to
$JPY$. $ATS$ joins the 'more regular ones' in the $USD$ ExR case.

Finally, {\it no correlation} between the $EUR$ and the 11 
participating currencies
exist from the point of view of $DKK$ ExR (Fig. 7b).

\section{Conclusions}

We have thus studied a few aspects of the $EUR$ exchange rates from 
the point of
view of the fluctuations of the $EUR$ and the 11 currencies forming the latter.
We have examined here the exchange rates toward $CHF$, $DKK$, $GBP$, $JPY$ and
$USD$.

In examining various $reconstructed$ exchange rates for the currencies forming
the $EUR$ before Jan. 01, 1999 we have searched at whether correlations would
confirm historical and financial view points and so called standard knowledge.
The fluctuation distribution density as examined confirms that the foreign
exchange markets do not follow Gaussian distributions. The distribution of the
fluctuations is close to a Gaussian one only for small fluctuations, with power-law
distribution\cite{MantegnaStanleybook} for large fluctuations. However the
correlation between fluctuations were close to the Brownian case, and 
the more so
after the $EUR$ was introduced. There is no doubt that speculators have not yet
found the best way to play on the $EUR$ exchange rates, as done if the
DFA-$\alpha$ exponent is markedly different from 0.5.\cite{nvma} It has been
noticed that the introduction of the $EUR$ tends to smoothen the 
fluctuations and
their correlations, as well gather together in a main stream most of 
the European
currencies, forming the $EUR$. Nevertheless it has been pointed out in the
previous sections that some currencies are more tied to external ones 
than others
due to economic and geographic-like conditions.

Several remarks follow from the correlations observed by examining 
the structural
diagrams (Figs.7). While the structural diagrams for $ATS$, $PTE$, 
and $ITL$ with
respect to $EUR$ show weak or no correlation at all, the structural diagram for
$DEM$ shows very strong correlations between the time dependent
$\alpha$-exponents. It is clear that the leading currencies {\it from the point
of view of the exchange rate fluctuations} were $DEM$, $FRF$, together with
$ATS$, $BEF$ and $NLG$, while $PTE$ and $IEP$ are far away from the 
main stream.

It is known that the median is sometimes a better representation of the main
behavior of a system, being more sensitive to the contributions from far away
events. Therefore we consider the ratio between the $\alpha_{mean}$ and the
$\alpha_{median}$. This is done in Fig. 8 for the mean and median values of the
$\alpha$ exponents FxR with respect to $CHF$, $GBP$, $JPY$, and $USD$. The case
of $DKK$ is seen in Fig. 6. The fact that the $\alpha_{mean}$/$\alpha_{median}$
ratio departs from unity indicates that the distribution of $\alpha$ for
$C_i/B_j$ ($j$=fixed) departs from Gaussian, for which $\alpha_{mean}=
\alpha_{median}$. Therefore one may expect that the ratio
$\alpha_{mean}$/$\alpha_{median}$ could also be used in speculation cases, and
improve the way gains can be obtained.\cite{nvma,kimalg}

We have observed that the $DEM$ is the strongest currency that has 
dominated the
correlations of the fluctuations in $EUR$ exchange rates with respect 
to the top
world leading currencies, $JPY$, $USD$, and even $GBP$ and $CHF$, 
while $PTE$ was
the most extreme one in the other direction.

Can one gain or loose in investing in EUR {\it per se} ? .. is still an open
question, though the behavior of fluctuations is not in favor of 
speculators. Is
the $EUR$ a gain for humanity or not? Is the $DEM$ really the strong 
currency? If
one had removed all currencies and established that the $DEM$ was to become the
currency in the $DEM$-zone, with the Deutsche Bundesbank (BuBa) in charge of
financial policy, most would have shouted loudly. In the present 
European policy,
all central banks, including the BuBa lost control of their financial 
policy ...
somewhat. Is that better ?

A fundamental question pertains to the trend and the next one to the 
fluctuations
in $EUR$ with respect to other moneys.

The answer is found in both cases in the behavior of $\alpha$ local, and in a
previous observation by Ref.\onlinecite{nvma} that the change in $\alpha$ as a
function of time is directly linked to the interest rate of the involved
currencies. Thus if $\alpha$ is too far away from 0.5 it is advisable 
for central
banks to modify the exchange rate in order to avoid gigantic 
speculation. Mr. A.
Greenspan just did it, without knowing the $local-\alpha$ behavior theory. The
same can be suggested to Mr. W. Duisenberg for the $EUR$ behavior. As long as
$\alpha$ is not too far away from 0.5 (the closest is $EUR/USD$) 
there is no fear
to have. The trend is governed by more general European economy considerations.
However if it is noticed that $local-\alpha$ significantly deviates 
from 0.5 (the
cases of $DKK$ and $GBP$) a major investigation on the $EUR$ and 
$DKK$ and $GBP$
rate evolution should be made. A change in interest rate should be swift if
necessary. Notice that to wait a long time, as in the Plaza agreement, case is
detrimental.\cite{nvma}

Floating exchange rates (and high capital mobility) presently describe the
monetary regime in most countries. In the early 1960s however almost all
countries were linked together by fixed exchange rates within the so-called
Bretton Woods System. International capital movements were highly curtailed, in
particular by extensive capital and exchange rate controls. Yet in 1961 Mundell
in his article on {\it optimum currency areas}\cite{Mundell1961} 
asked whether it
is advantageous to relinquish monetary sovereignty in favor of a 
common currency?
It was noticed that due to high capital mobility in the world economy, regimes
with a temporarily fixed, but adjustable, exchange rate become fragile. A
currency union or a floating exchange rate was proposed to be a relevant
alternative, as with the common European currency.  Under a fixed 
exchange rate,
the central bank must intervene on the currency market in order to satisfy the
public's demand for foreign currency at this exchange rate. As a result, the
central banks loose control of the money supply, which adjusts itself to the
domestic liquidity. To implement independent national monetary policy 
by means of
so-called {\it open market operations} becomes futile because neither the
interest rate nor the exchange rate could be affected. From Fig.1 it 
is observed
that  already before 1999 the separate currencies in the then European Monetary
Union were indeed fluctuating in rather similar ways.

How this was implemented is nevertheless irrelevant for our purpose: complex
monetary and fiscal policy of many (sorts of) governments have been surely
influencing national economic activity, and correlation between FExCR and other
economic data is out of our intent but could be usefully investigated. Yet the
speed of adjustment on FExCR markets and some temporarily $overshoot$ are well
seen in the various figures; they could be related to the wake of 
some political
$regional$ disturbances.  Whether high labor mobility (supposed to be 
the remedy
in order to offset such economic policy disturbances) was effectively intrinsic
is hard to believe at this time.\footnote {It is known that some of the main
advantages of a common currency rests in low transaction costs in 
trade and also
less uncertainty about relative prices. Indeed  an increased demand in
one part of the currency union can cause an increased employment there which no
local government bank could counteract by increasing interests rates to prevent
inflation. If one feels employment is valuable, and low inflation is also
valuable, then a major drawback of a currency union is that emerging local
imbalances within the union cannot be counteracted only by changing interest or
exchange rates for the local currency. Another of the major drawbacks is the
difficulty of maintaining employment when reduction in demand for goods or more
generally {\it so-called asymmetric shocks} exist. This usually requires a
lowering of wages or worker lay-offs. }  Yet (after 1999) no supposedly strong
currency was sacrificed for the $EUR$, and the $EUR$ did not save any 
supposedly
weak currency. Even though in 1996 several countries were far away from the
Maastricht criteria,\cite{stern} the $EUR$ seems to have been an interesting
vaccine.\cite{ecu}

All inventors of moneys surely thought about stability.  Economic conditions
showed the lure of such an ideal.  New Franc, new zlotys have been invented
recently, in order to avoid filling papers with a huge list of zeroes. Carolus
Magnus in order to unify his Empire invented a Carolingian Pound, a piece of
gold, indeed weighing half a pound.  Still the Pound is used in 
Great Britain
as a monetary unit, even though it is reduced to a light sheet  of paper. In
Roman time (200 BC), the coins also served to measure weight and/or buy slaves.
Supposed to weigh 12 ounces (321.45 g) the Pound had a weight of only 10 ounces in 
order to take
care for taxes and emission rights, and costs.  During the years, its gold
equivalent weight decreased, was made of copper, .. leather, and disappeared.
What will be the future of $EUR$? It will surely depend on its rate in foreign
exchanges.

\vskip 1cm

{\noindent \large Acknowledgements} \vskip 0.6cm We are very grateful to D.
Stauffer for enticing us in further examining several aspects of these foreign
exchange rate behaviors.

\newpage {\bf Figure Captions}

\vskip 1.0cm {\noindent \bf Figure 1} --- Normalized $EUR$ and currency forming
the $EUR$ exchange rates with respect to the (a-e) $CHF$, $DKK$, 
$GBP$, $JPY$ and
$USD$ between Jan. 01, 1993 and June 30, 2000. The data are artificially
multiplied by two and then displaced along the vertical axis in order 
to make the
fluctuations noticeable.

\vskip 1.0cm {\noindent \bf Figure 2} --- Distributions of the exchange rate
fluctuations for (a-e) $EUR/CHF$, $EUR/DKK$, $EUR/GBP$, $EUR/JPY$, 
and $EUR/USD$
for data in Fig. 1 (a-e).

\vskip 1.0cm {\noindent \bf Figure 3} --- Log-log plot of the DFA function
showing how to obtain the $\alpha$ exponent for the 11 exchange rates 
of interest
for (a-e) $EUR/CHF$, $EUR/DKK$, $EUR/GBP$, $EUR/JPY$, and $EUR/USD$. The fit
slope being only of interest, the DFA function data has been arbitrarily
displaced along the vertical axis. The arrows indicate the best scaling ranges.

\vskip 1.0cm {\noindent \bf Figure 4} --- Time dependence of the DFA $local$
$\alpha$-exponent for $EUR$ and each currency (which forms the $EUR$) exchange
rate with respect to (a-e) $DKK$, $CHF$, $GBP$, $JPY$, and $USD$. The
$\alpha$-values are artificially multiplied by two and then displaced along the
vertical axis in order to make the fluctuations noticeable. For each time
dependent $\alpha$ a horizontal dashed line is drawn to indicate a reference to
Brownian fluctuations.

\vskip 1.0cm {\noindent \bf Figure 5} --- Time evolution of the mean, 
median and
standard deviation of the $\alpha$ exponents for the currencies forming the
$EUR$, compared to that of the $EUR$ and $DEM$ for $CHF$, $GBP$, 
$JPY$ and $USD$
as reference currencies (a-d). The $\alpha_{mean}$ and $\sigma(\alpha)$ curves
are not displaced. The $\alpha_{median}$ curves are displaced by -0.25. The
$\alpha_{EUR/}$ curves are displaced by +0.25 and the $\alpha_{DEM/}$ 
curves are
displaced by +0.5 in (a-d). Horizontal dashed lines mark Brownian motion 0.5
level for each $\alpha$-curve.

\vskip 1.0cm {\noindent \bf Figure 6} --- Time evolution of the mean, 
median and
standard deviation of the $\alpha$ exponent for the currency exchange 
rates with
respect to $DKK$, currencies forming the $EUR$, compared to that of 
the $EUR$ and
$DEM$. The curves are displaced for readability though the y-axis scale is
constant and the $\alpha_{median}$ and $\sigma(\alpha)$ curves are 
not displaced.
The $\alpha_{mean}$ curve is displaced by +0.3. The $\alpha_{EUR/DKK}$ curve is
displaced by +0.6 and the $\alpha_{DEM/DKK}$ curve is displaced by +0.9. The
horizontal dashed lines mark Brownian motion 0.5 level for each $\alpha$ curve.
The $\alpha_{mean}/\alpha_{median}$ ratio extracted from the $\alpha$ exponents
is also shown. Horizontal dashed line at y=2.2 corresponds to
$\alpha_{mean}/\alpha_{median}=1$.

\vskip 1.0cm {\noindent \bf Figure 7} --- Graphical representation of the
so-called correlation matrix elements for the time interval Jan. 01, 1995 till
Dec. 31, 1999 for the various $local$ $\alpha_{C_i/B_j}$ {\it vs.}
$\alpha_{EUR/B_j}$ exponents, where $C_i$ are the ten $EUR$ currencies of
interest ($i$=1,10), and (a-e) $B_j$ are the five foreign currencies ($j$=1,5)
considered in the text.

\vskip 1.0cm {\noindent \bf Figure 8} --- Time evolution of the
$\alpha_{mean}/\alpha_{median}$ ratio extracted from the $\alpha$ exponents for
the currency exchange rates, currencies forming the $EUR$, with 
respect to $CHF$,
$GBP$, $JPY$ and $USD$. The curves are displaced for readability though the
y-axis scale is constant and the $GBP$ curve is not displaced. The horizontal
dashed lines correspond to $\alpha_{mean}/\alpha_{median}=1$.

\newpage

\begin{table}[ht] \caption{Indicative values, for normalization purposes of all
currencies exchange rates on Oct 2, 1996; e.g. 1 $ATS$ = 0.93 $USD$.}
\begin{center} \begin{tabular}{||c||r|r|r|r|r||} \hline \hline $C_i /$ & CHF &
DKK & GBP & JPY & USD\\ \hline ATS & 0.1167 & 0.5448 & 0.0593 & 10.3913 &
0.9300\\

BEF & 0.0399 & 0.1861 & 0.0203 & 3.5497 & 0.0318\\

DEM & 0.8211 & 3.8326 & 0.4171 & 73.1140 & 0.6545\\

ESP & 0.0098 & 0.0456 & 0.0050 & 0.8696 & 0.0078\\

FIM & 0.2750 & 1.2847 & 0.1397 & 24.4880 & 0.2192\\

FRF & 0.2423 & 1.1327 & 0.1231 & 21.5770 & 0.1932\\

IEP & 2.0819 & 9.6202 & 1.0213 & 179.0200 & 1.6026\\

ITL & 0.0008 & 0.0039 & 0.0004 & 0.0739 & 0.0007\\

NLG & 0.7317 & 3.4179 & 0.3717 & 65.1580 & 0.5833\\

PTE & 0.0081 & 0.0380 & 0.0041 & 0.7196 & 0.0064 \\ \hline EUR & 
1.6142 & 7.5324
& 0.8174 & 143.2853 & 1.2827\\ \hline \hline \end{tabular} \end{center}
\end{table}

\begin{table}[ht] \caption{Numerical values of DFA-$\alpha$ exponent for all
$EUR$-forming currency exchange rates; the single scaling time interval is
usually $ca.$ one year, except for $DKK$ having two distinguishable processes}
\begin{center} \begin{tabular}{||c||c||c|c||c||c||c||} \hline \hline $C_i /$ &
CHF & \multicolumn{2}{c||}{DKK}&GBP&JPY&USD\\ \hline $range$[weeks]& [1-52]
&[1-12] &[12-52] &[1-52] &[1-52] &[1-52] \\ \hline

ATS&$0.38 \pm0.02$& $0.22\pm0.01$&$0.48\pm0.04$ &$0.47\pm0.02$& $0.50 \pm
0.03$&$0.51\pm0.02$\\

BEF&$0.45 \pm0.02$& \multicolumn{2}{c||}{$0.37\pm0.03$} 
&$0.47\pm0.02$& $0.50 \pm
0.03$&$0.49\pm0.02$\\

DEM&$0.50 \pm0.02$& $0.41\pm0.03$&$0.58\pm0.05$ &$0.50\pm0.02$& $0.51 \pm
0.03$&$0.50\pm0.02$\\

ESP&$0.50 \pm0.02$& \multicolumn{2}{c||}{$0.47\pm0.02$} 
&$0.51\pm0.02$& $0.51 \pm
0.03$&$0.52\pm0.02$\\

FIM&$0.48 \pm0.02$& $0.40\pm0.01$&$0.48\pm0.03$ &$0.46\pm0.02$& $0.52 \pm
0.03$&$0.53\pm0.02$\\

FRF&$0.50 \pm0.02$& \multicolumn{2}{c||}{$0.41\pm0.03$} 
&$0.47\pm0.02$& $0.51 \pm
0.03$&$0.49\pm0.02$\\

IEP&$0.47 \pm0.03$& \multicolumn{2}{c||}{$0.48\pm0.01$} 
&$0.41\pm0.03$& $0.53 \pm
0.02$&$0.45\pm0.02$\\

ITL&$0.56 \pm0.03$& $0.46\pm0.02$&$0.60\pm0.04$ &$0.46\pm0.02$& $0.53 \pm
0.02$&$0.51\pm0.02$\\

NLG&$0.47 \pm0.02$& $0.32\pm0.02$&$0.53\pm0.05$ &$0.49\pm0.02$& $0.50 \pm
0.03$&$0.50\pm0.02$\\

PTE&$0.41 \pm0.03$& \multicolumn{2}{c||}{$0.29\pm0.02$} 
&$0.45\pm0.02$& $0.48 \pm
0.03$&$0.48\pm0.02$ \\ \hline EUR&$0.51 \pm0.03$& $0.45\pm0.02$&$0.46\pm0.02$
&$0.46\pm0.02$& $0.52 \pm 0.03$&$0.50\pm0.02$ \\ \hline \hline \end{tabular}
\end{center} \end{table}

\begin{figure}[ht]
\begin{center}
\leavevmode
\epsfysize=6cm
\epsffile{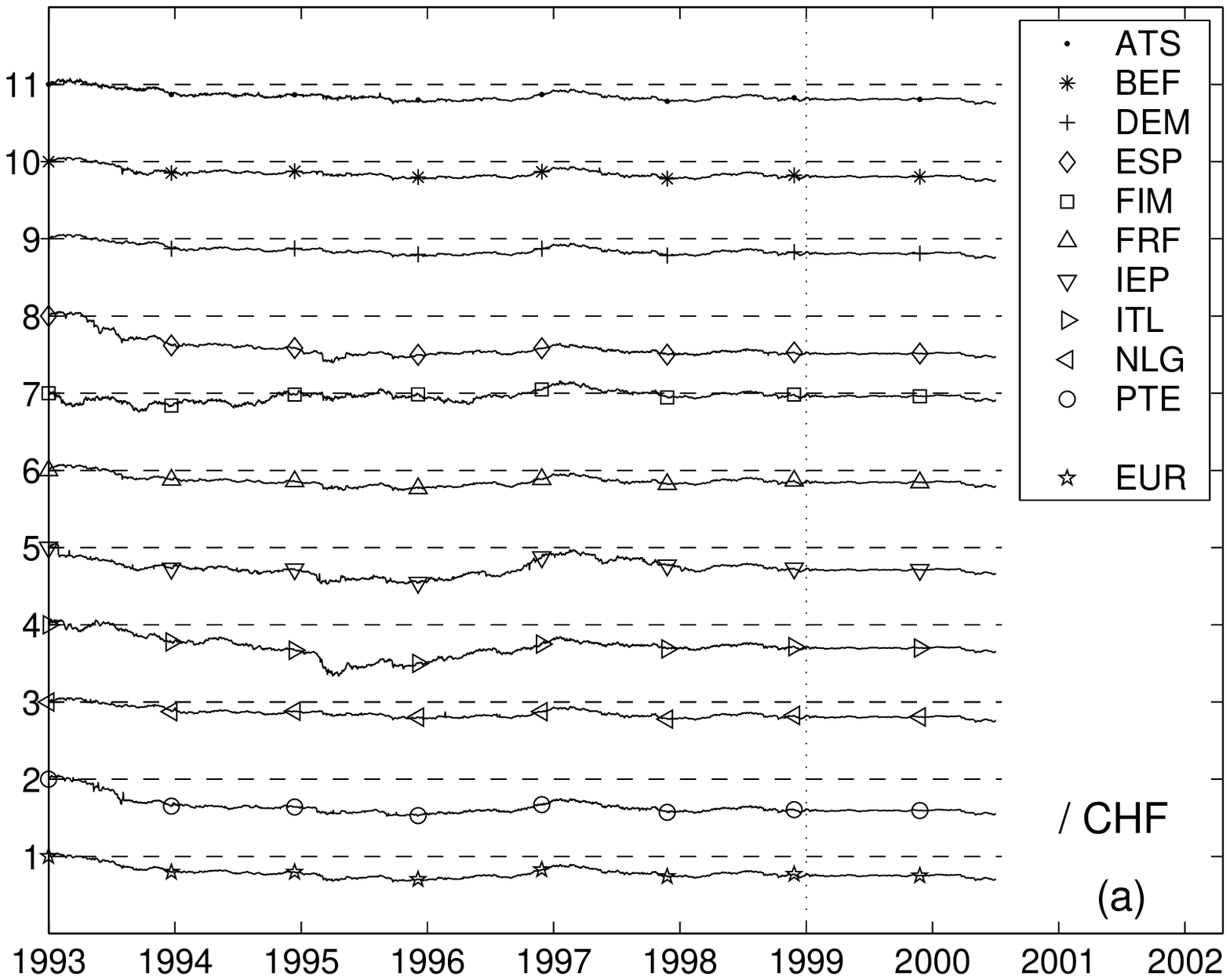}
\hfill
\leavevmode
\epsfysize=6cm
\epsffile{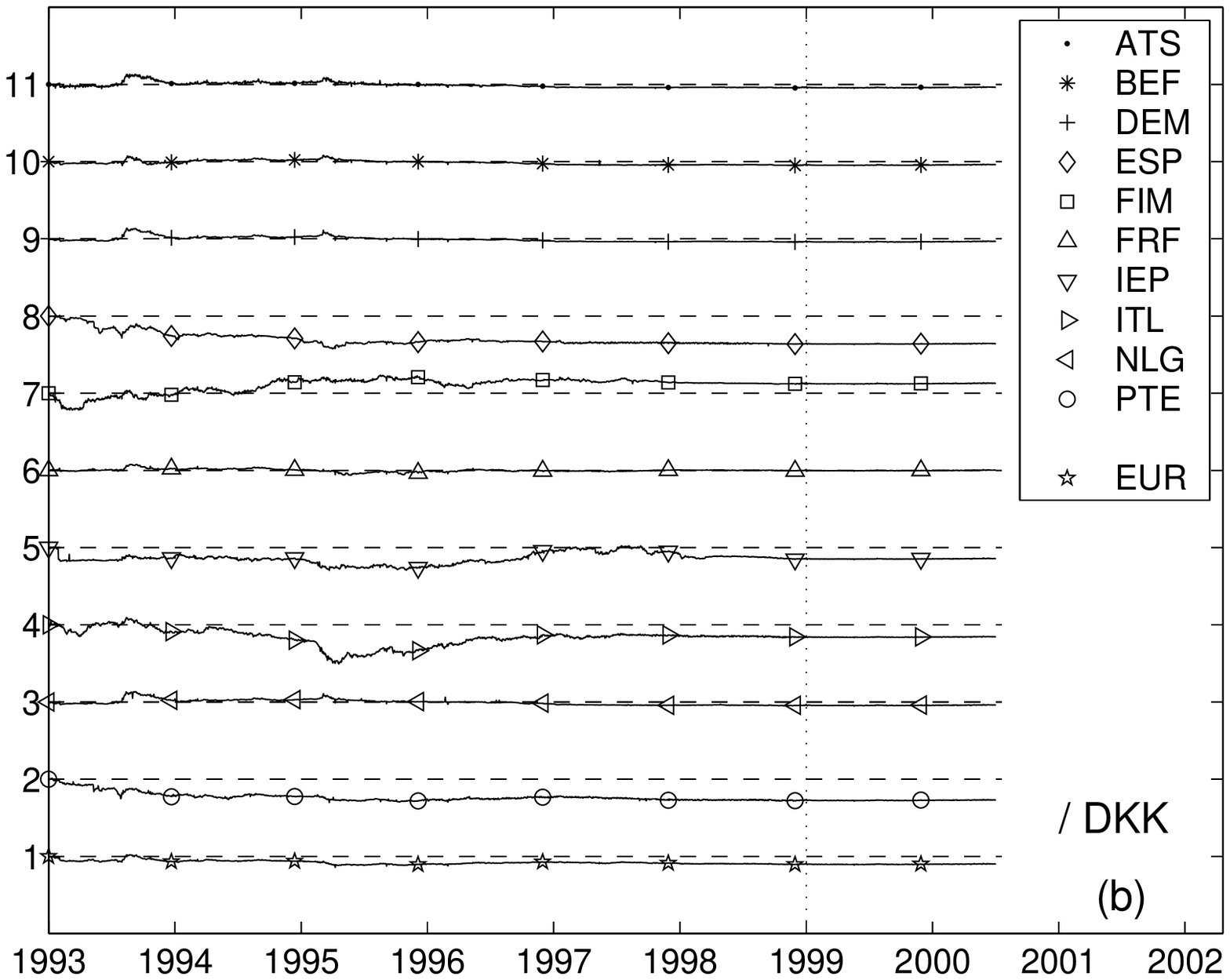}
\vfill
\leavevmode
\epsfysize=6cm
\epsffile{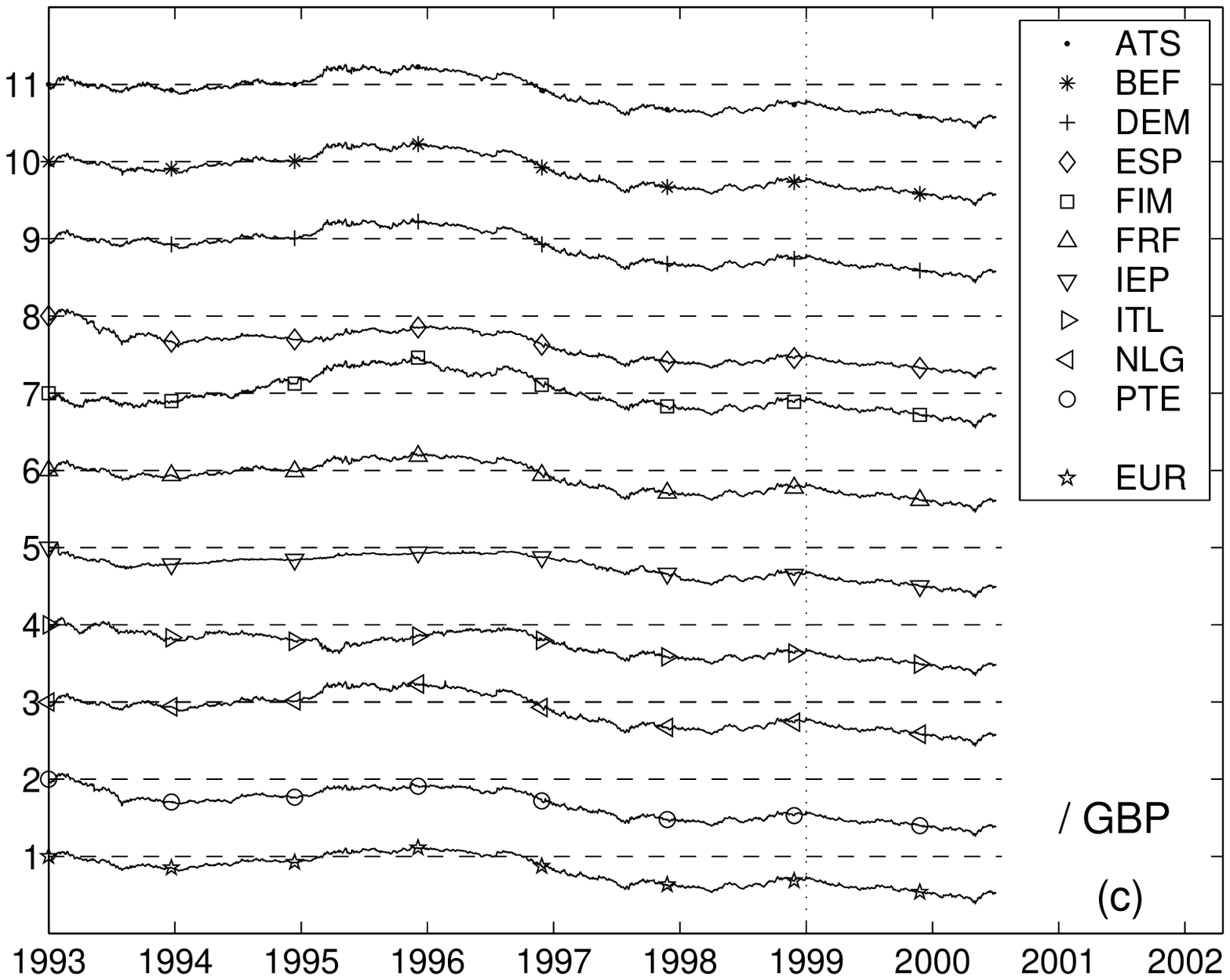}
\hfill
\leavevmode
\epsfysize=6cm
\epsffile{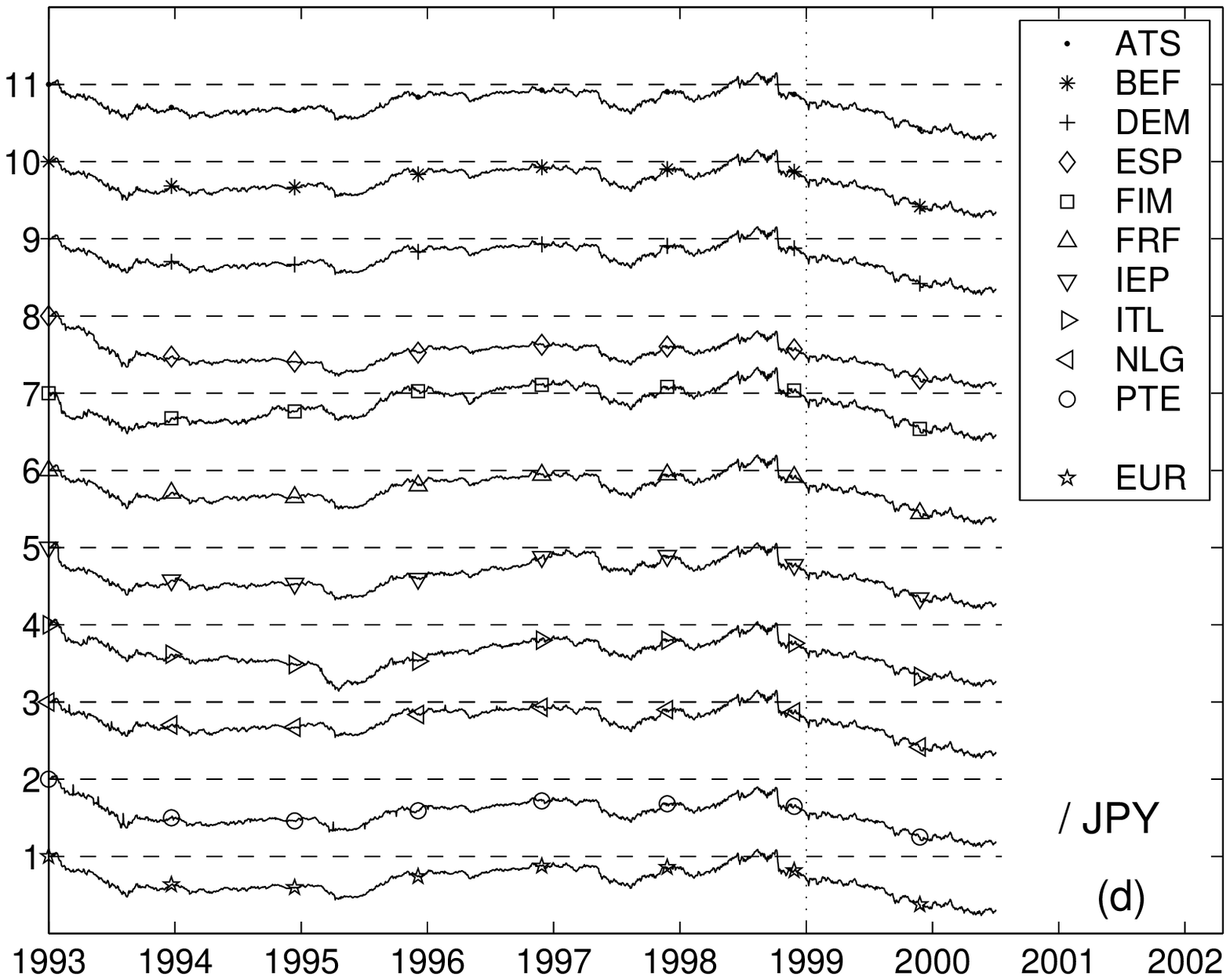}
\vfill
\leavevmode
\epsfysize=6cm
\epsffile{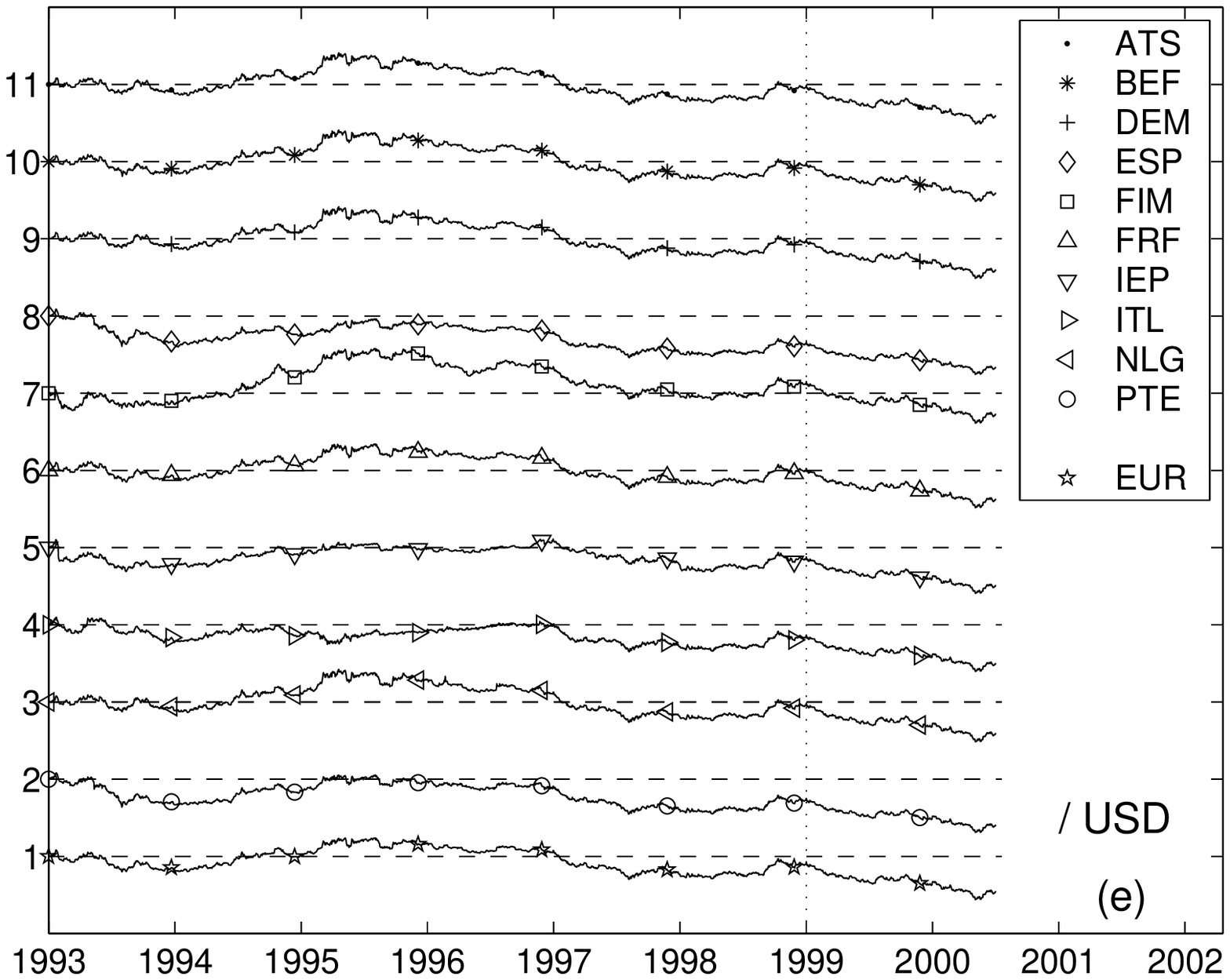}
\end{center}
\caption{Normalized $EUR$ and currency forming
the $EUR$ exchange rates with respect to the (a-e) $CHF$, $DKK$, 
$GBP$, $JPY$ and
$USD$ between Jan. 01, 1993 and June 30, 2000. The data are artificially
multiplied by two and then displaced along the vertical axis in order 
to make the
fluctuations noticeable.
} \label{fig1} \end{figure}

\begin{figure}[ht]
\begin{center}
\leavevmode
\epsfysize=6cm
\epsffile{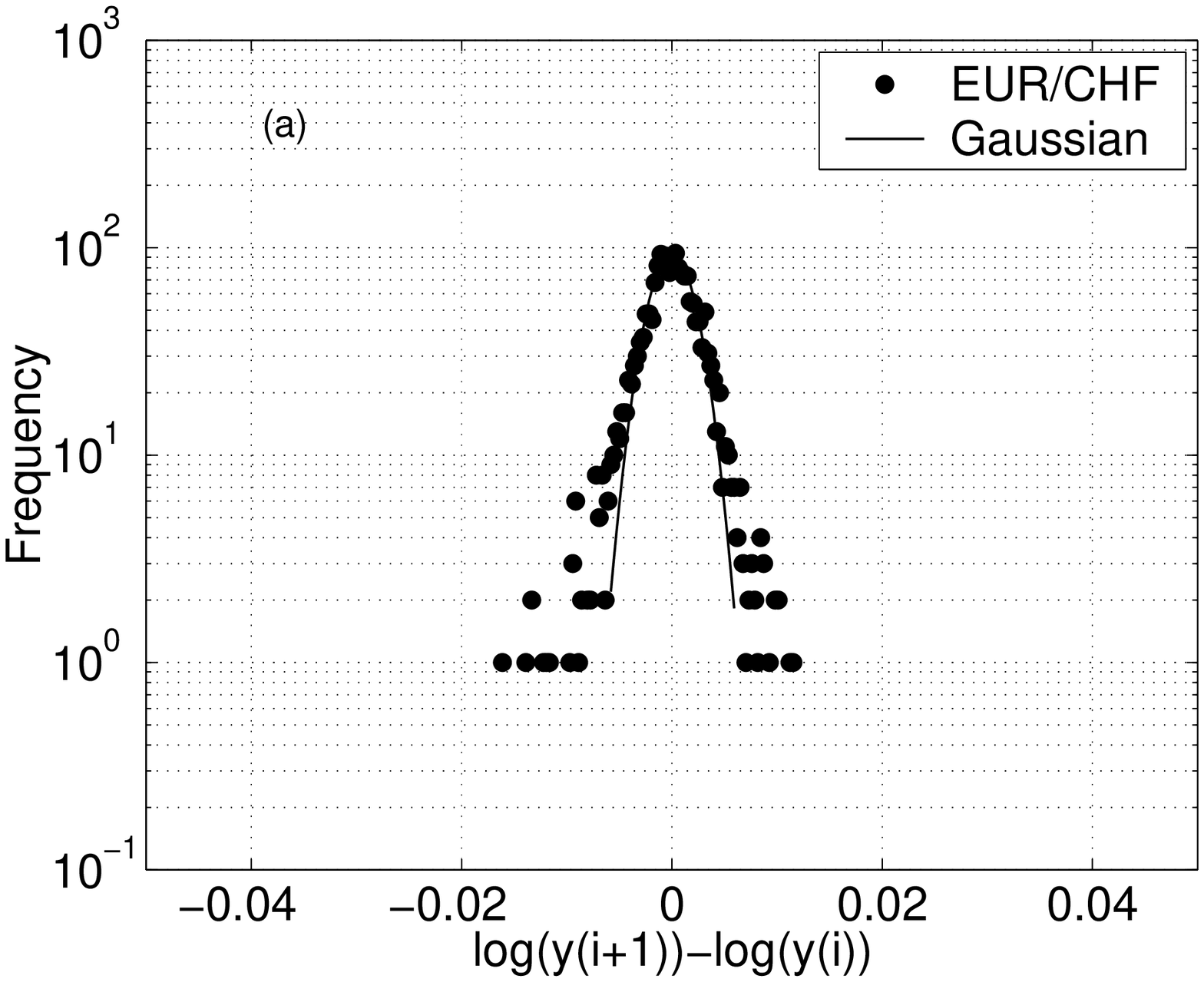}
\hfill
\leavevmode
\epsfysize=6cm
\epsffile{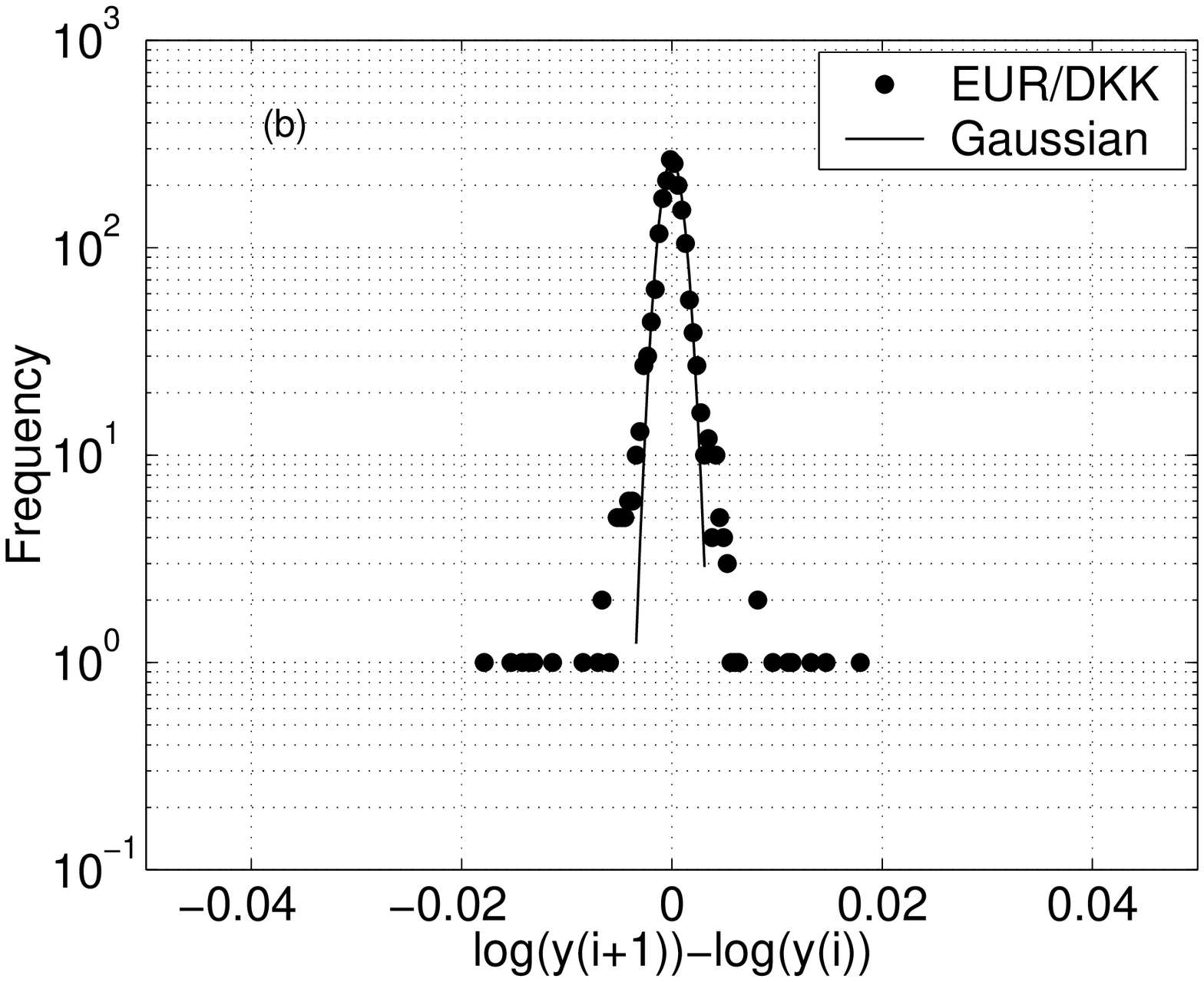}
\vfill
\leavevmode
\epsfysize=6cm
\epsffile{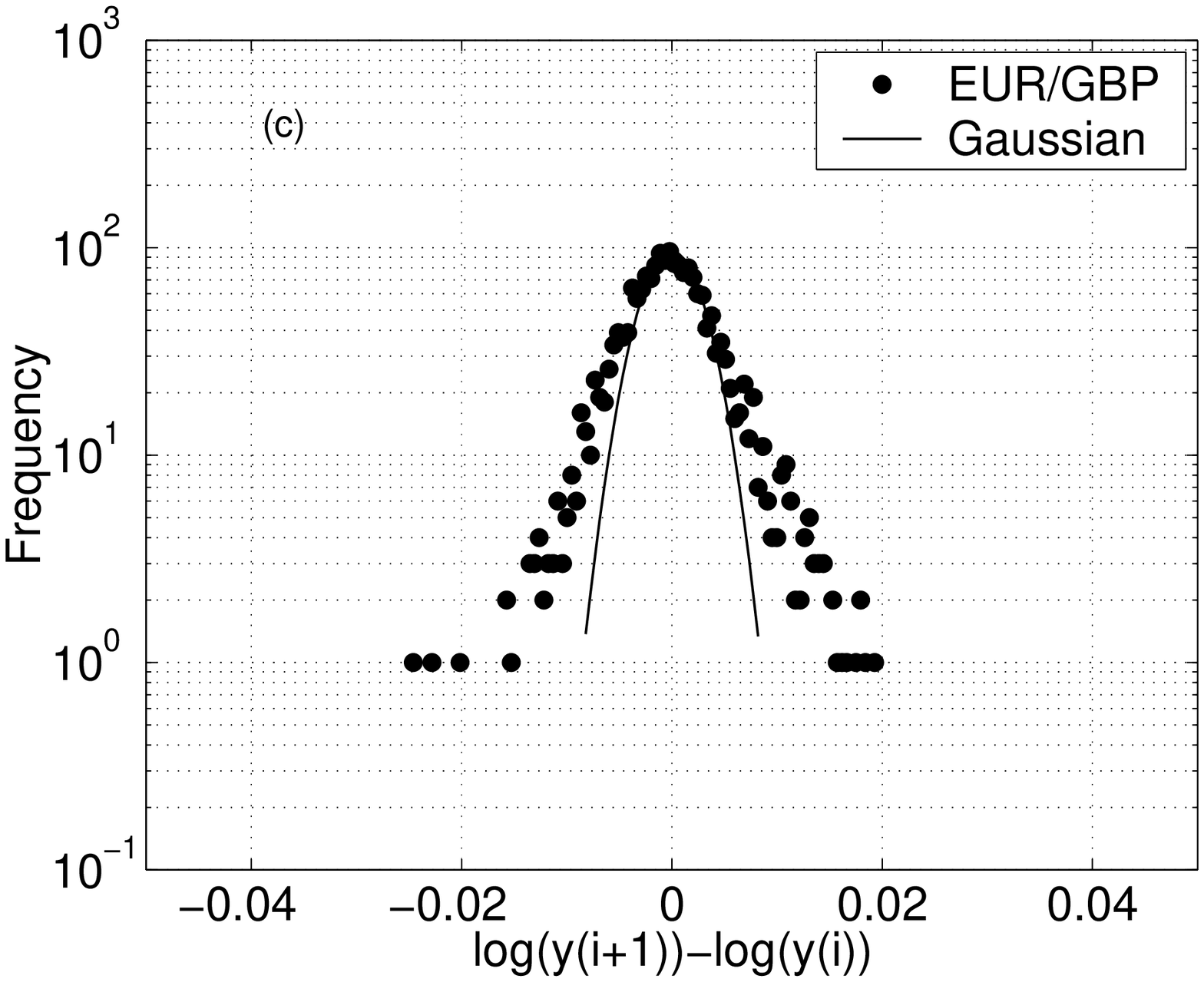}
\hfill
\leavevmode
\epsfysize=6cm
\epsffile{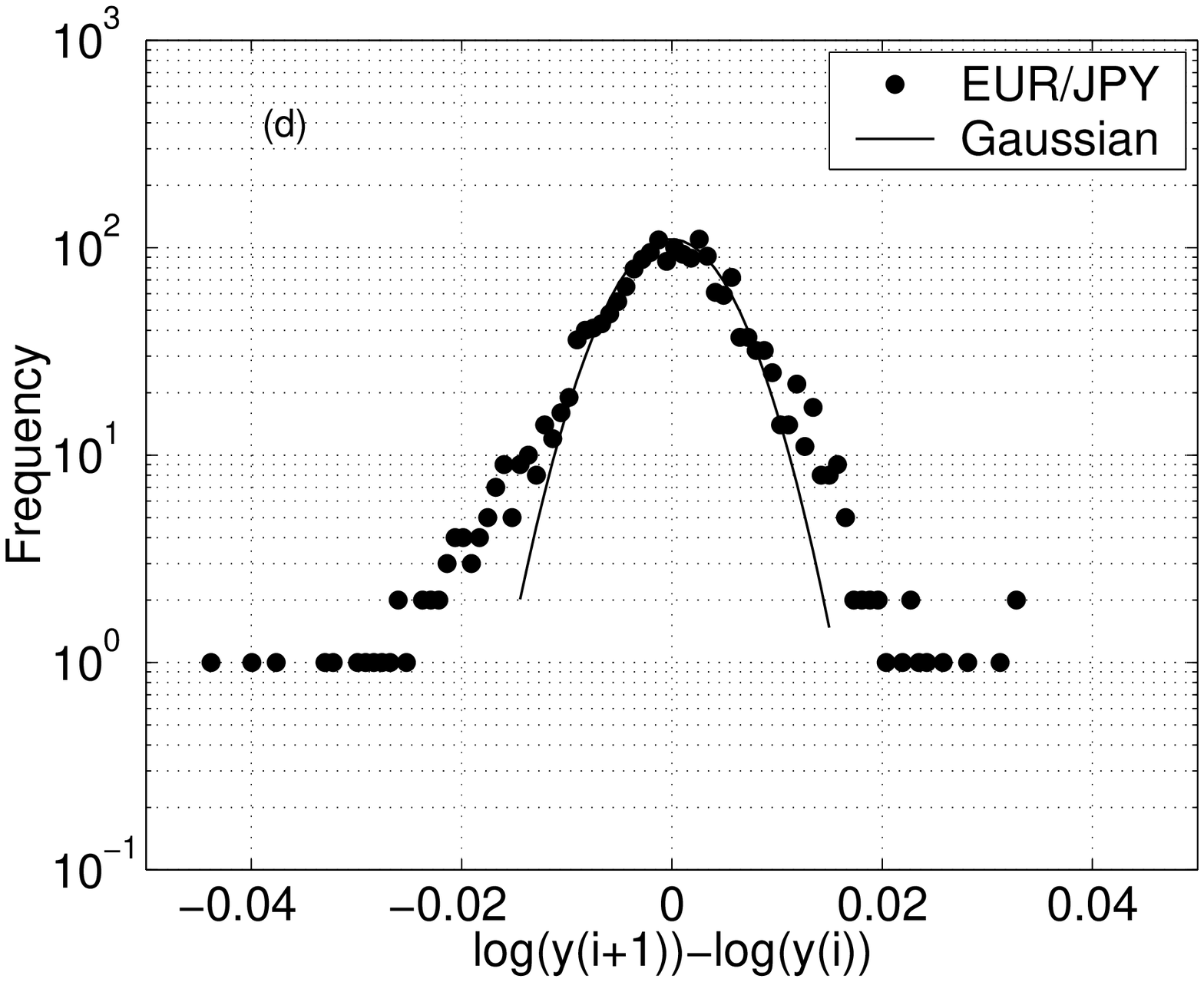}
\vfill
\leavevmode
\epsfysize=6cm
\epsffile{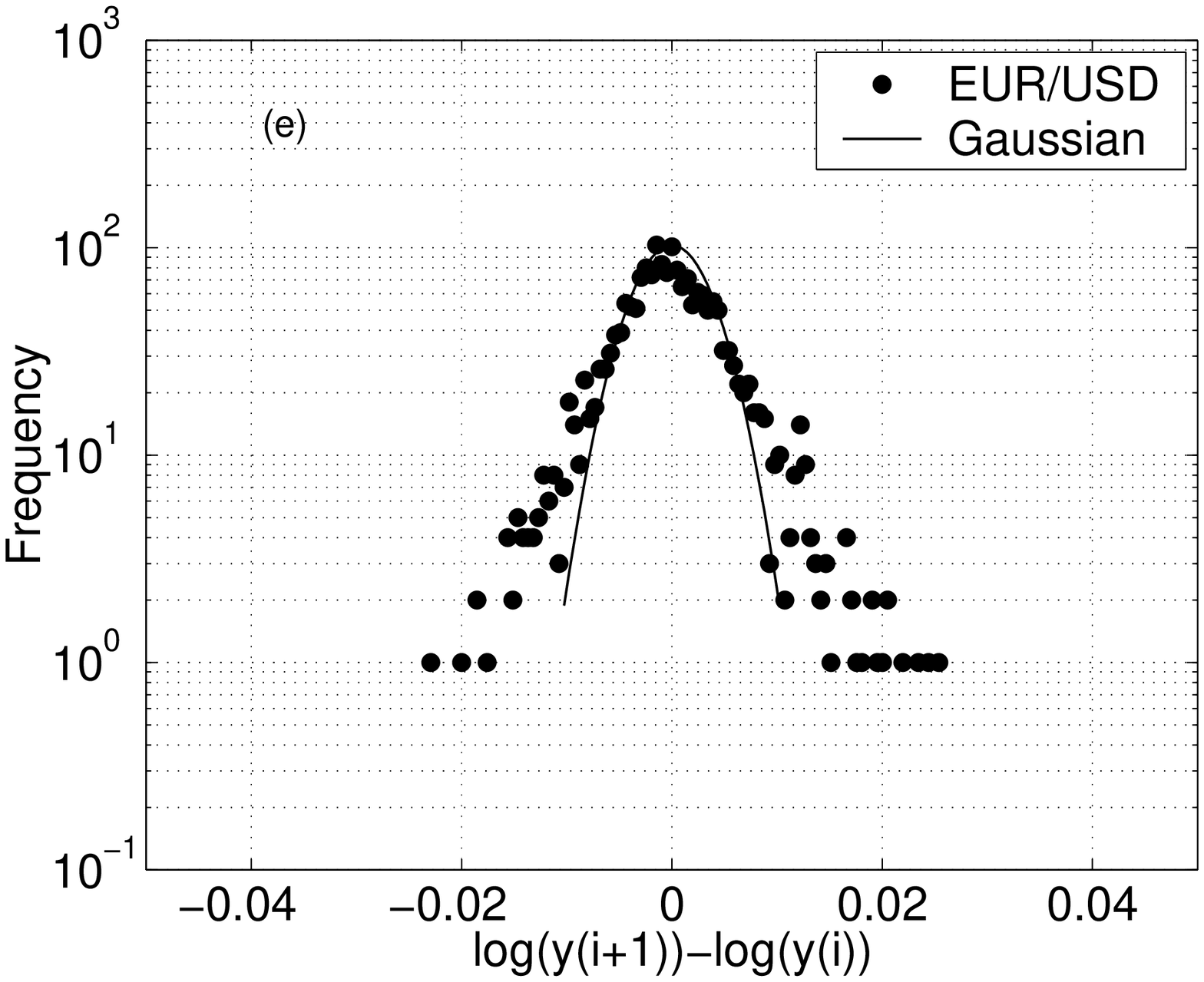}
\end{center}
\caption{Distributions of the exchange rate
fluctuations for (a-e) $EUR/CHF$, $EUR/DKK$, $EUR/GBP$, $EUR/JPY$, 
and $EUR/USD$
for data in Fig. 1 (a-e).
} \label{fig2} 
\end{figure}

\begin{figure}[ht]
\begin{center}
\leavevmode
\epsfysize=6cm
\epsffile{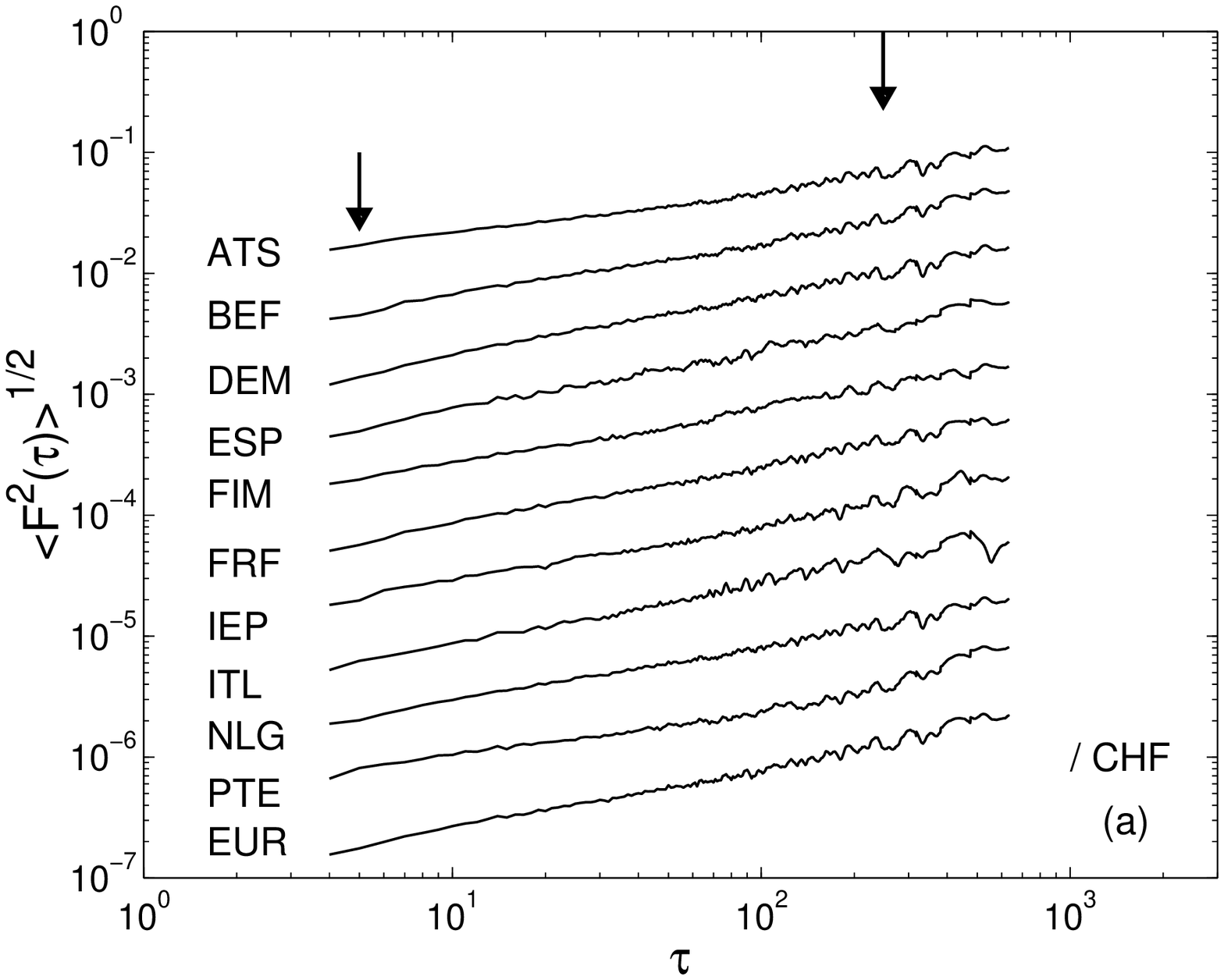}
\hfill
\leavevmode
\epsfysize=6cm
\epsffile{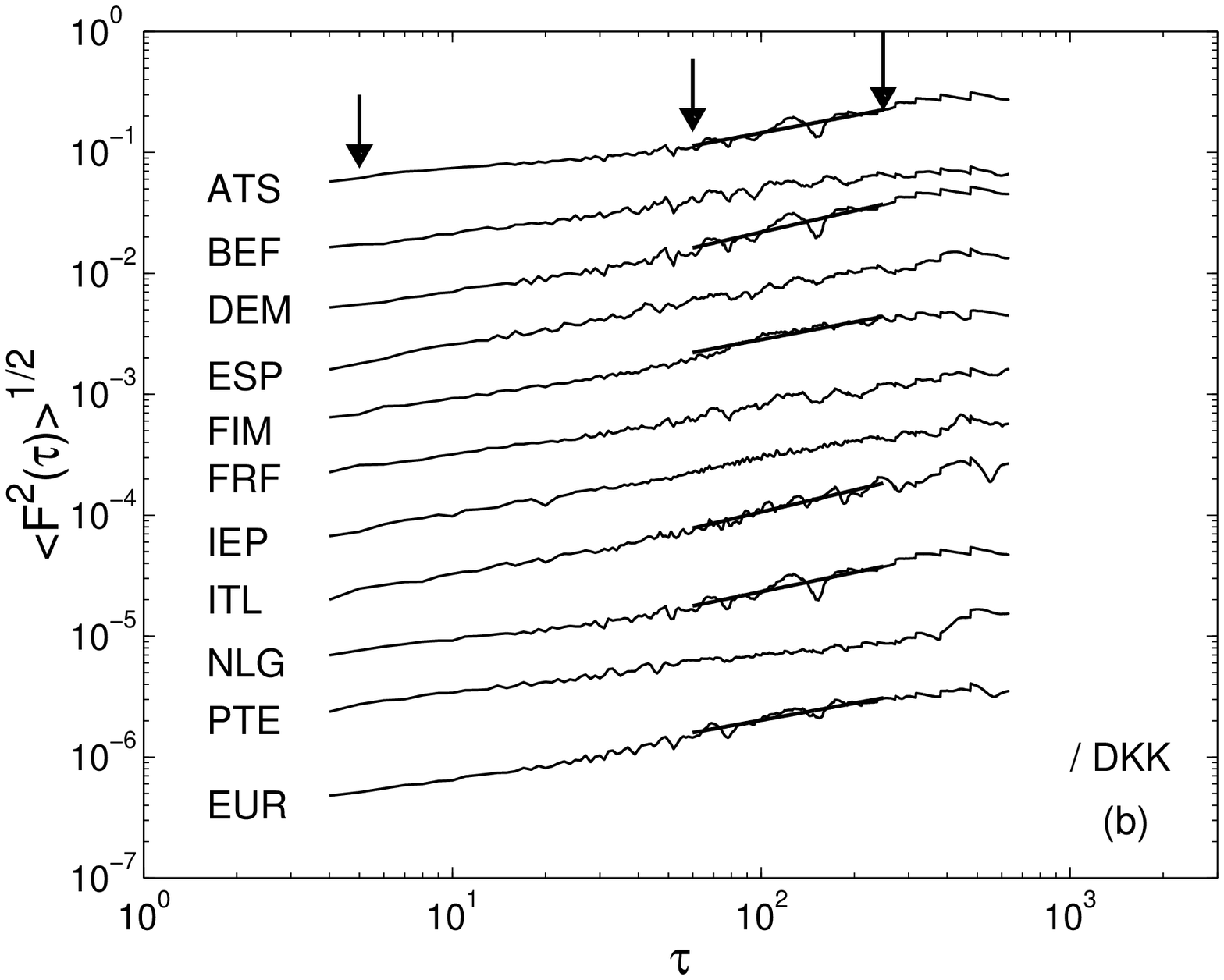}
\vfill
\leavevmode
\epsfysize=6cm
\epsffile{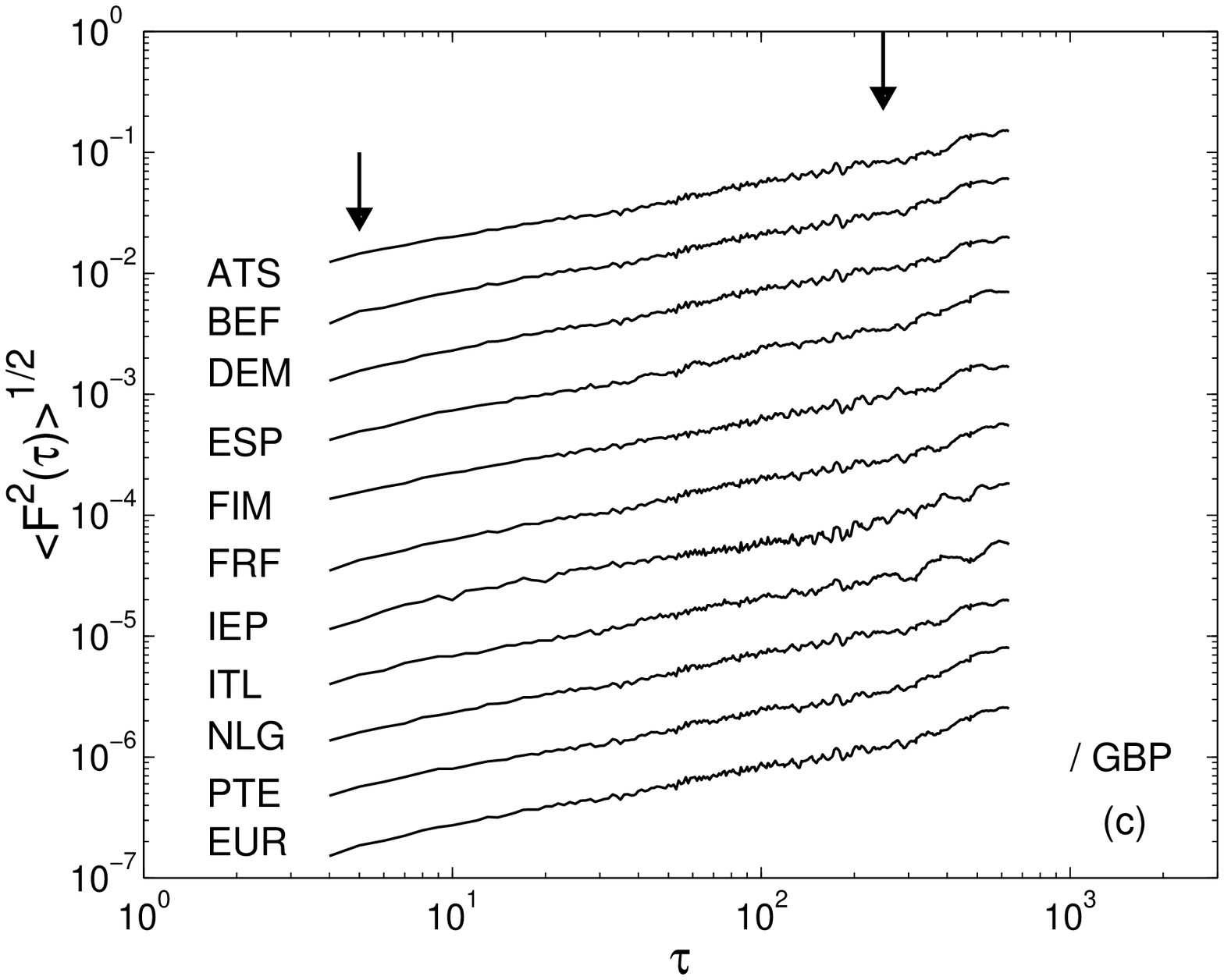}
\hfill
\leavevmode
\epsfysize=6cm
\epsffile{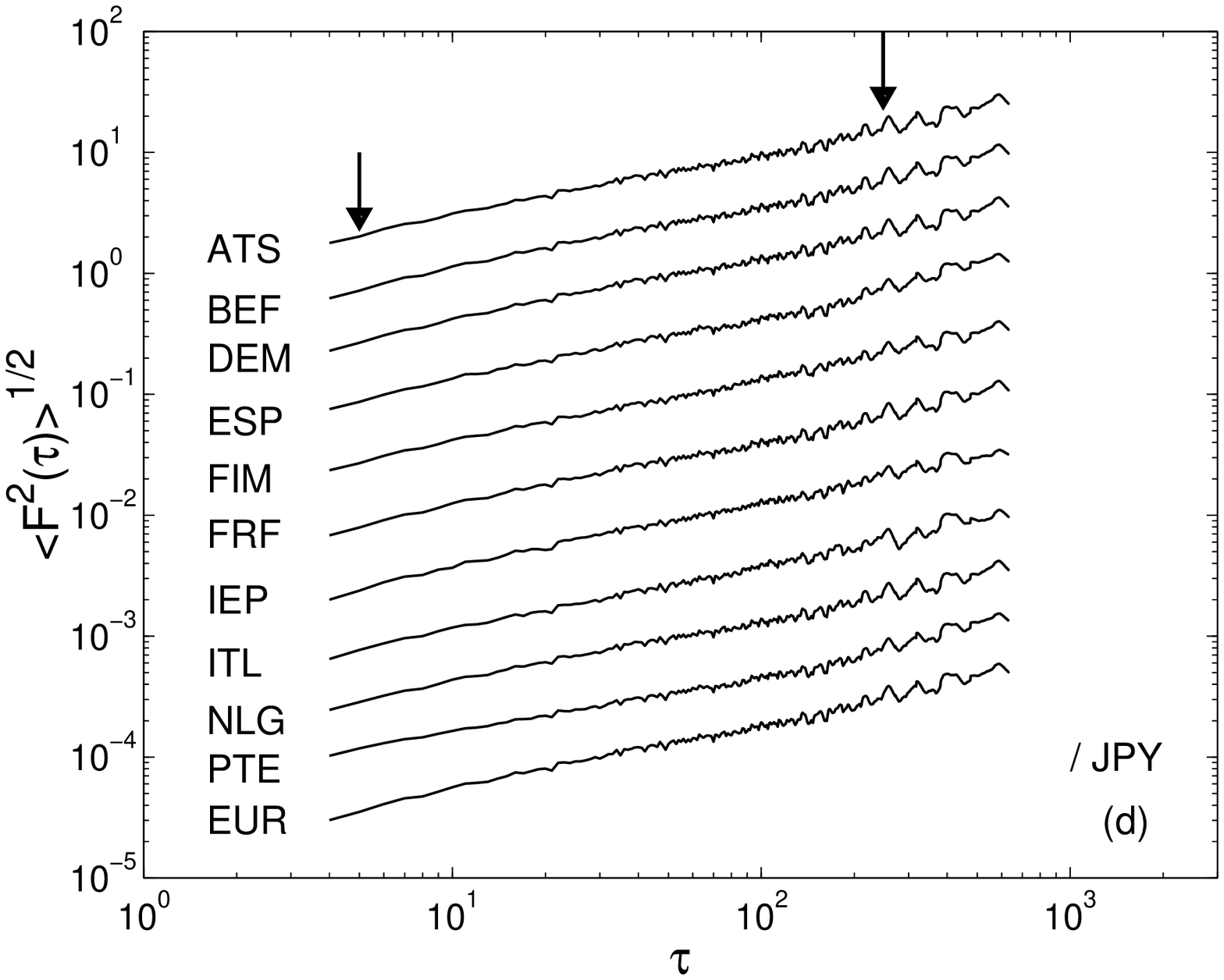}
\vfill
\leavevmode
\epsfysize=6cm
\epsffile{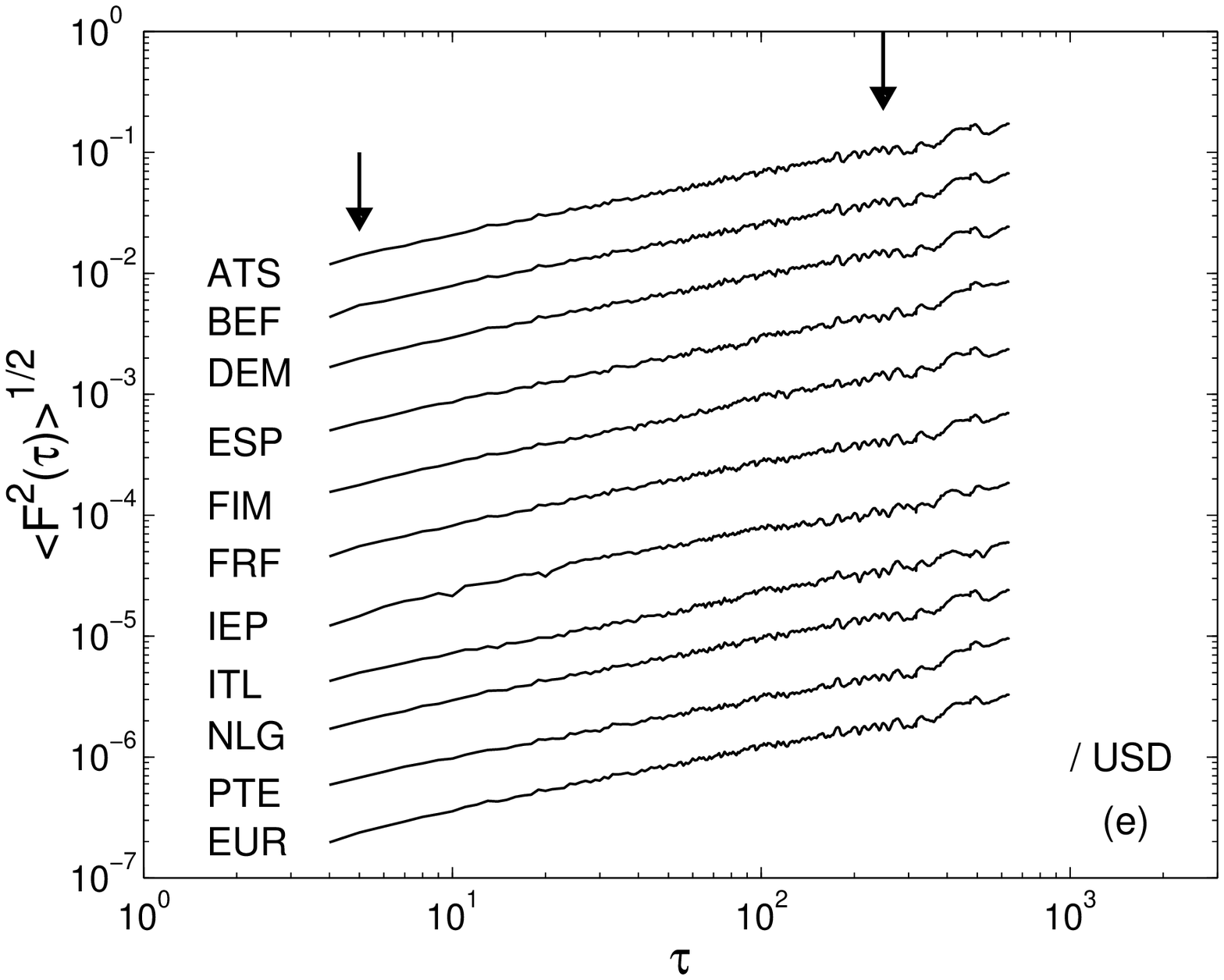}
\end{center}
\caption{Log-log plot of the DFA function
showing how to obtain the $\alpha$ exponent for the 11 exchange rates 
of interest
for (a-e) $EUR/CHF$, $EUR/DKK$, $EUR/GBP$, $EUR/JPY$, and $EUR/USD$. The fit
slope being only of interest, the DFA function data has been arbitrarily
displaced along the vertical axis. The arrows indicate the best scaling ranges.}
\label{fig3} \end{figure}

\begin{figure}[ht]
\begin{center}
\leavevmode
\epsfysize=6cm
\epsffile{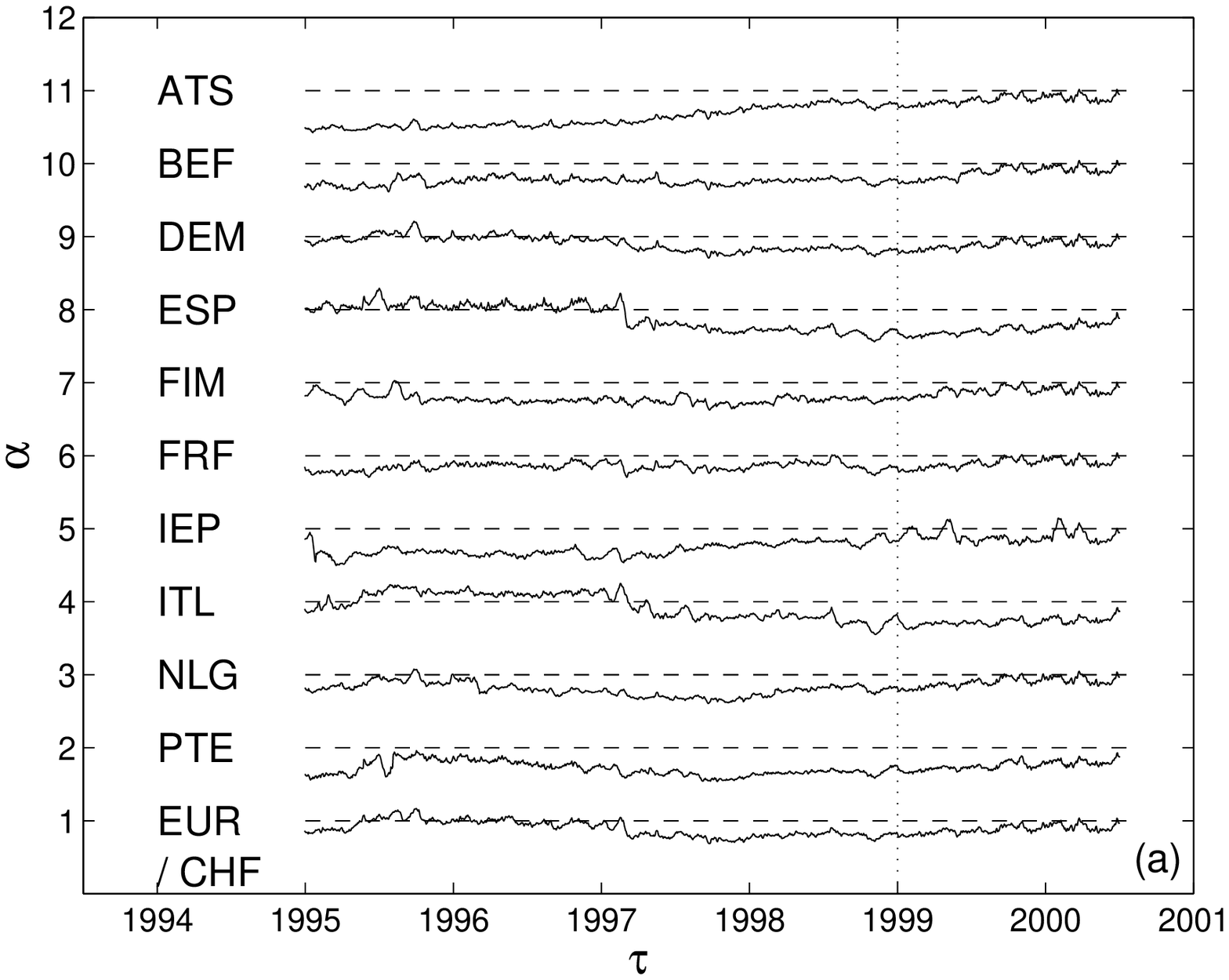}
\hfill
\leavevmode
\epsfysize=6cm
\epsffile{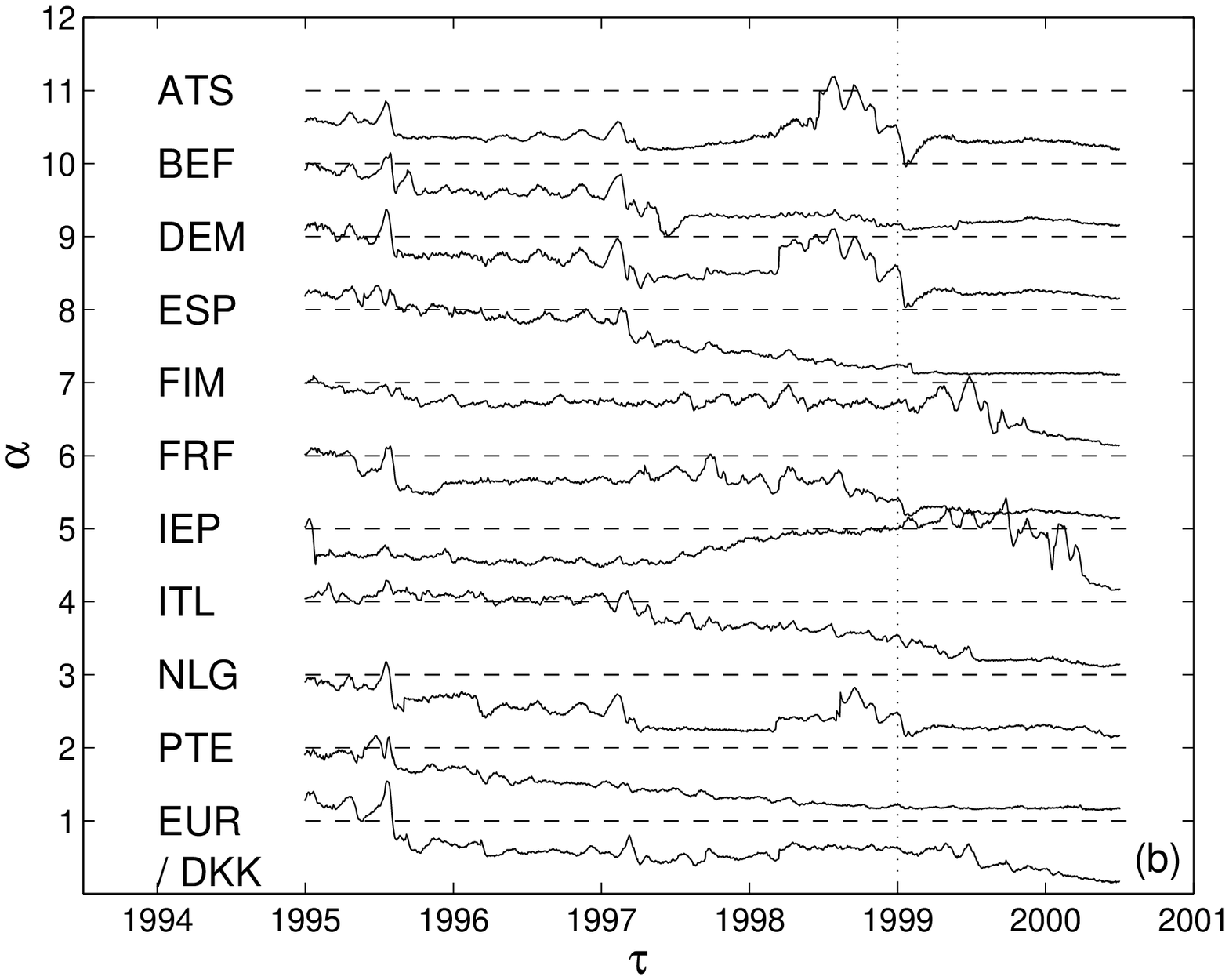}
\vfill
\leavevmode
\epsfysize=6cm
\epsffile{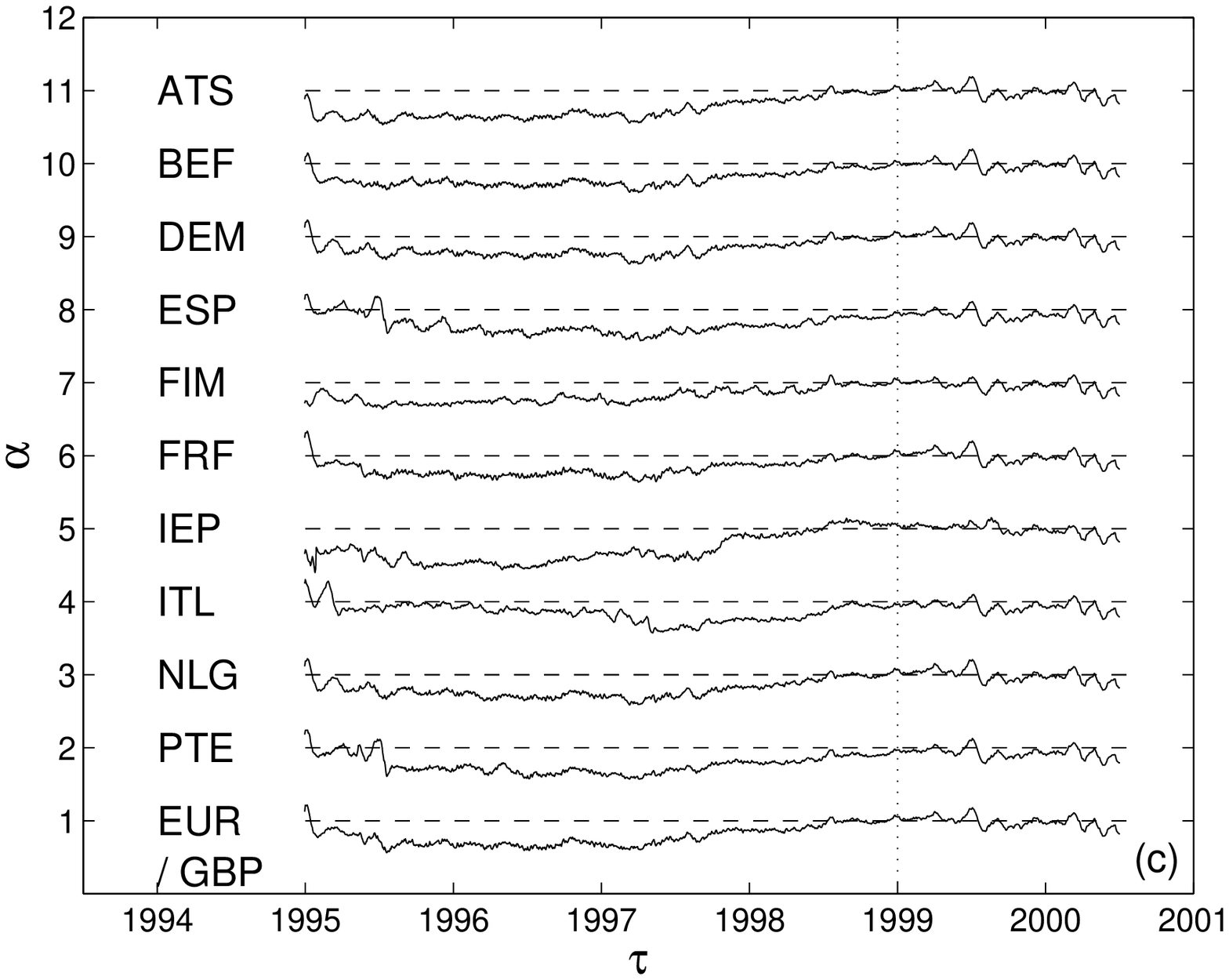}
\hfill
\leavevmode
\epsfysize=6cm
\epsffile{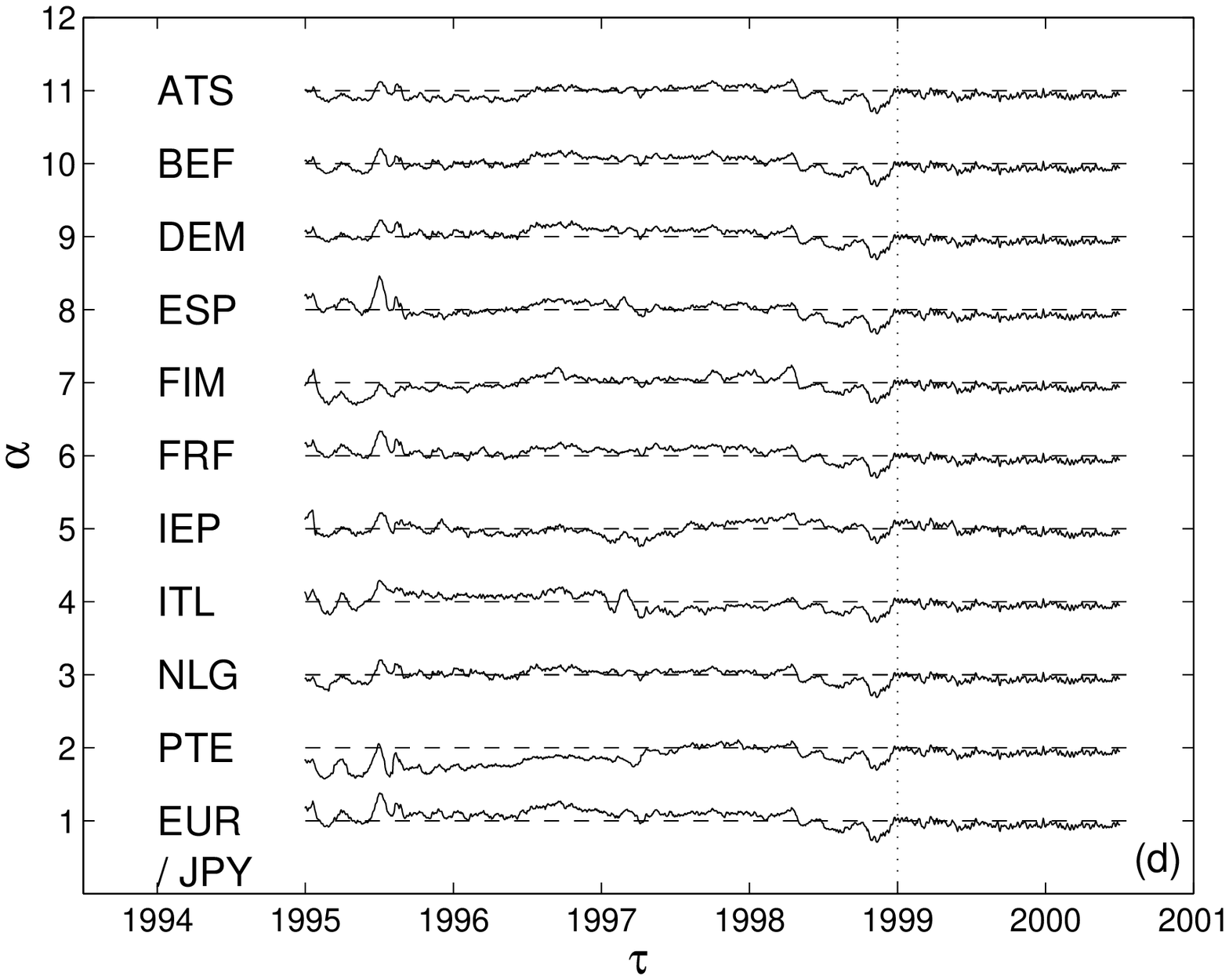}
\vfill
\leavevmode
\epsfysize=6cm
\epsffile{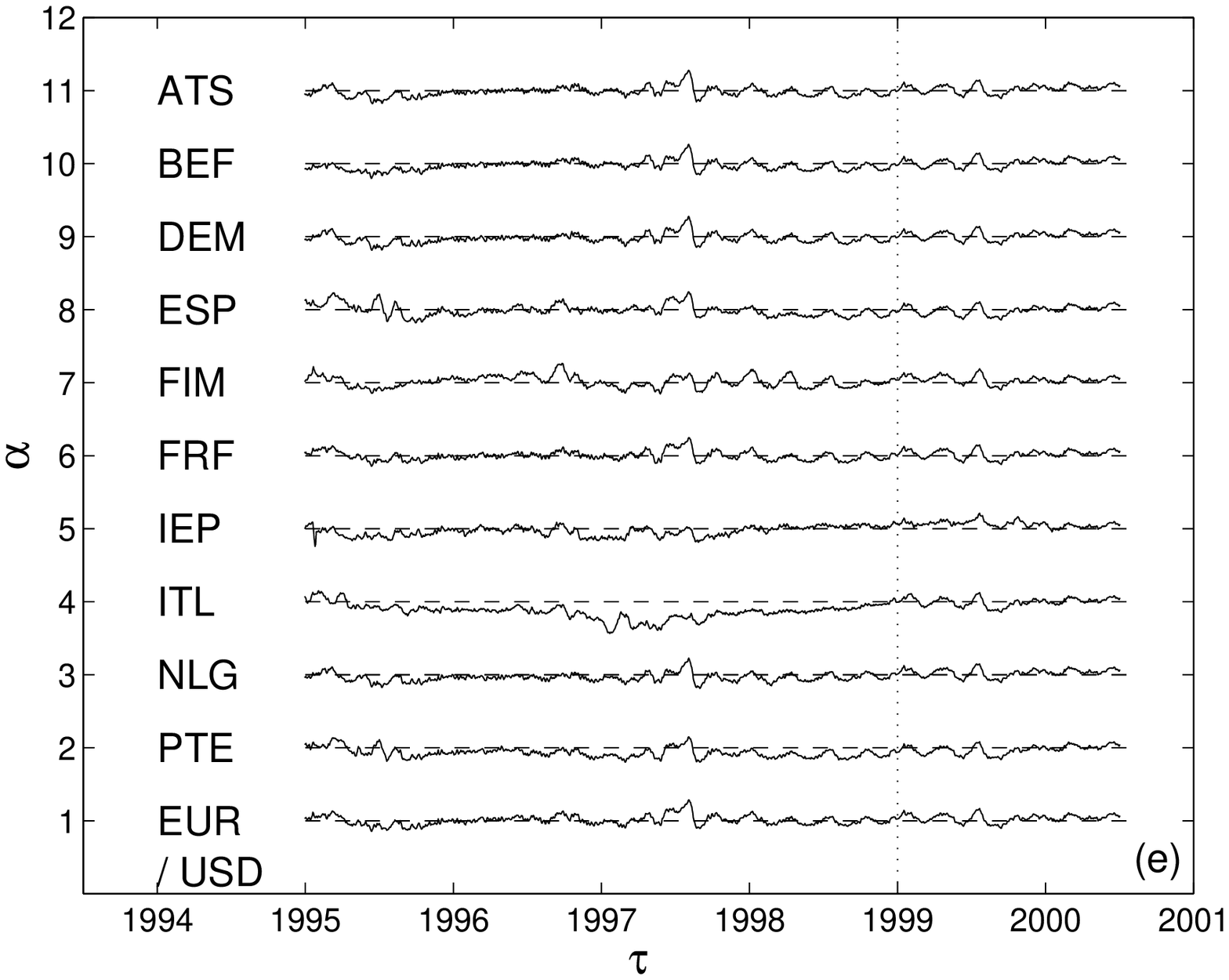}
\end{center}
\caption{Time dependence of the DFA $local$
$\alpha$-exponent for $EUR$ and each currency (which forms the $EUR$) exchange
rate with respect to (a-e) $DKK$, $CHF$, $GBP$, $JPY$, and $USD$. The
$\alpha$-values are artificially multiplied by two and then displaced along the
vertical axis in order to make the fluctuations noticeable. For each time
dependent $\alpha$ a horizontal dashed line is drawn to indicate a reference to
Brownian fluctuations.
} \label{fig4} \end{figure}

\begin{figure}[ht]
\begin{center}
\leavevmode
\epsfysize=6cm
\epsffile{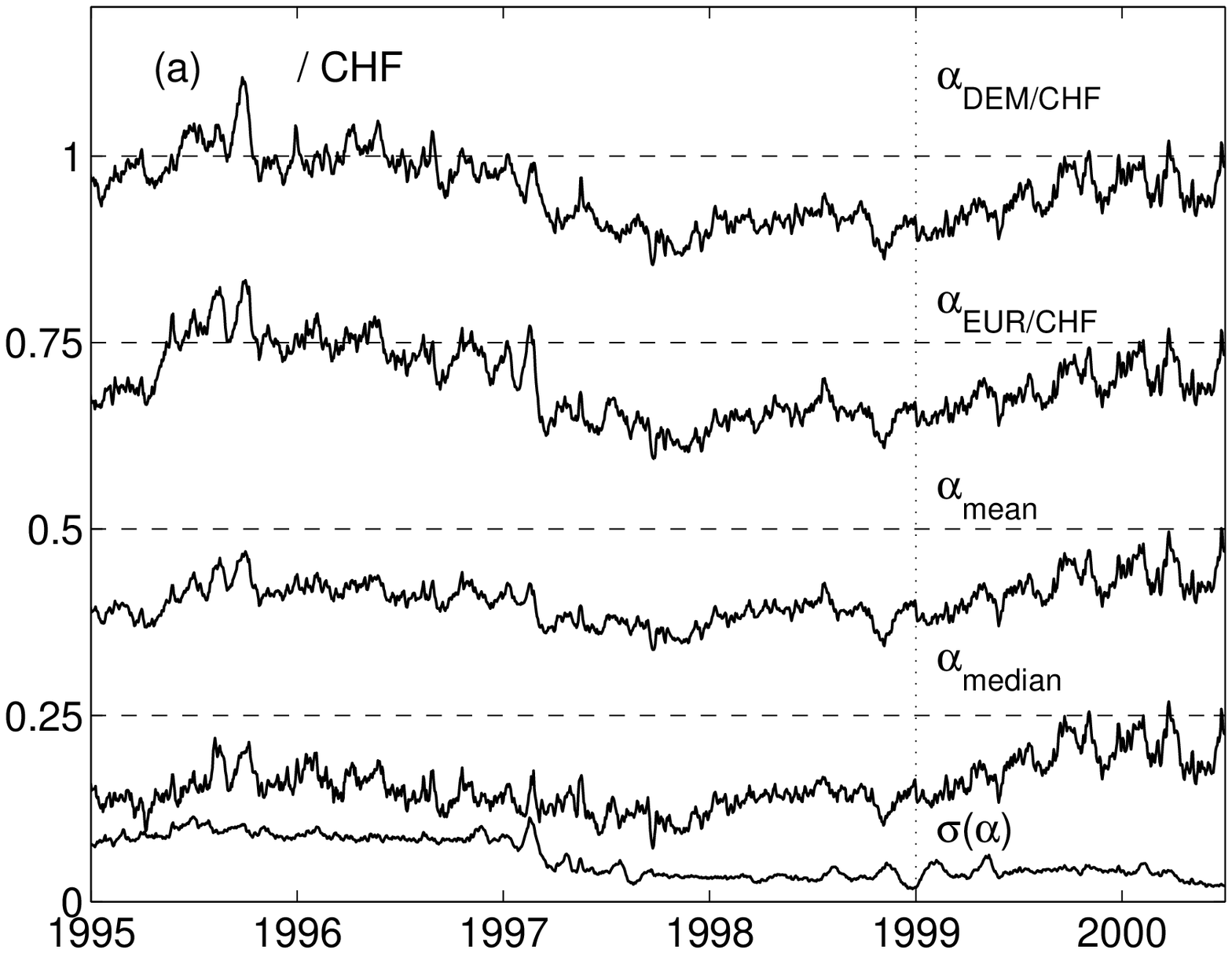}
\hfill
\leavevmode
\epsfysize=6cm
\epsffile{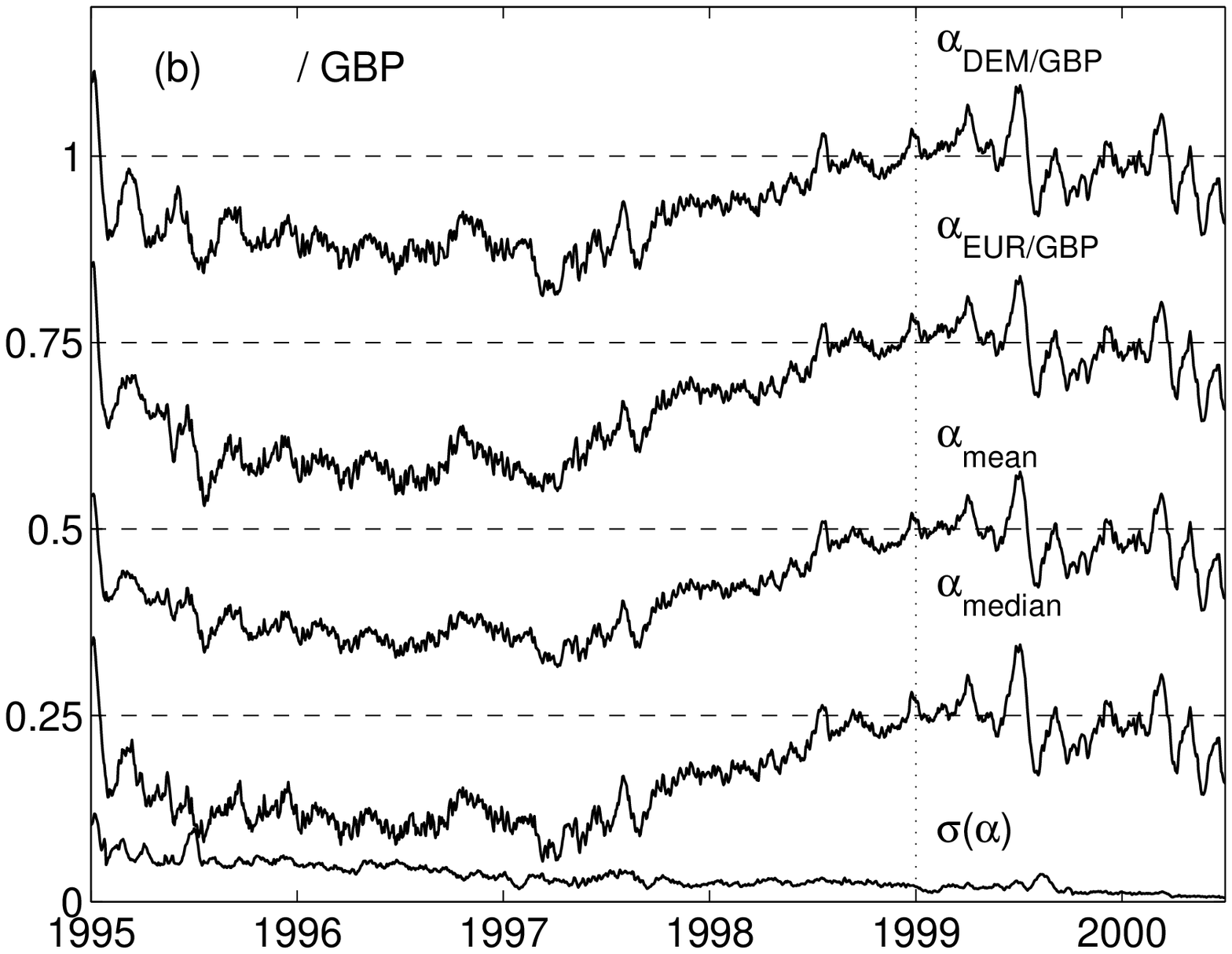}
\vfill
\leavevmode
\epsfysize=6cm
\epsffile{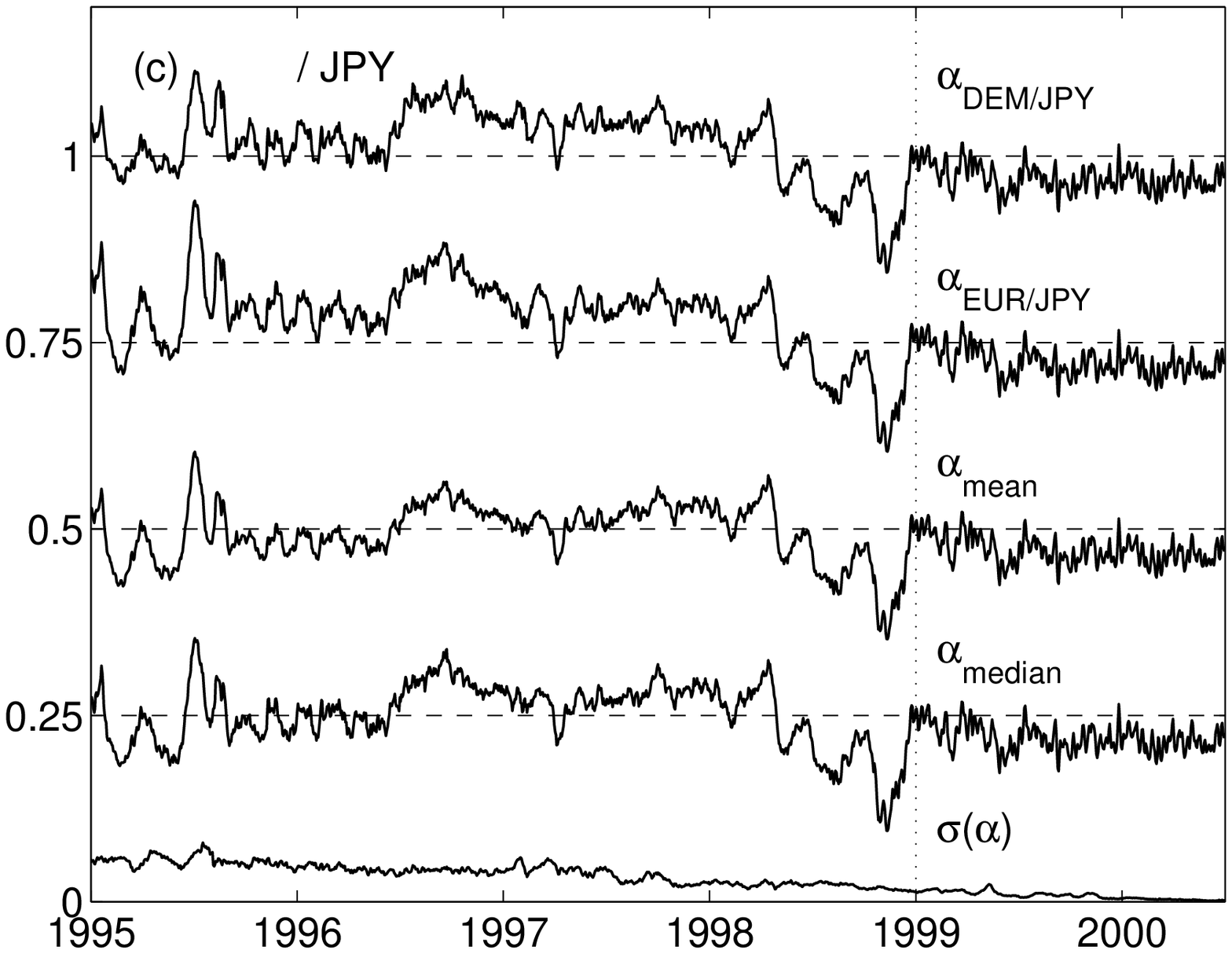}
\hfill
\leavevmode
\epsfysize=6cm
\epsffile{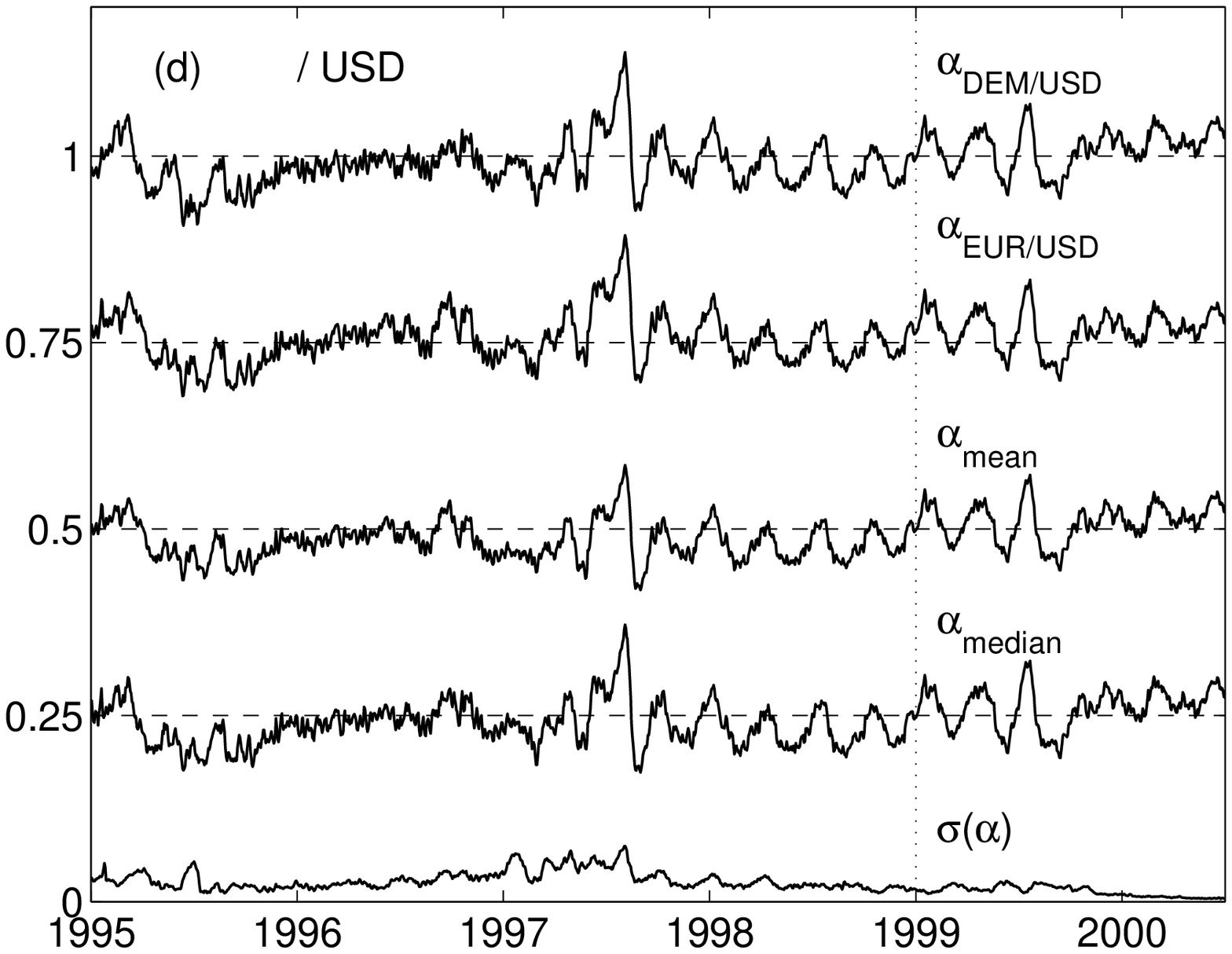}
\end{center}
\caption{Time evolution of the mean, 
median and
standard deviation of the $\alpha$ exponents for the currencies forming the
$EUR$, compared to that of the $EUR$ and $DEM$ for $CHF$, $GBP$, 
$JPY$ and $USD$
as reference currencies (a-d). The $\alpha_{mean}$ and $\sigma(\alpha)$ curves
are not displaced. The $\alpha_{median}$ curves are displaced by -0.25. The
$\alpha_{EUR/}$ curves are displaced by +0.25 and the $\alpha_{DEM/}$ 
curves are
displaced by +0.5 in (a-d). Horizontal dashed lines mark Brownian motion 0.5
level for each $\alpha$-curve.} \label{fig5} \end{figure}

\begin{figure} 
\begin{center}
\leavevmode
\epsfysize=8cm
\epsffile{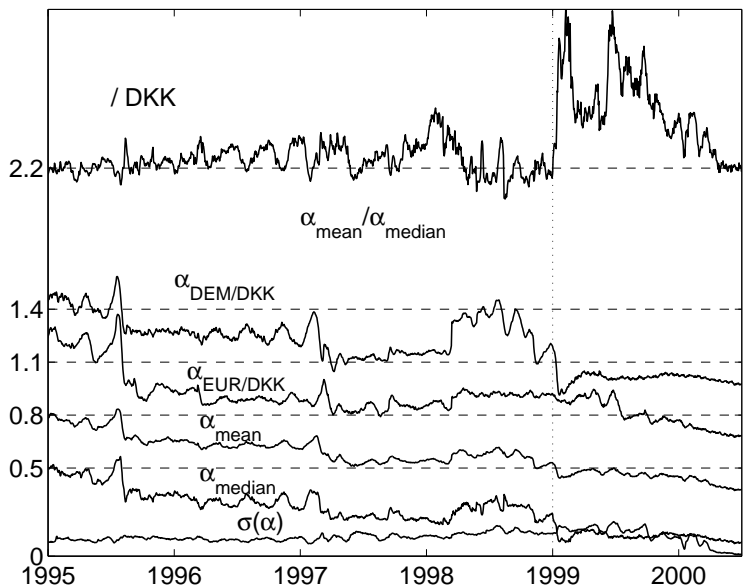}
\end{center}
\caption{Time evolution of the mean, 
median and
standard deviation of the $\alpha$ exponent for the currency exchange 
rates with
respect to $DKK$, currencies forming the $EUR$, compared to that of 
the $EUR$ and
$DEM$. The curves are displaced for readability though the y-axis scale is
constant and the $\alpha_{median}$ and $\sigma(\alpha)$ curves are 
not displaced.
The $\alpha_{mean}$ curve is displaced by +0.3. The $\alpha_{EUR/DKK}$ curve is
displaced by +0.6 and the $\alpha_{DEM/DKK}$ curve is displaced by +0.9. The
horizontal dashed lines mark Brownian motion 0.5 level for each $\alpha$ curve.
The $\alpha_{mean}/\alpha_{median}$ ratio extracted from the $\alpha$ exponents
is also shown. Horizontal dashed line at y=2.2 corresponds to
$\alpha_{mean}/\alpha_{median}=1$.} \label{fig6} \end{figure}

\newpage
\begin{figure}[ht]
\begin{center}
\leavevmode
\epsfysize=6cm
\epsffile{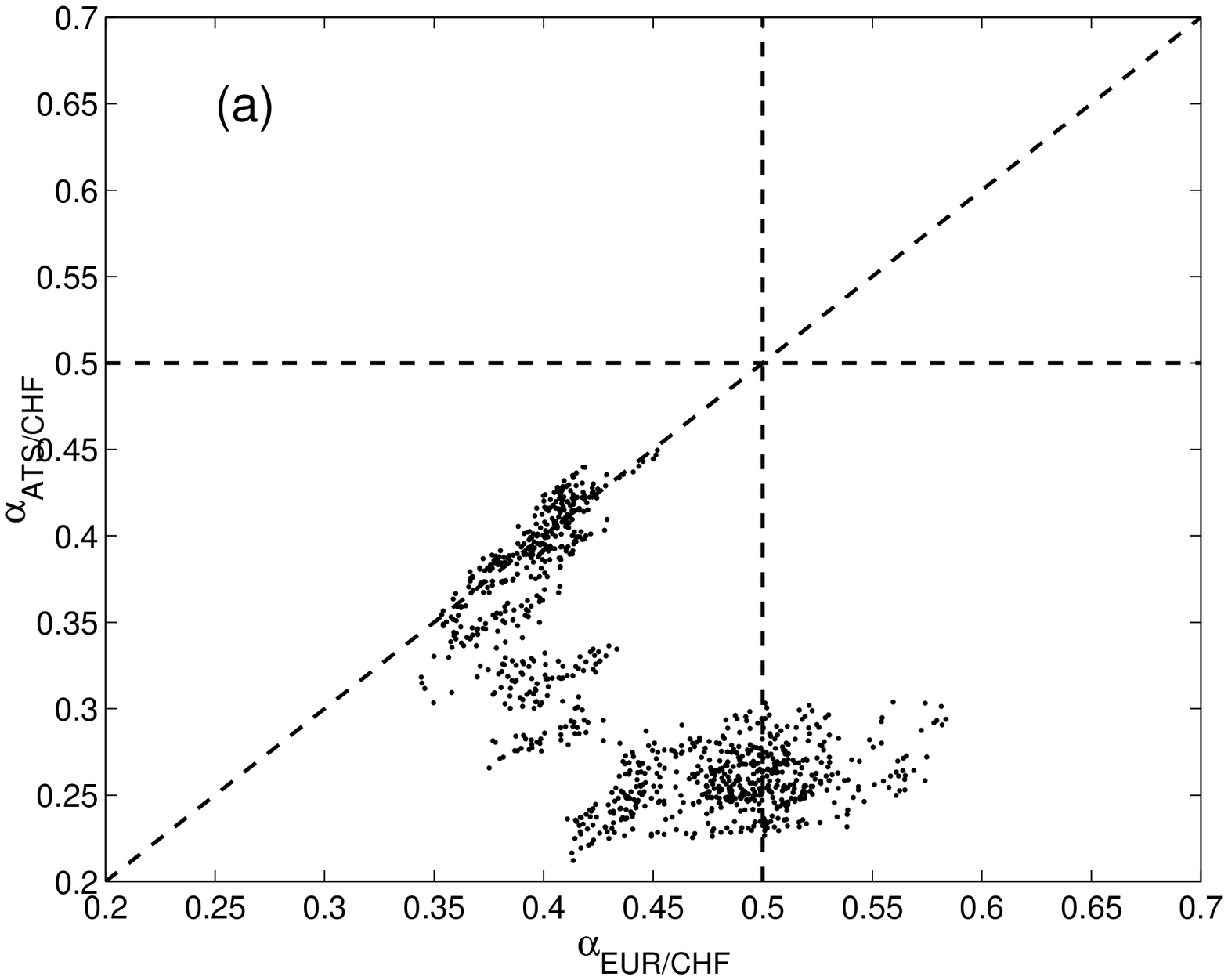}
\hfill
\leavevmode
\epsfysize=6cm
\epsffile{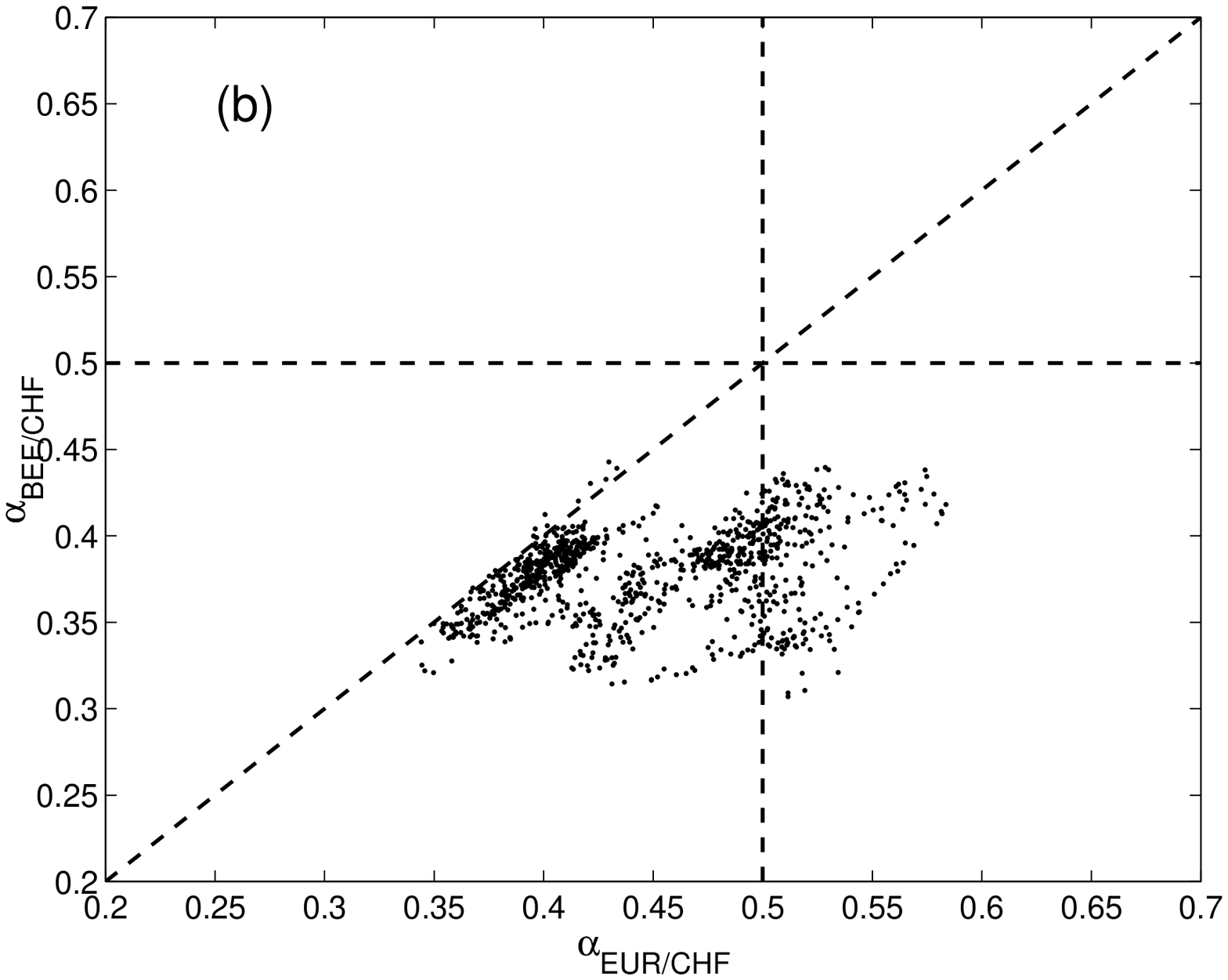}
\vfill
\leavevmode
\epsfysize=6cm
\epsffile{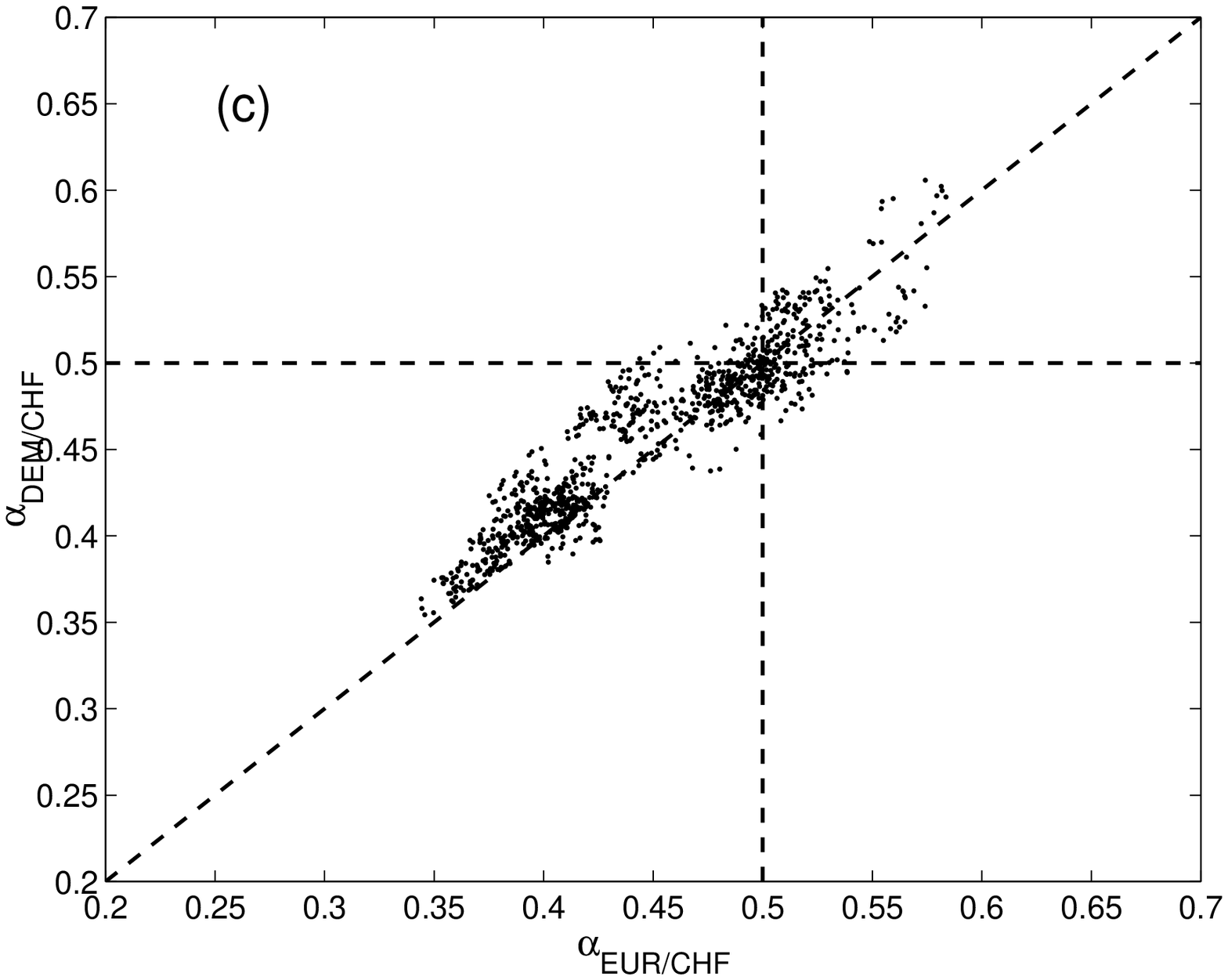}
\hfill
\leavevmode
\epsfysize=6cm
\epsffile{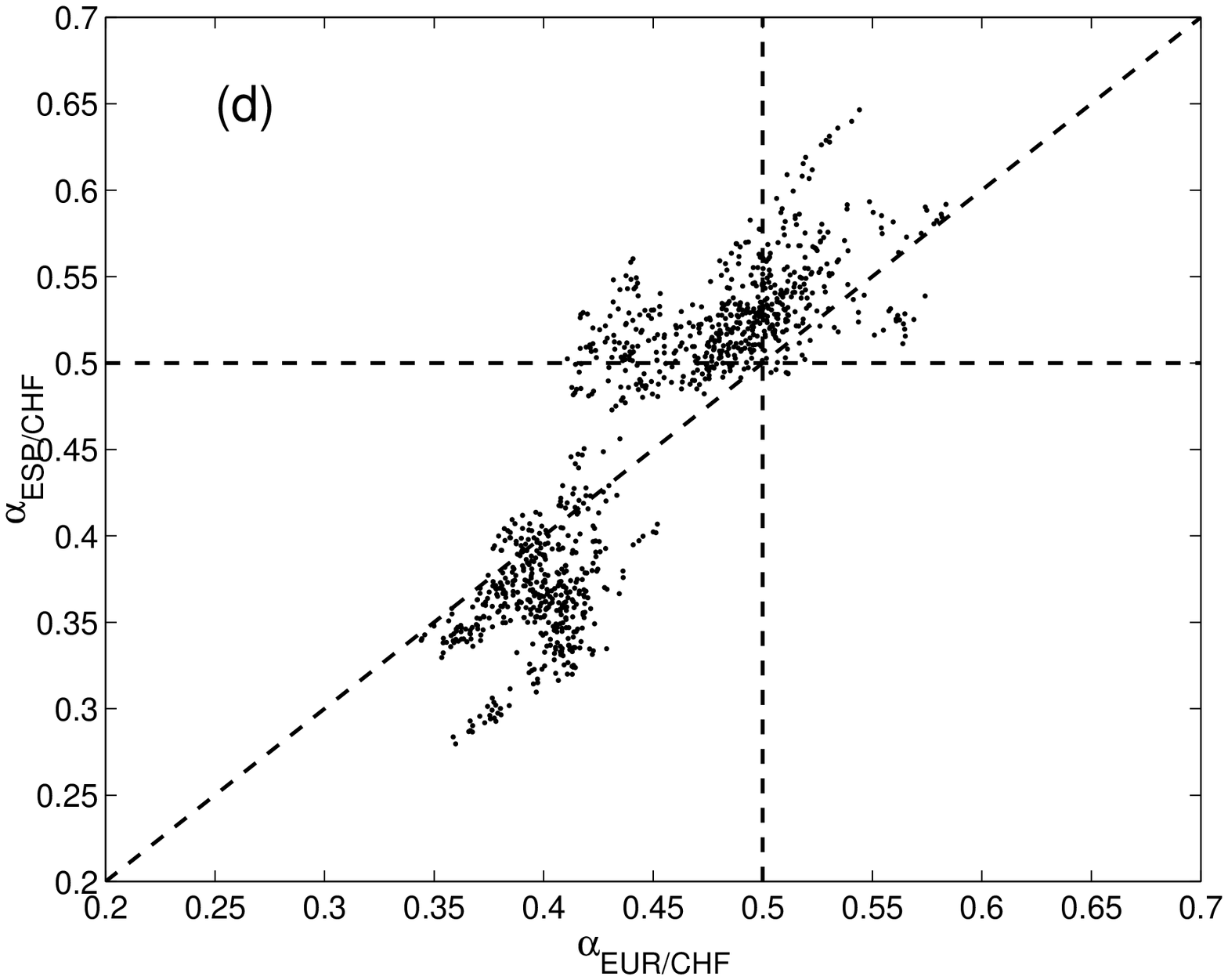}
\end{center}

\newpage
\begin{center}
\leavevmode
\epsfysize=6cm
\epsffile{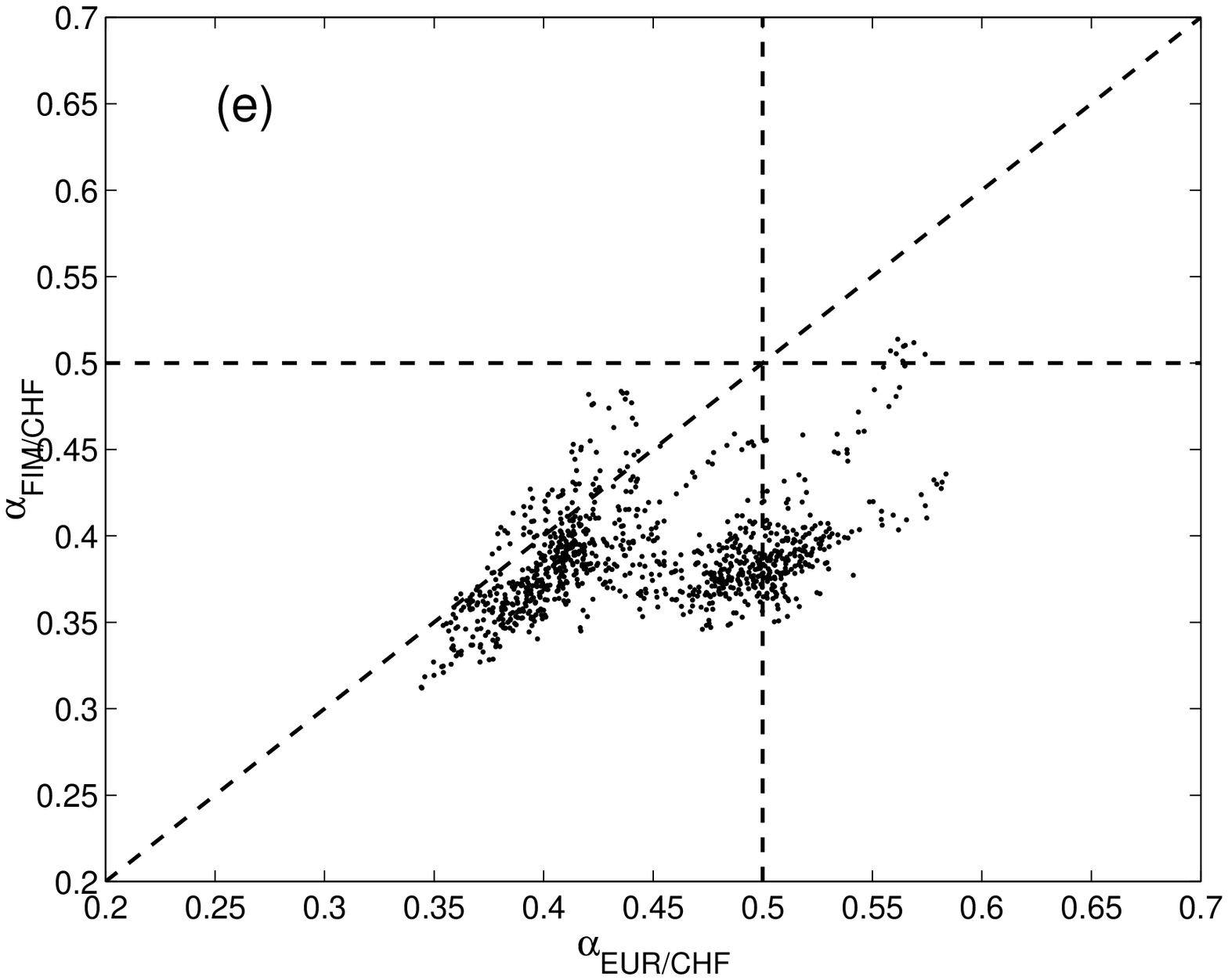}
\hfill
\leavevmode
\epsfysize=6cm
\epsffile{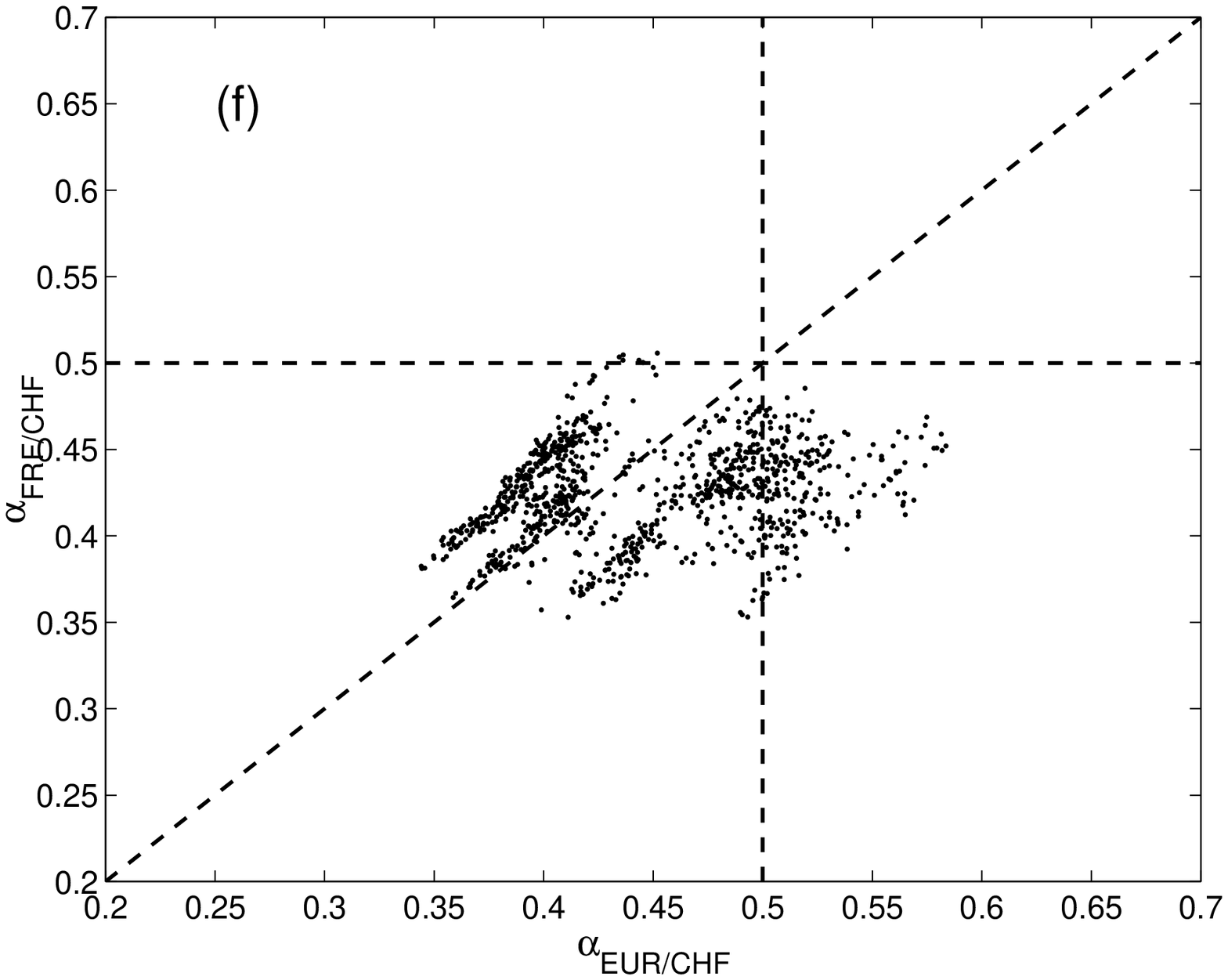}
\vfill
\leavevmode
\epsfysize=6cm
\epsffile{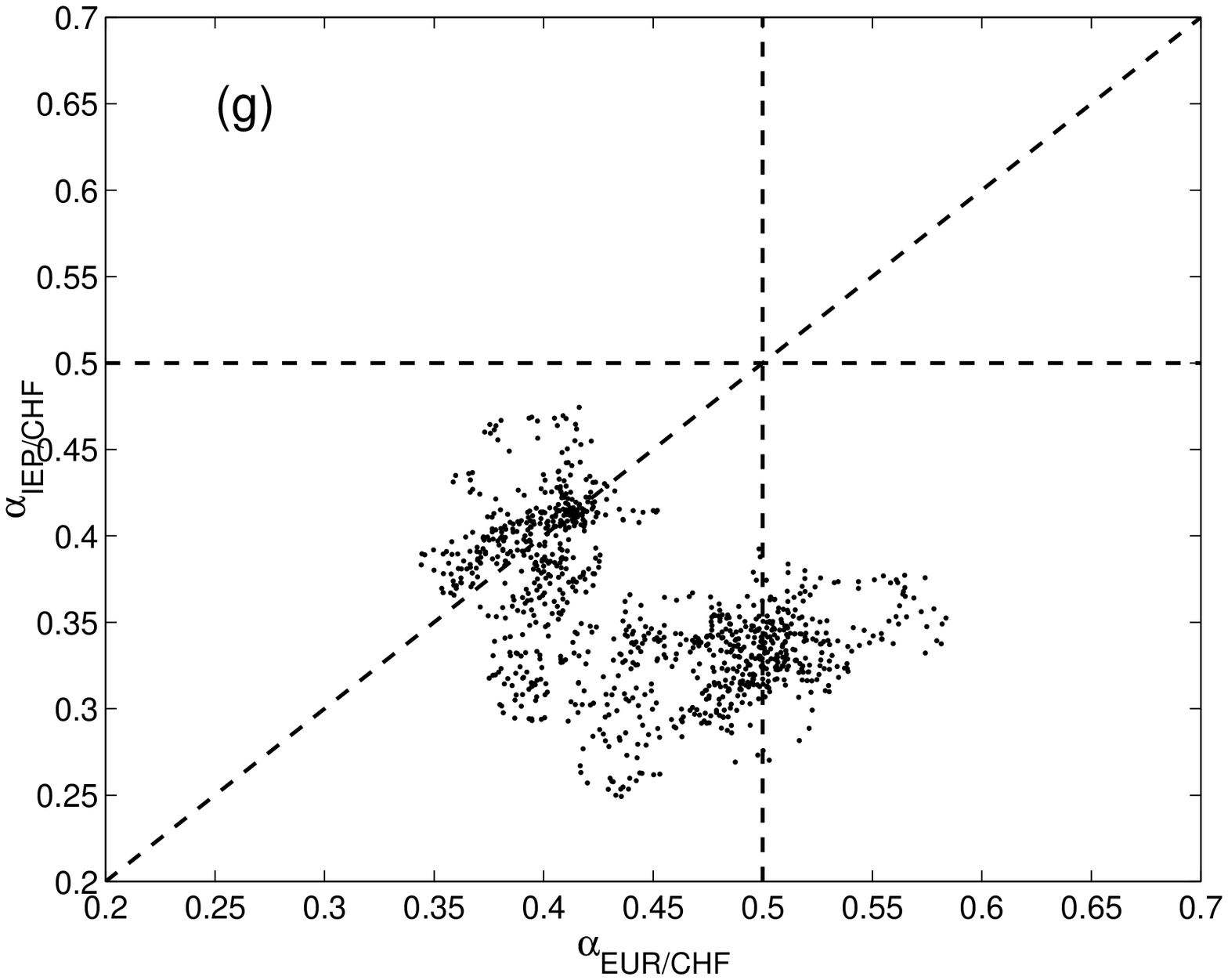}
\hfill
\leavevmode
\epsfysize=6cm
\epsffile{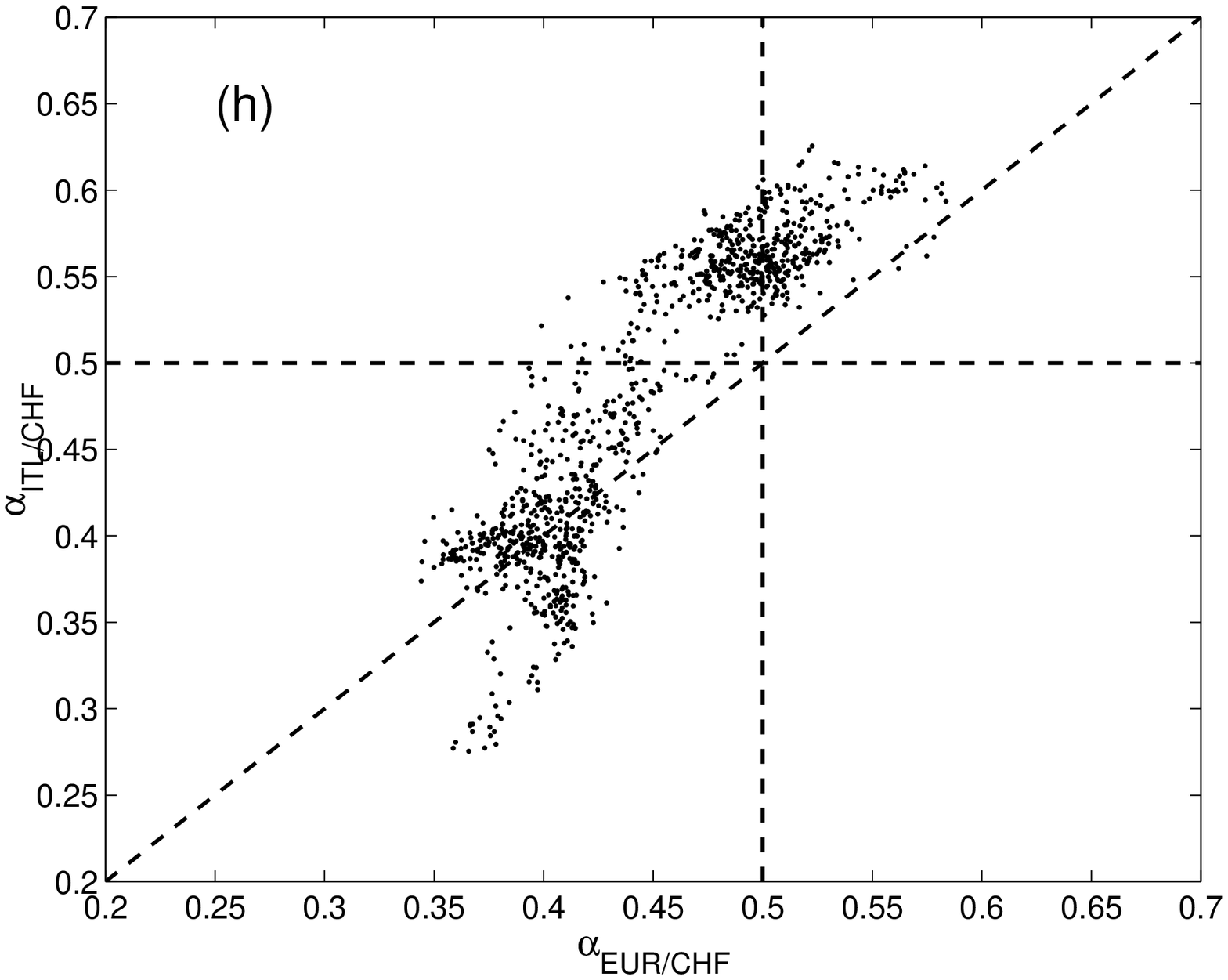}
\vfill
\leavevmode
\epsfysize=6cm
\epsffile{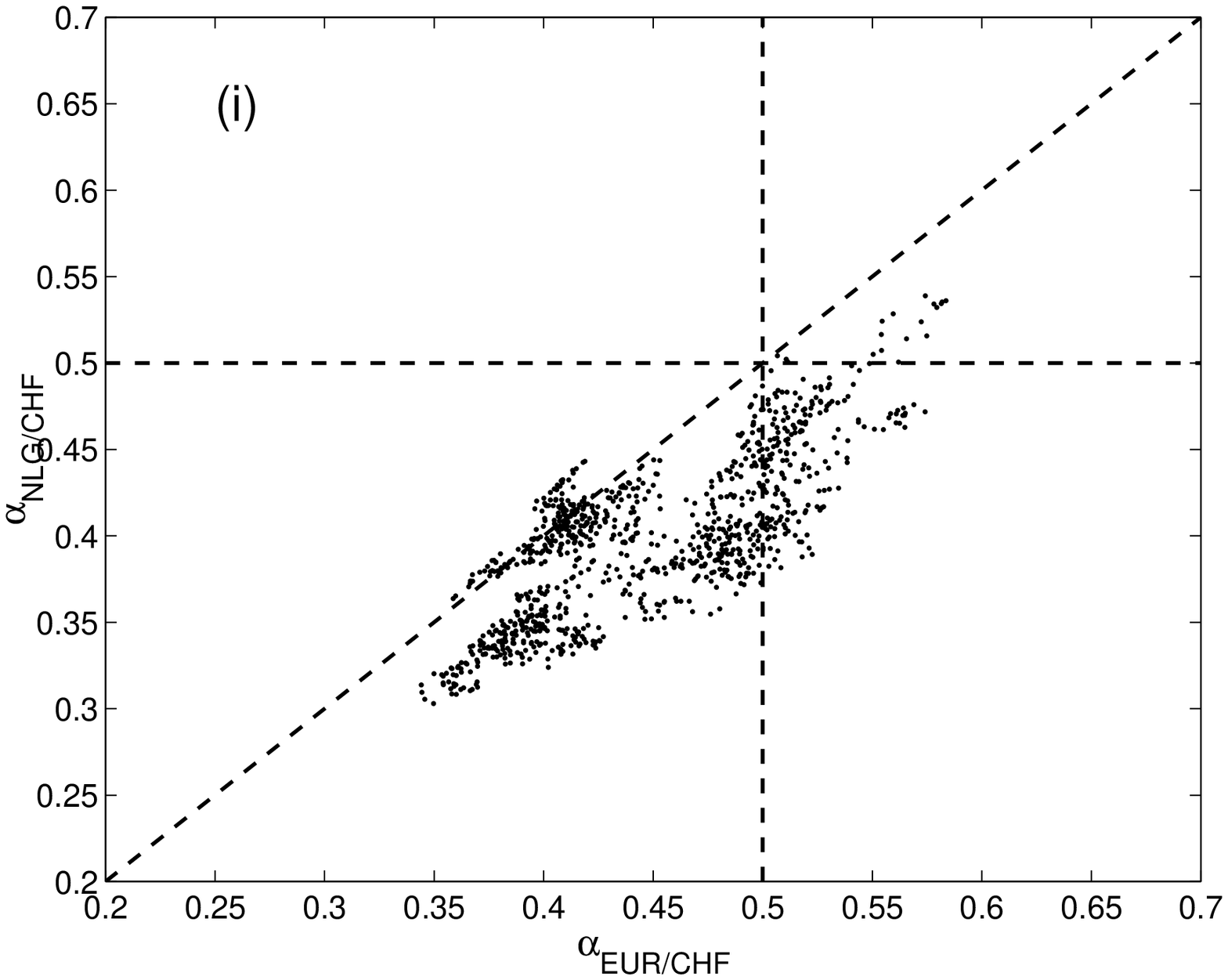}
\hfill
\leavevmode
\epsfysize=6cm
\epsffile{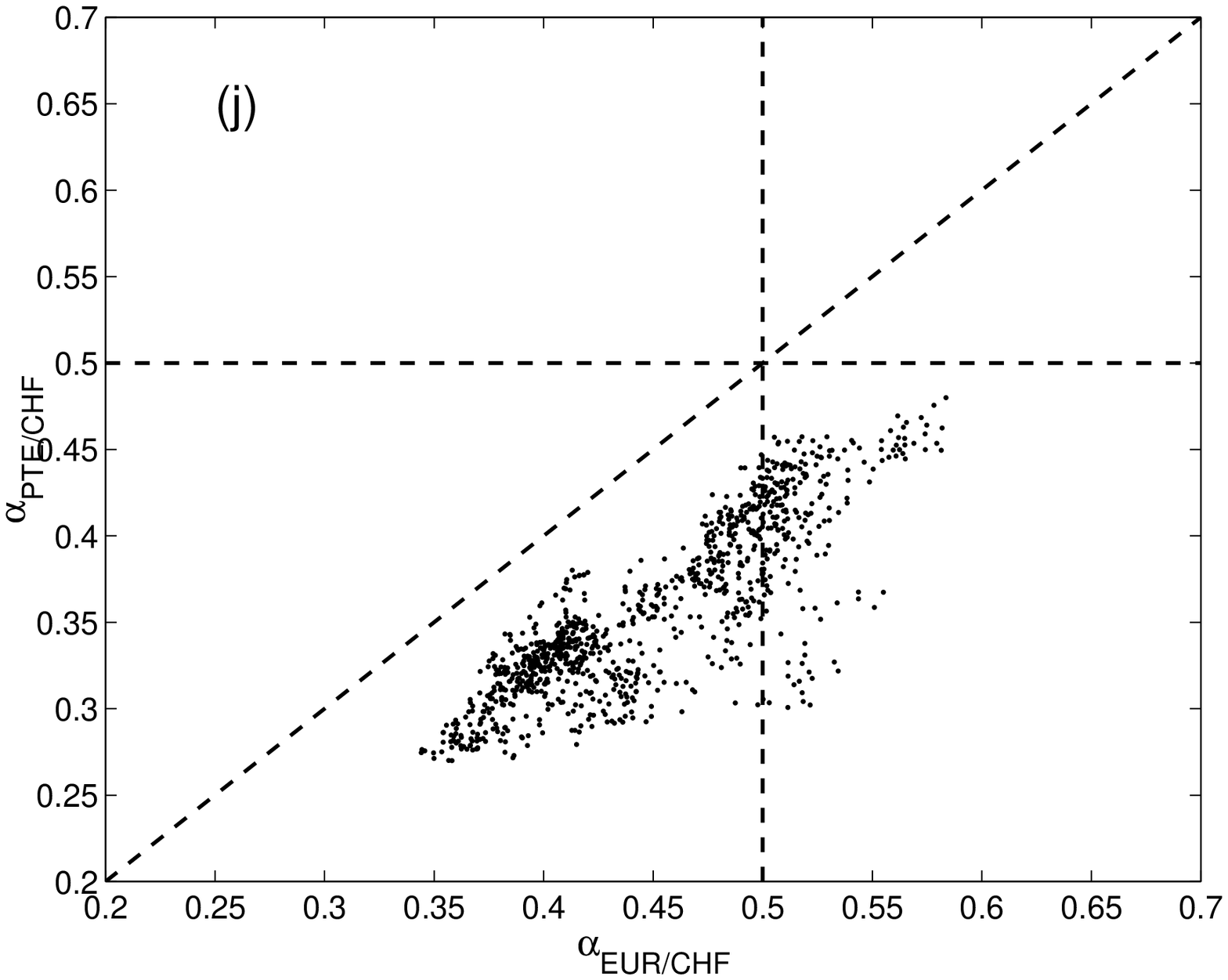}
\end{center}
\caption{Graphical representation of the
so-called correlation matrix elements for the time interval Jan. 01, 1995 till
Dec. 31, 1999 for the various $local$ $\alpha_{C_i/B_j}$ {\it vs.}
$\alpha_{EUR/B_j}$ exponents, where $C_i$ are the ten $EUR$ currencies of
interest ($i$=1,10), and (a-e) $B_j$ are the five foreign currencies ($j$=1,5)
considered in the text.} \label{fig7} \end{figure}

\begin{figure}[ht]
\begin{center}
\leavevmode
\epsfysize=6.5cm
\epsffile{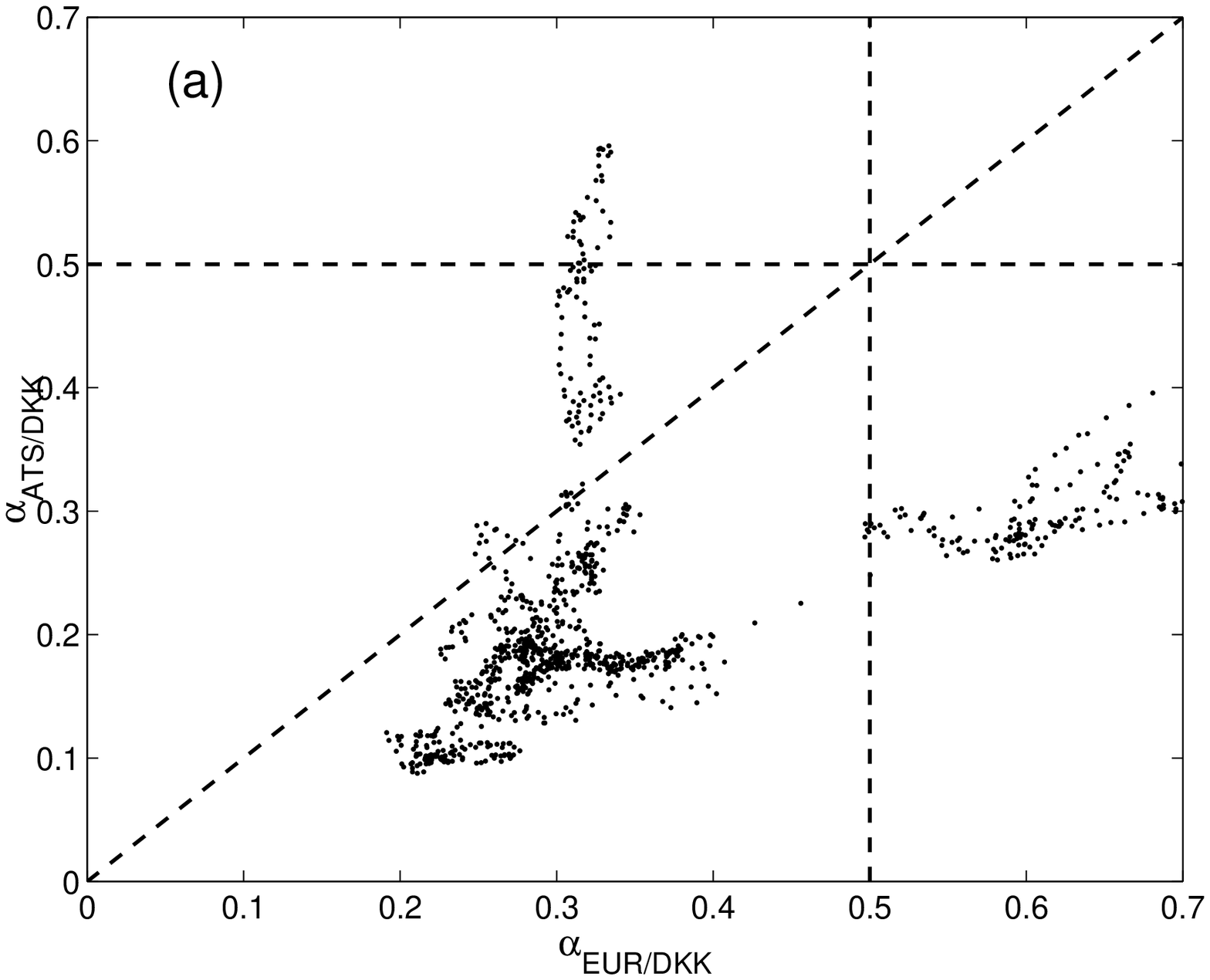}
\hfill
\leavevmode
\epsfysize=6.5cm
\epsffile{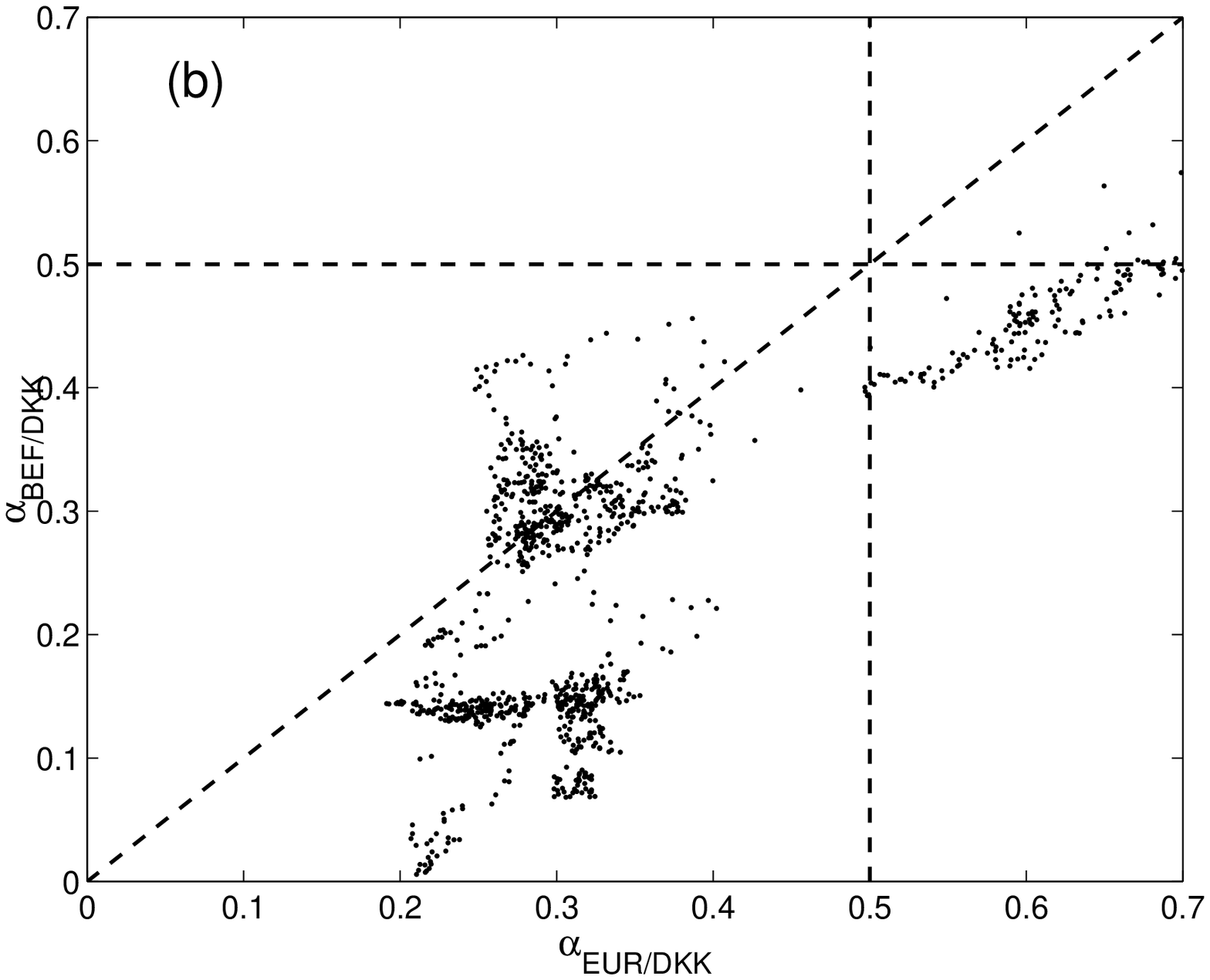}
\vfill
\leavevmode
\epsfysize=6.5cm
\epsffile{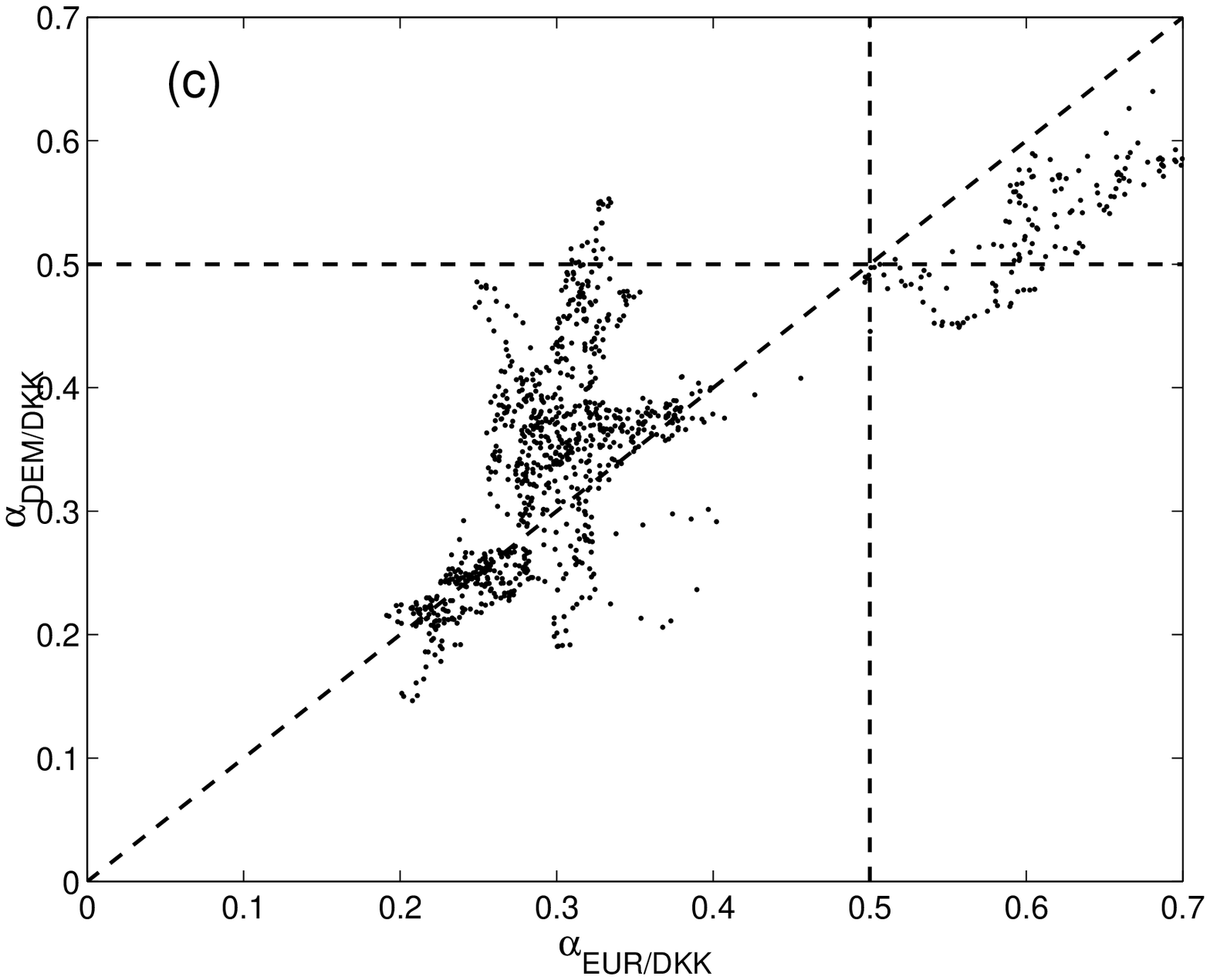}
\hfill
\leavevmode
\epsfysize=6.5cm
\epsffile{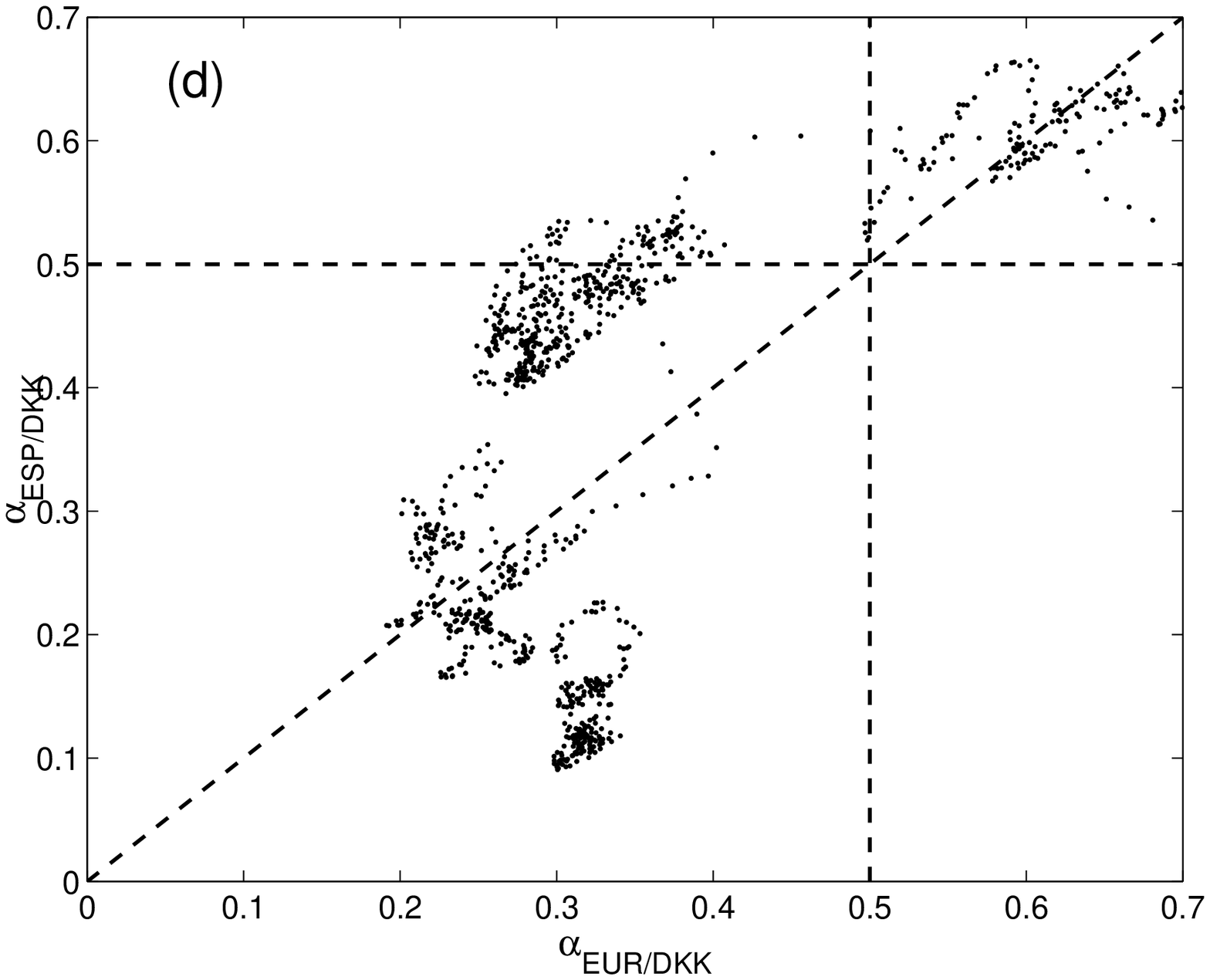}
\end{center}

\newpage
\begin{center}
\leavevmode
\epsfysize=6.5cm
\epsffile{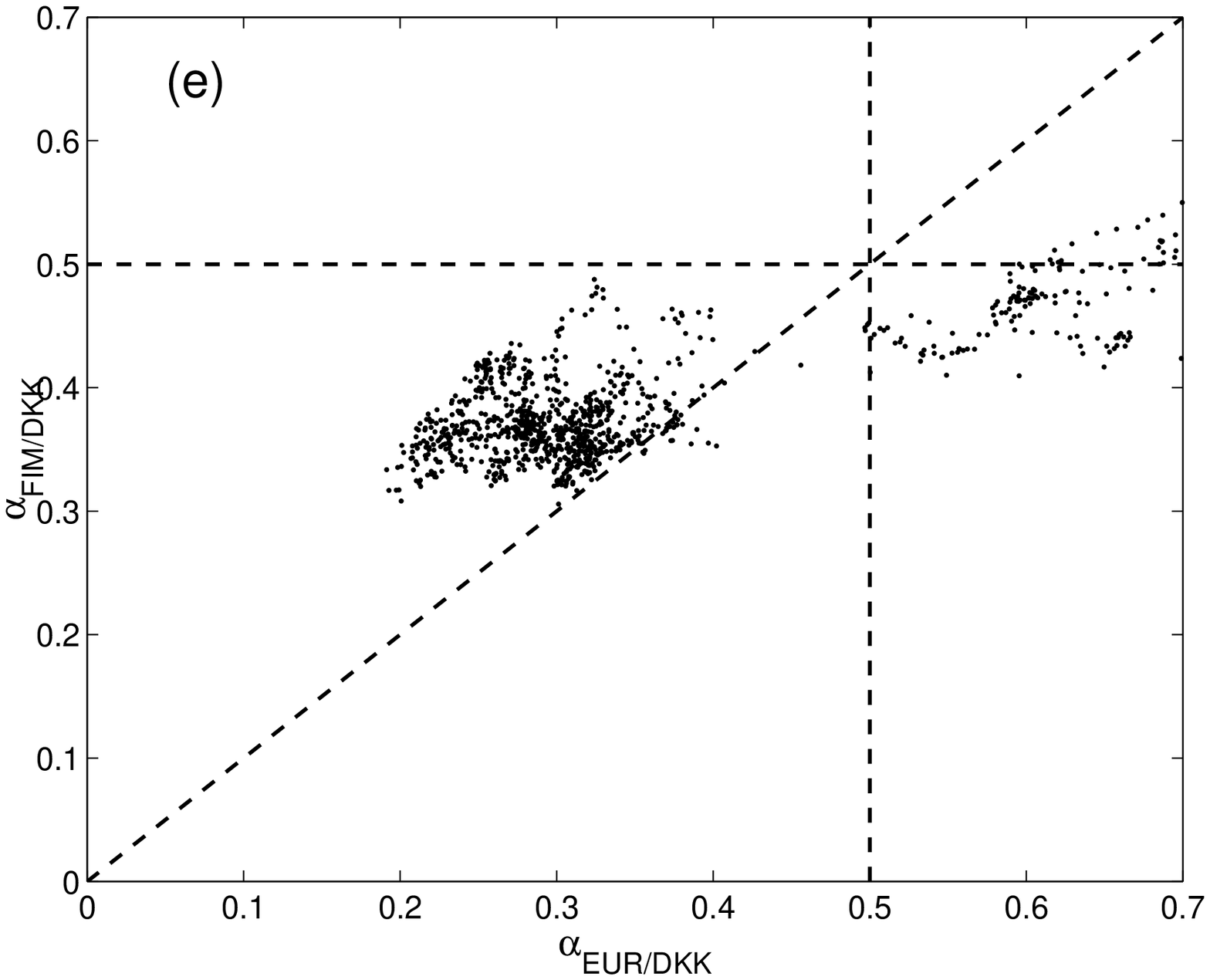}
\hfill
\leavevmode
\epsfysize=6.5cm
\epsffile{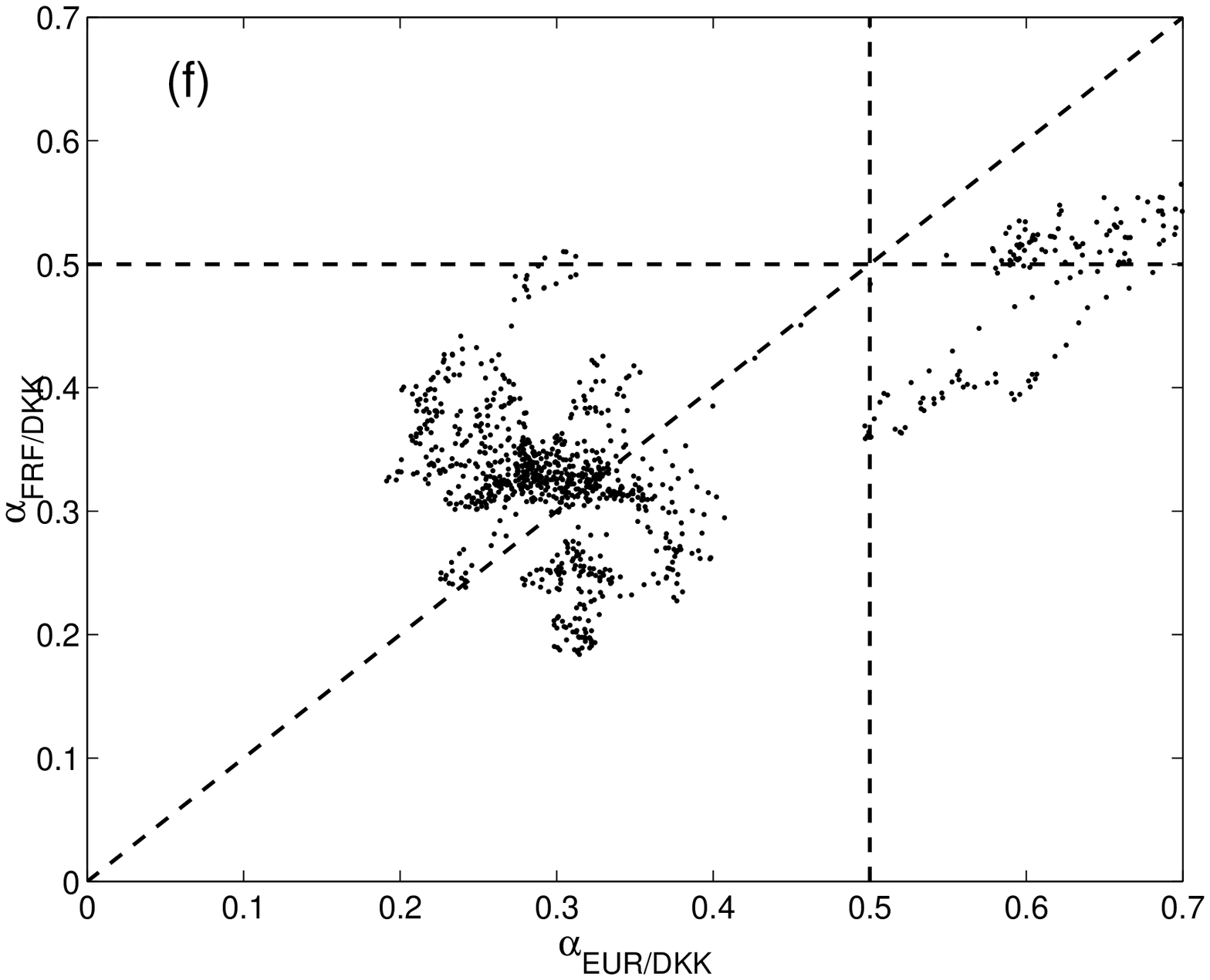}
\vfill
\leavevmode
\epsfysize=6.5cm
\epsffile{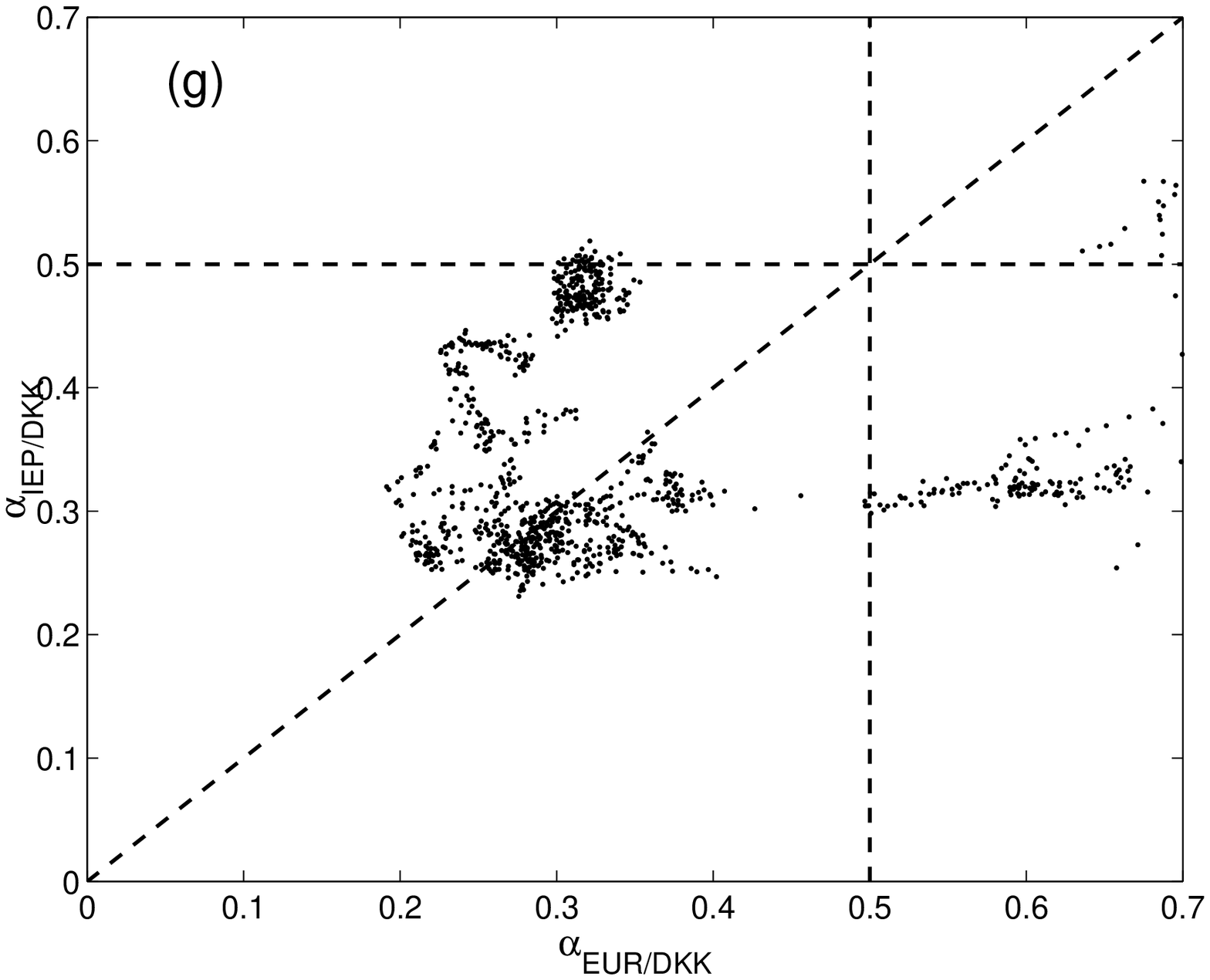}
\hfill
\leavevmode
\epsfysize=6.5cm
\epsffile{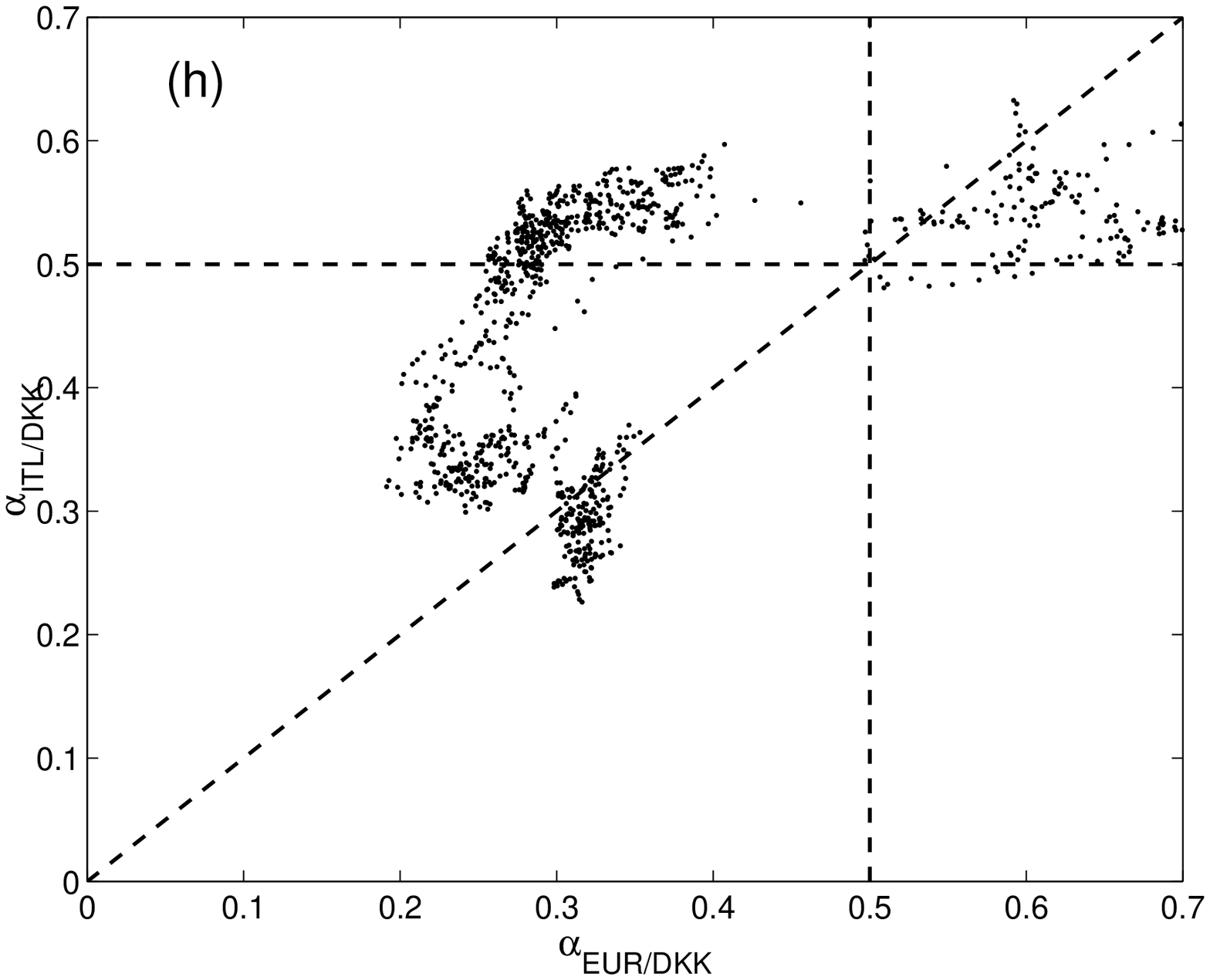}
\vfill
\leavevmode
\epsfysize=6.5cm
\epsffile{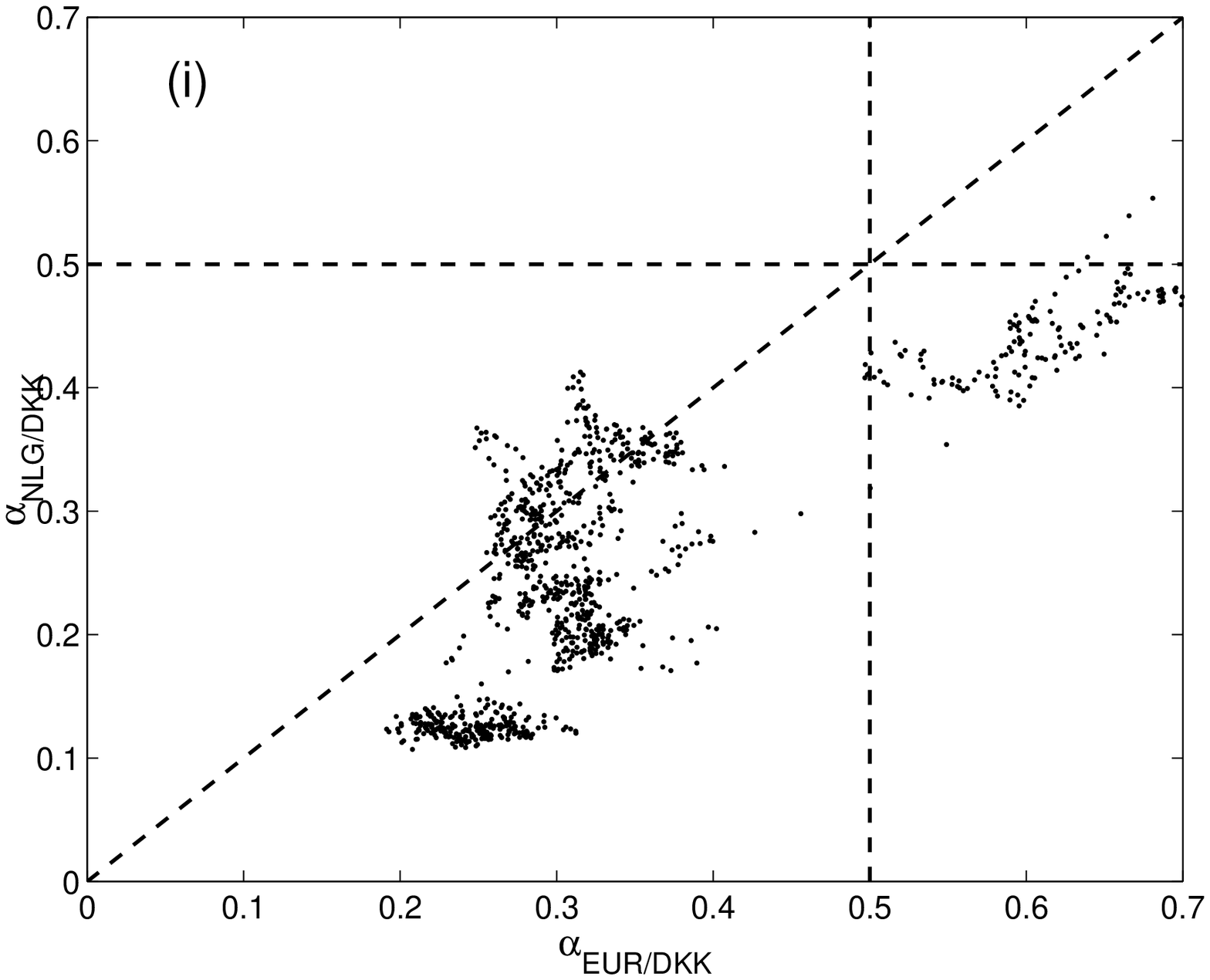}
\hfill
\leavevmode
\epsfysize=6.5cm
\epsffile{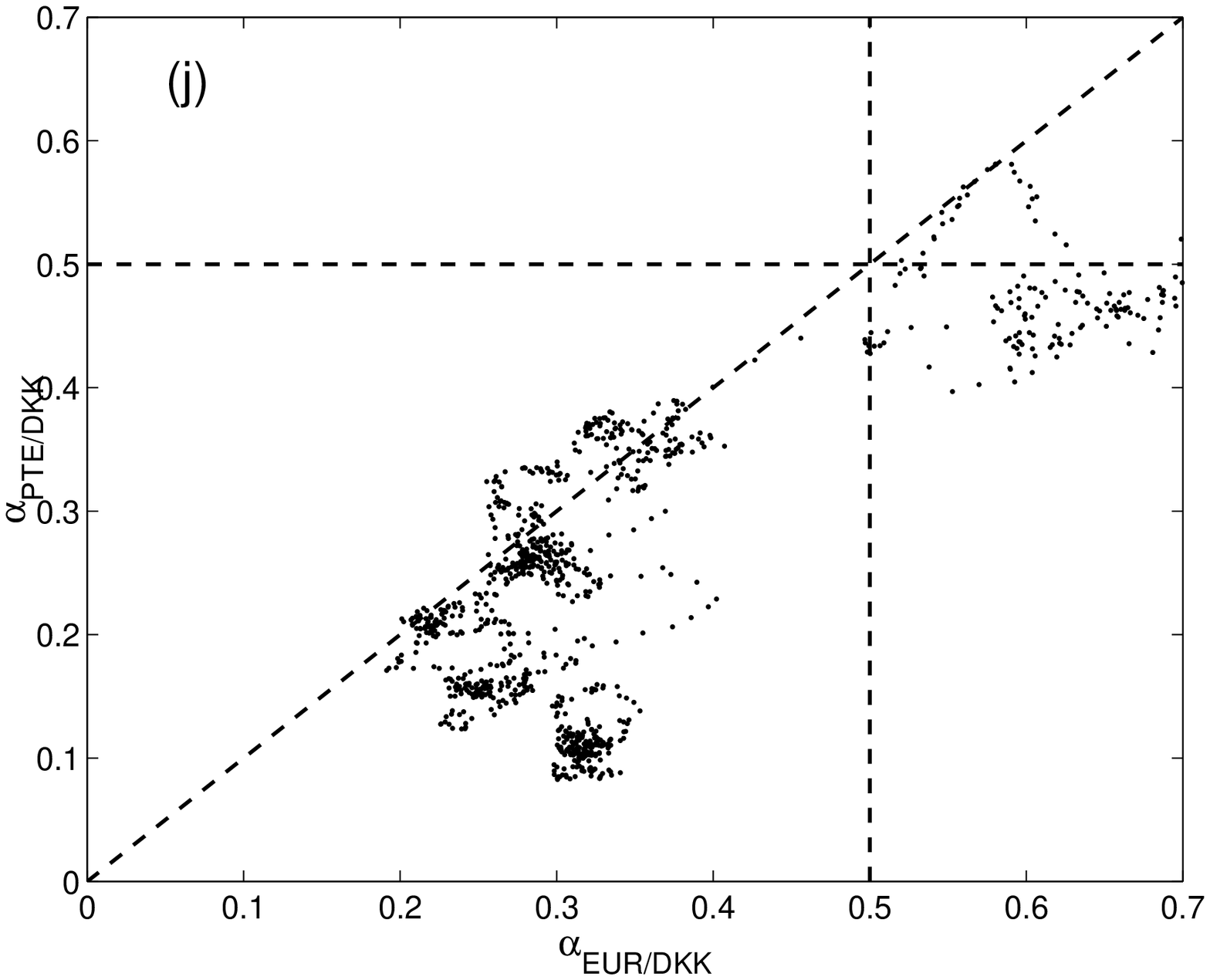}
\end{center}
\caption{Continue Fig 7b}
\label{fig7b}
\end{figure}

\begin{figure}[ht]
\begin{center}
\leavevmode
\epsfysize=6.5cm
\epsffile{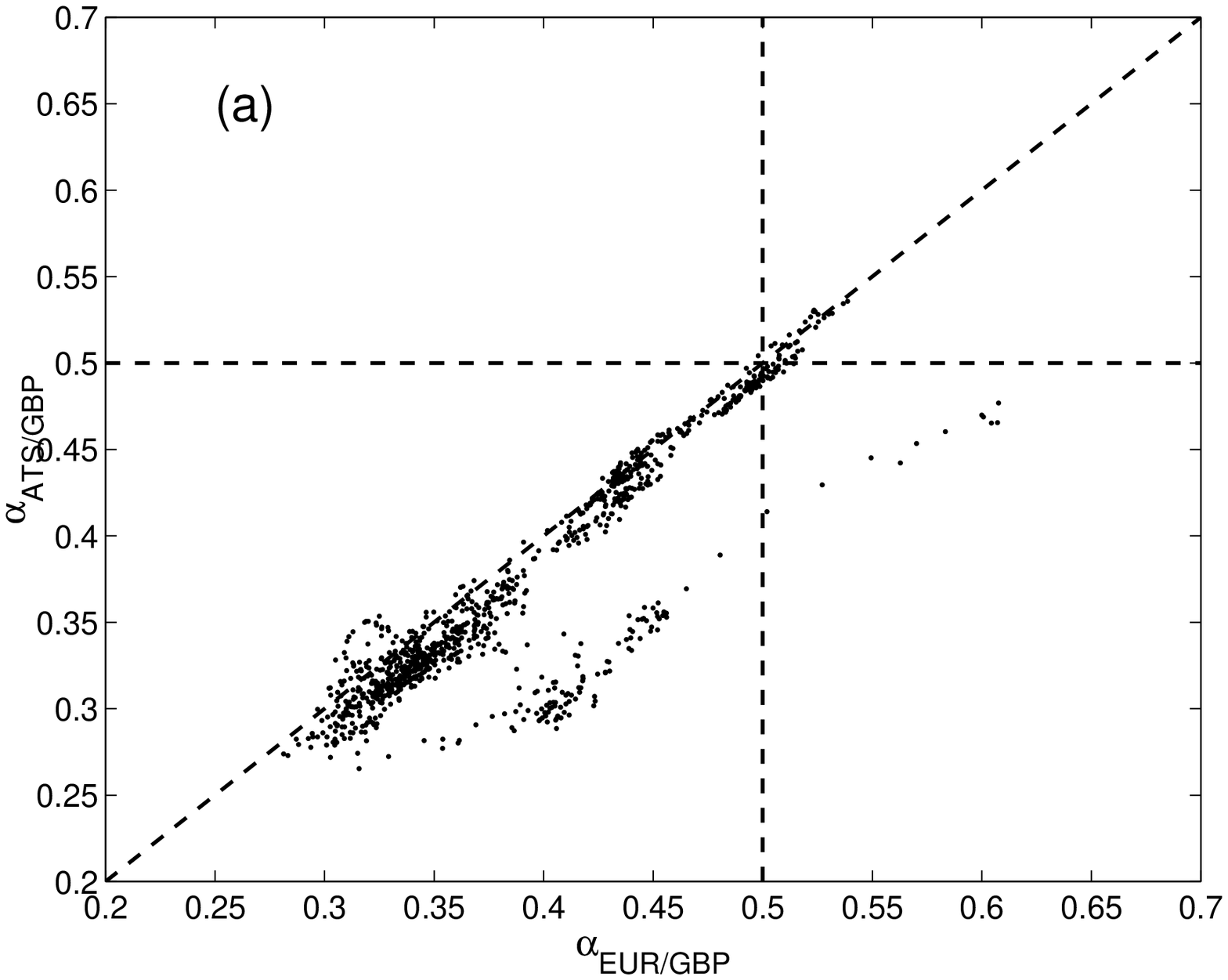}
\hfill
\leavevmode
\epsfysize=6.5cm
\epsffile{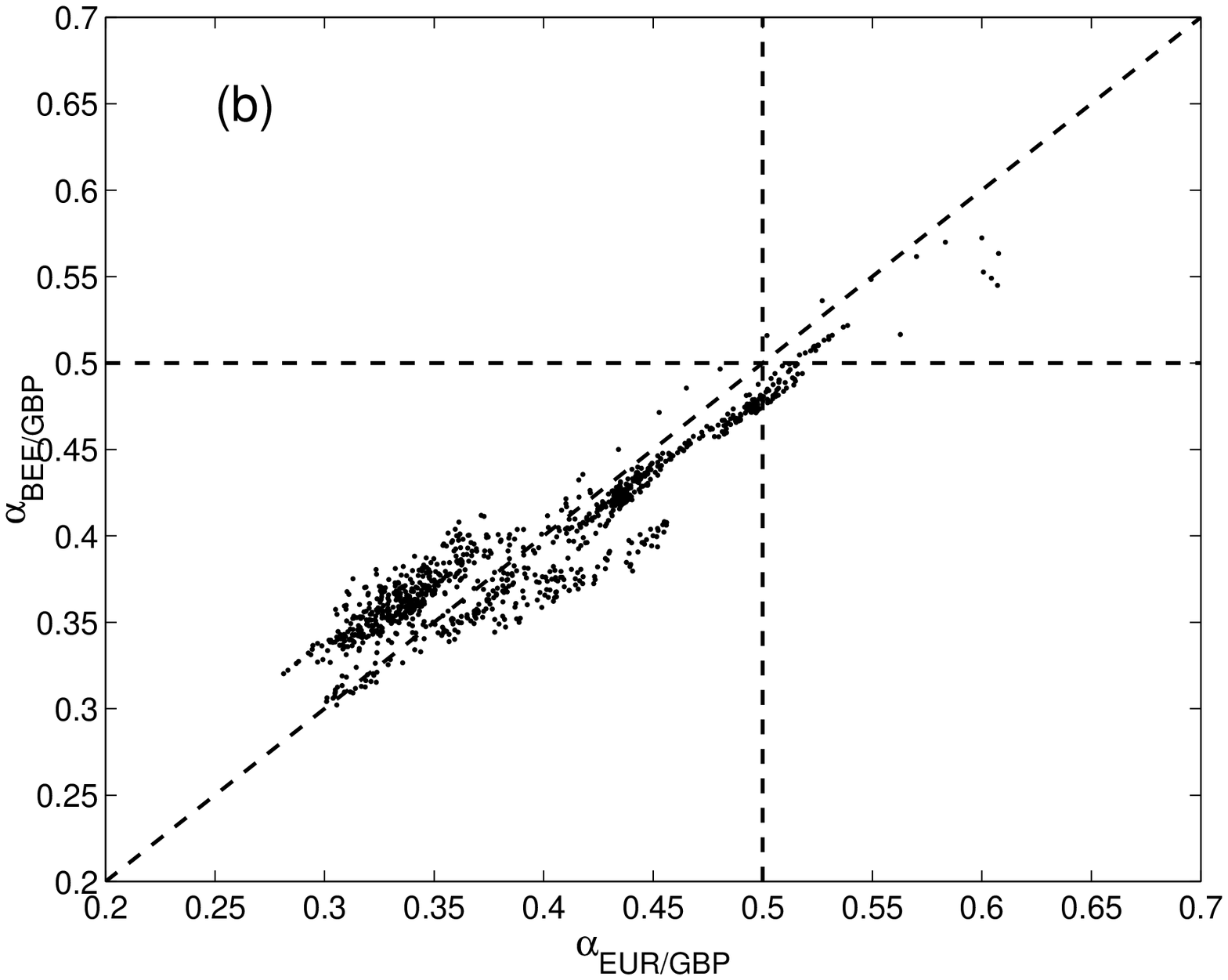}
\vfill
\leavevmode
\epsfysize=6.5cm
\epsffile{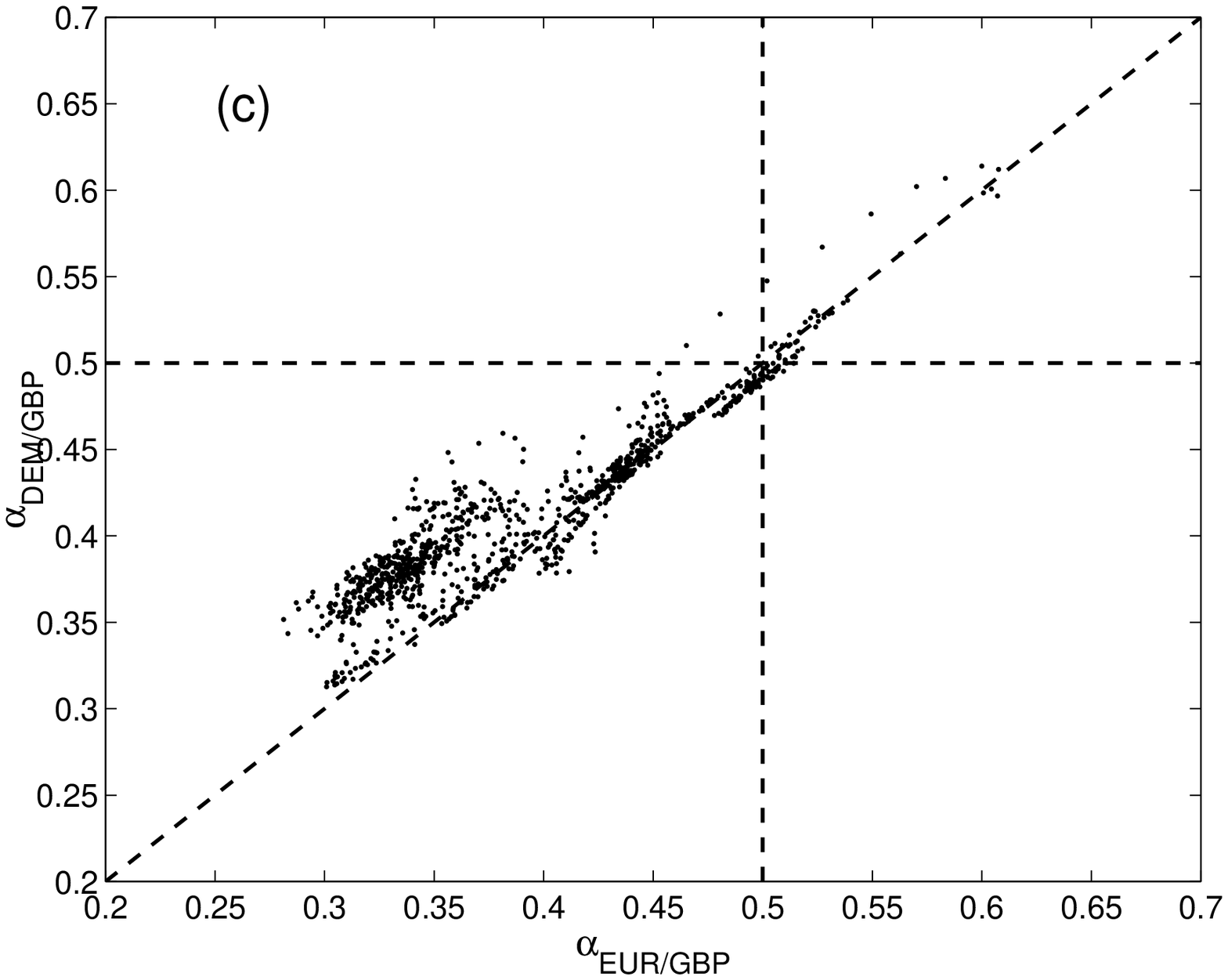}
\hfill
\leavevmode
\epsfysize=6.5cm
\epsffile{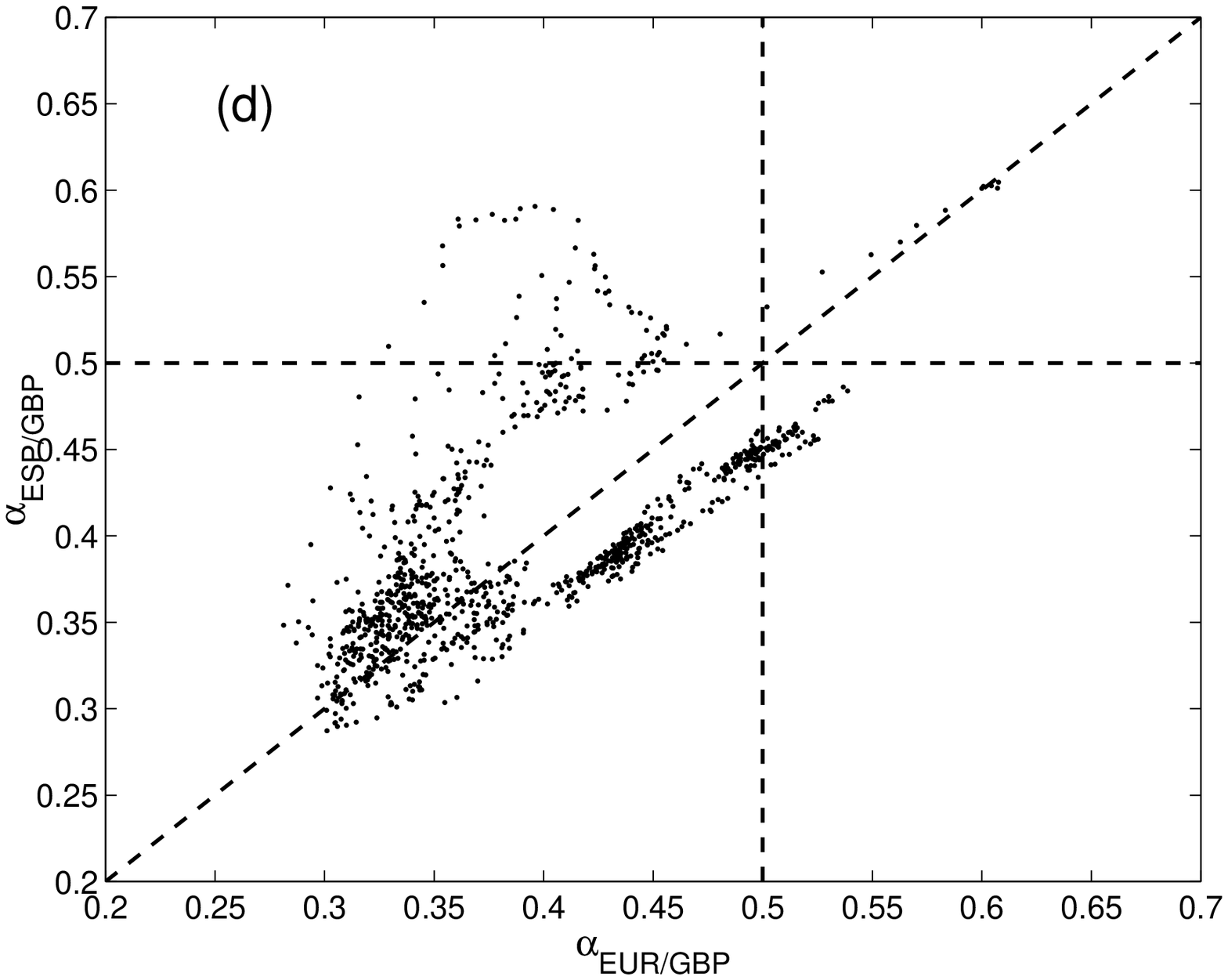}
\end{center}

\newpage
\begin{center}
\leavevmode
\epsfysize=6.5cm
\epsffile{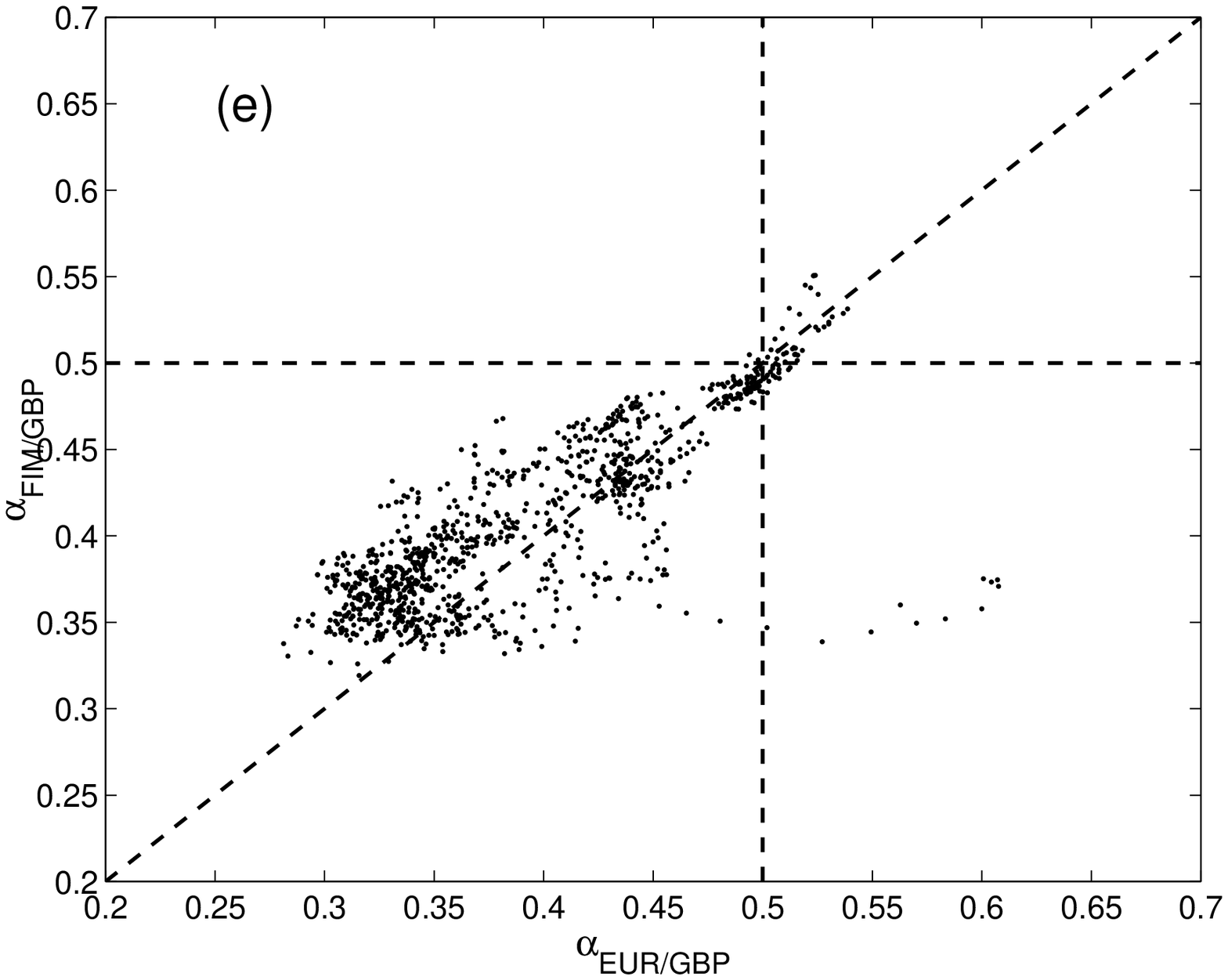}
\hfill
\leavevmode
\epsfysize=6.5cm
\epsffile{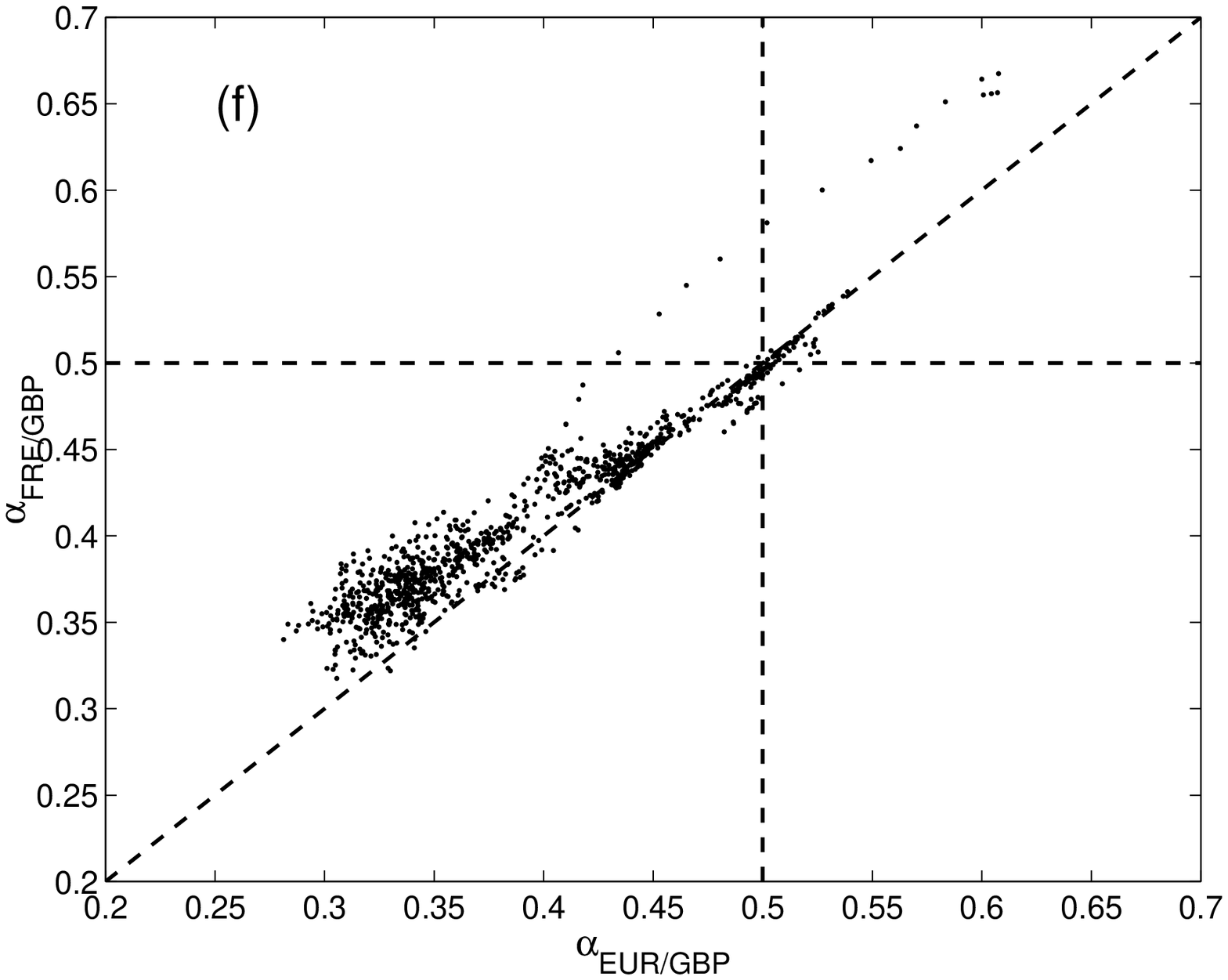}
\vfill
\leavevmode
\epsfysize=6.5cm
\epsffile{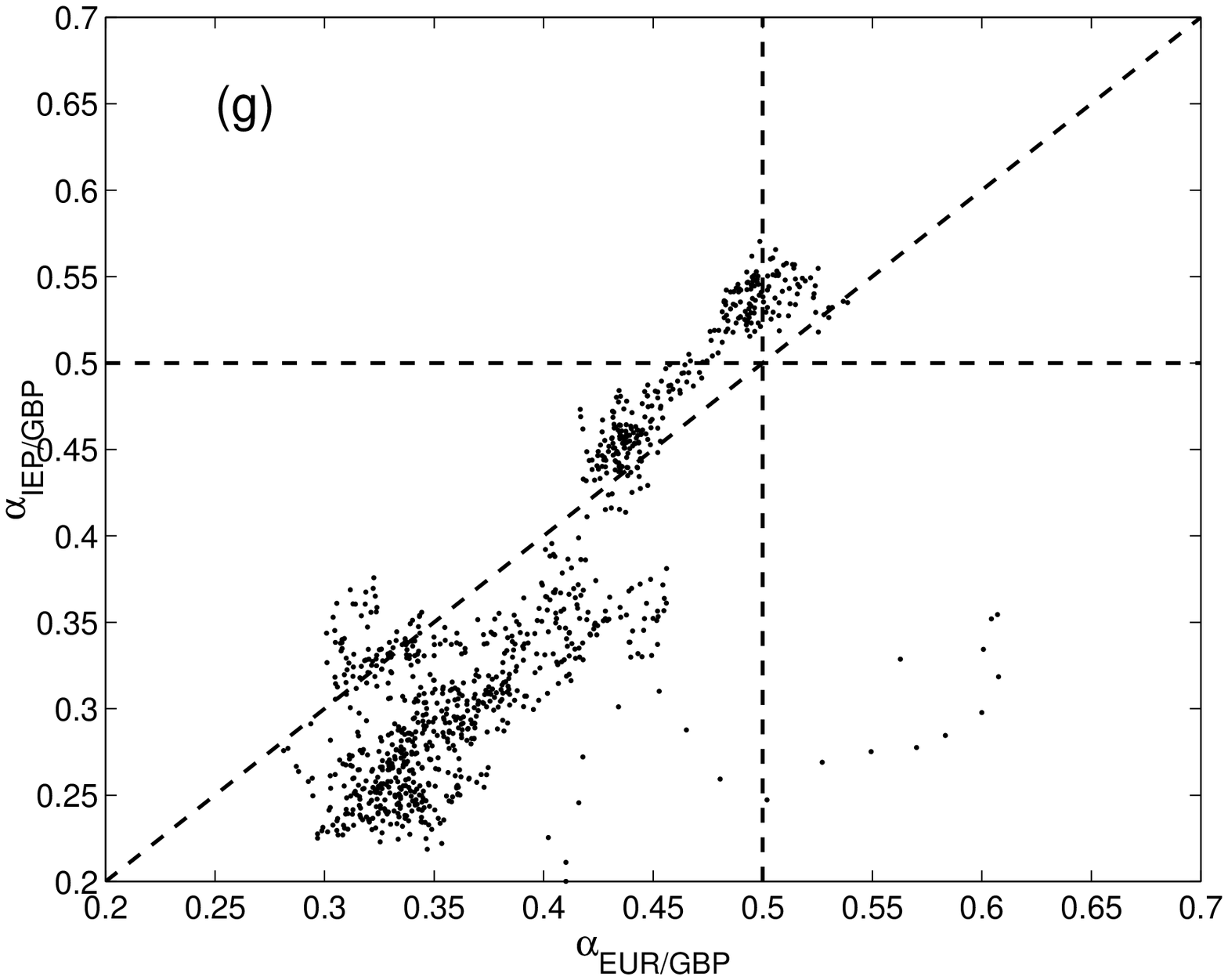}
\hfill
\leavevmode
\epsfysize=6.5cm
\epsffile{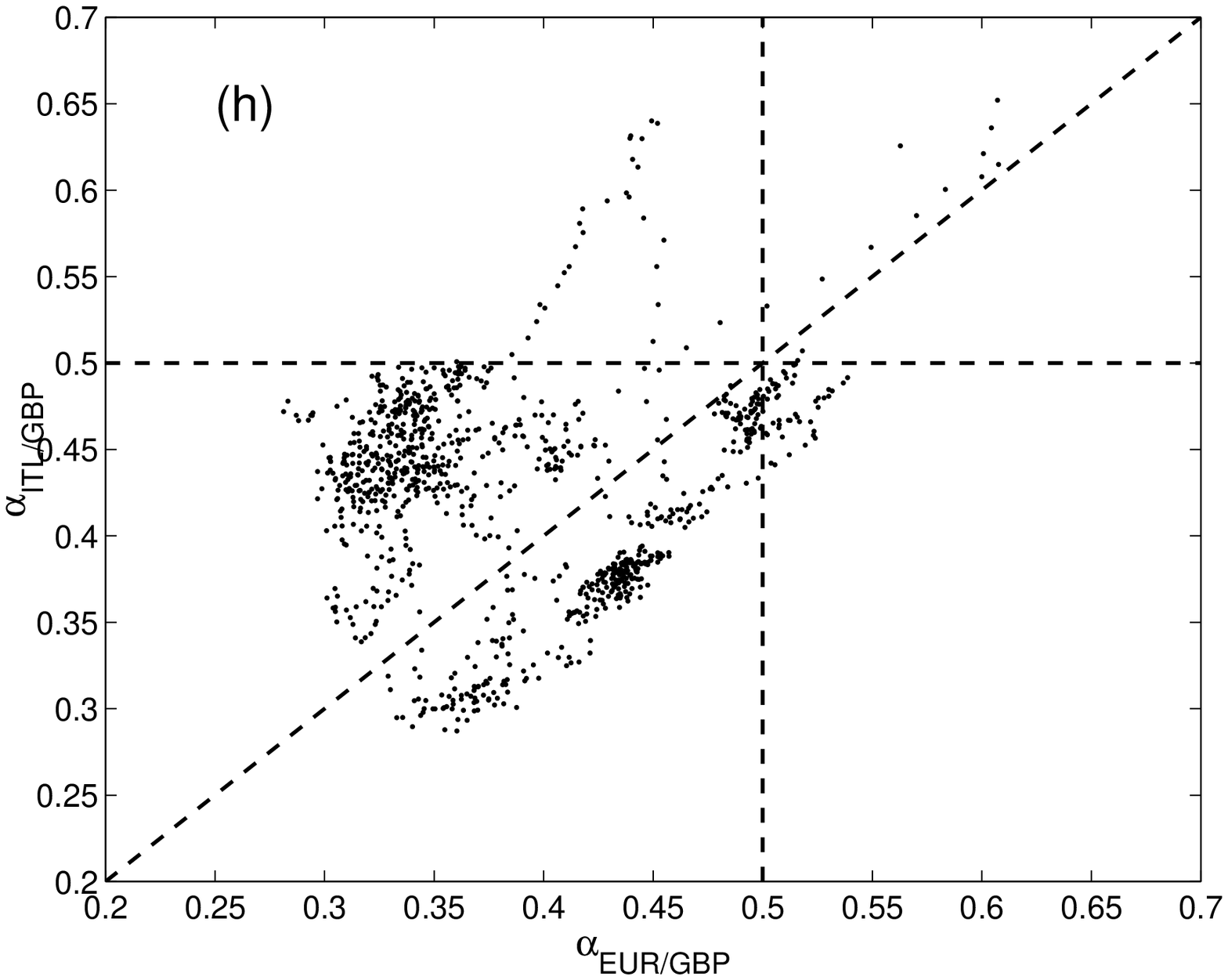}
\vfill
\leavevmode
\epsfysize=6.5cm
\epsffile{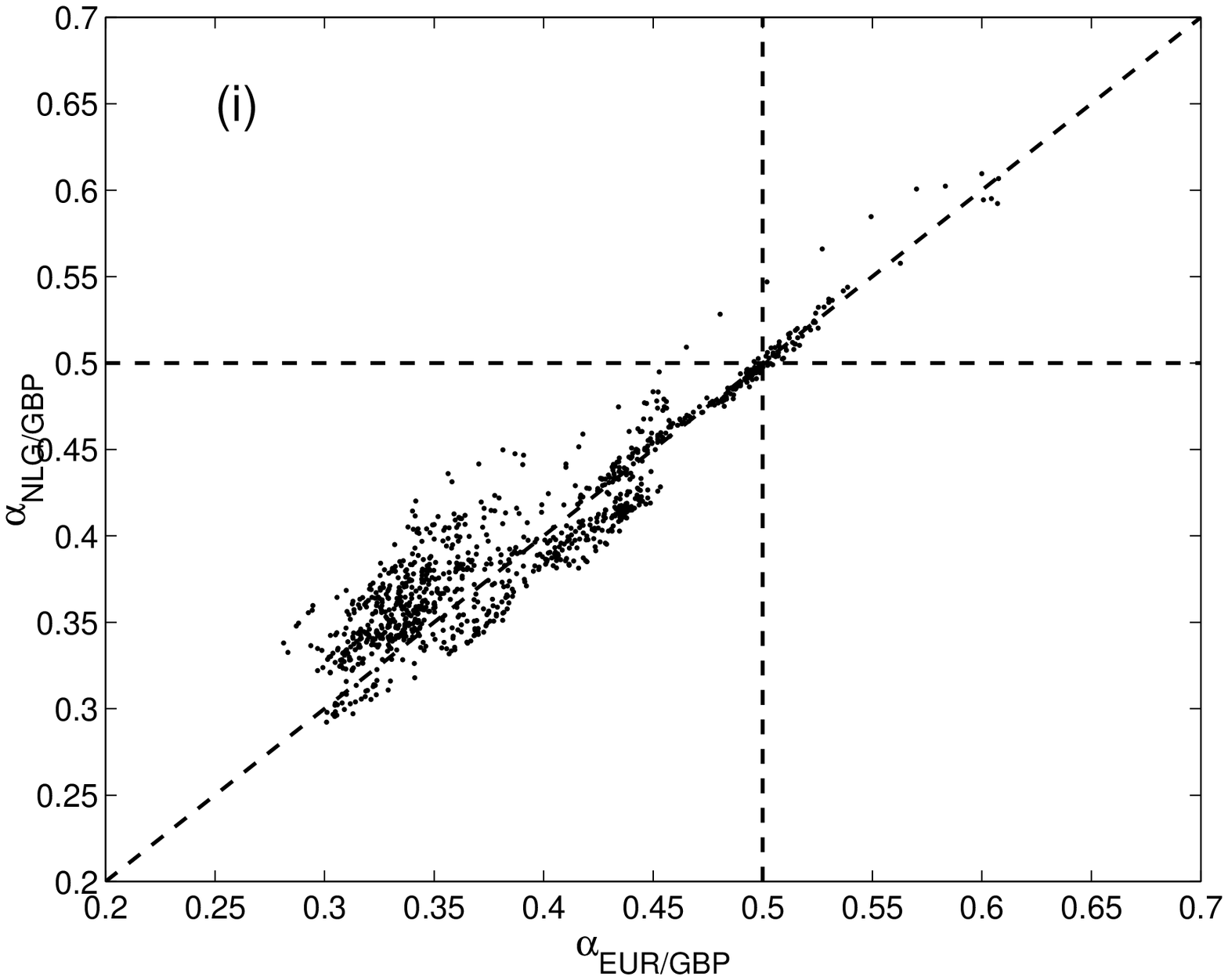}
\hfill
\leavevmode
\epsfysize=6.5cm
\epsffile{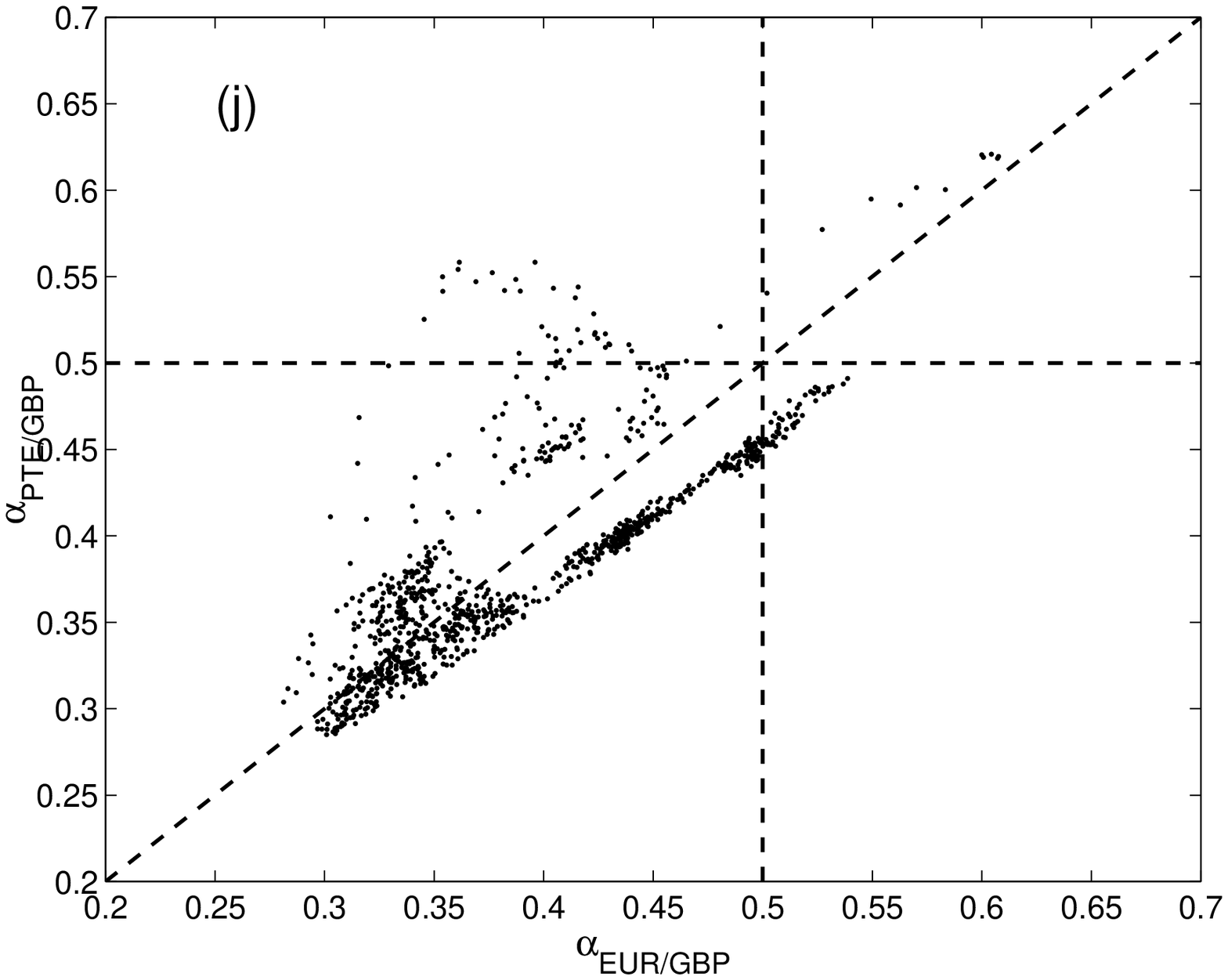}
\end{center}
\caption{Continue Fig 7c}
\label{fig7c}
\end{figure}

\begin{figure}[ht]
\begin{center}
\leavevmode
\epsfysize=6.5cm
\epsffile{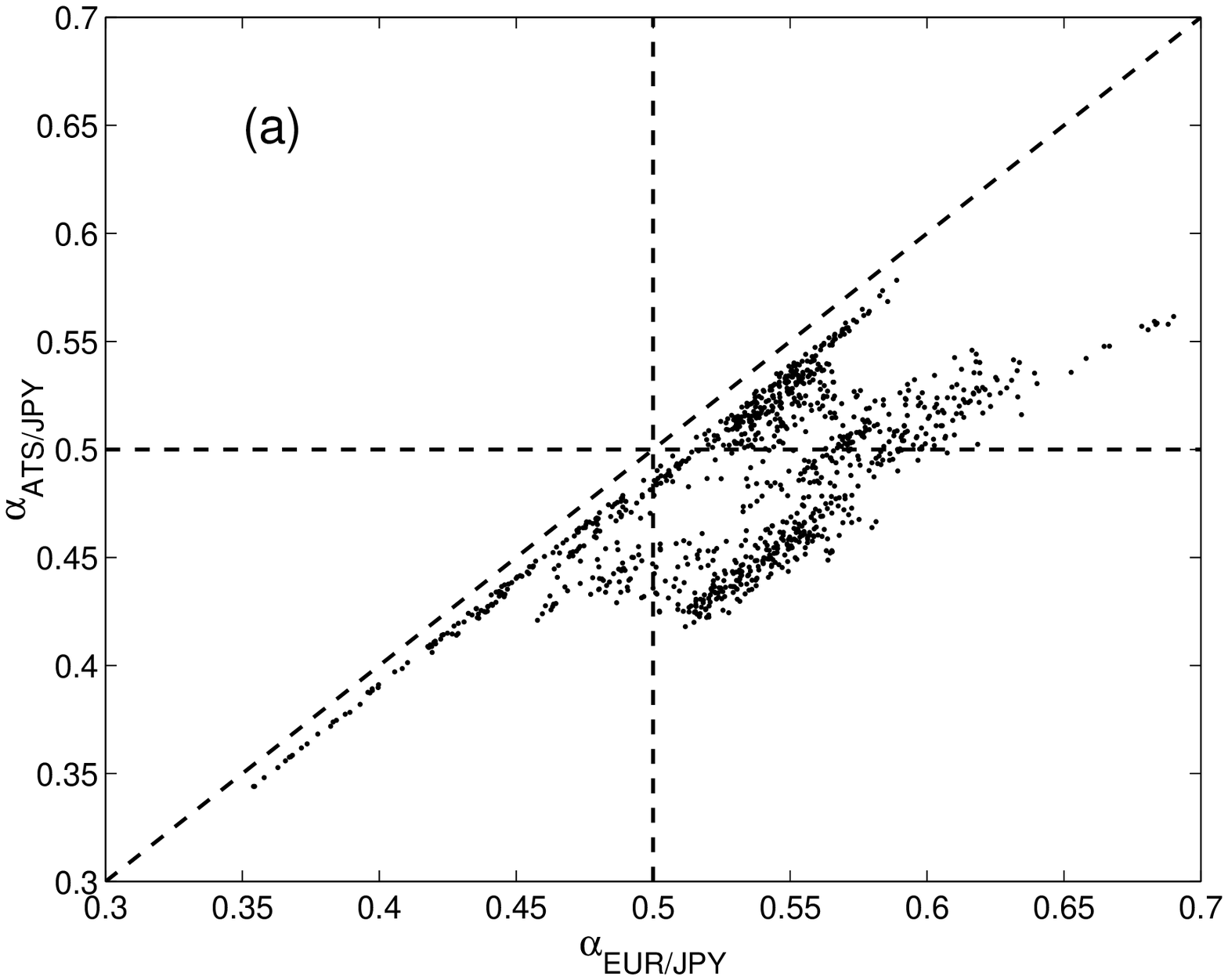}
\hfill
\leavevmode
\epsfysize=6.5cm
\epsffile{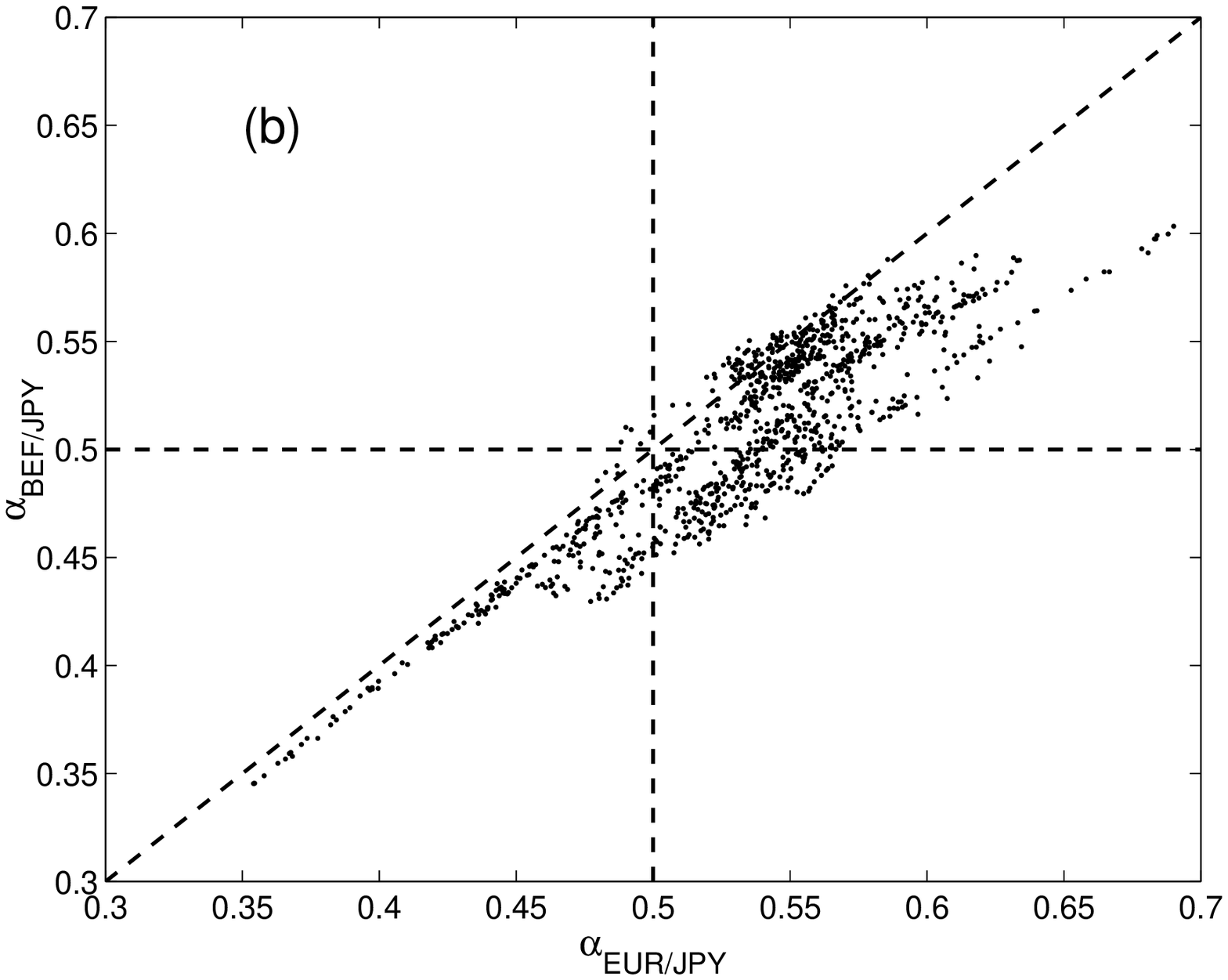}
\vfill
\leavevmode
\epsfysize=6.5cm
\epsffile{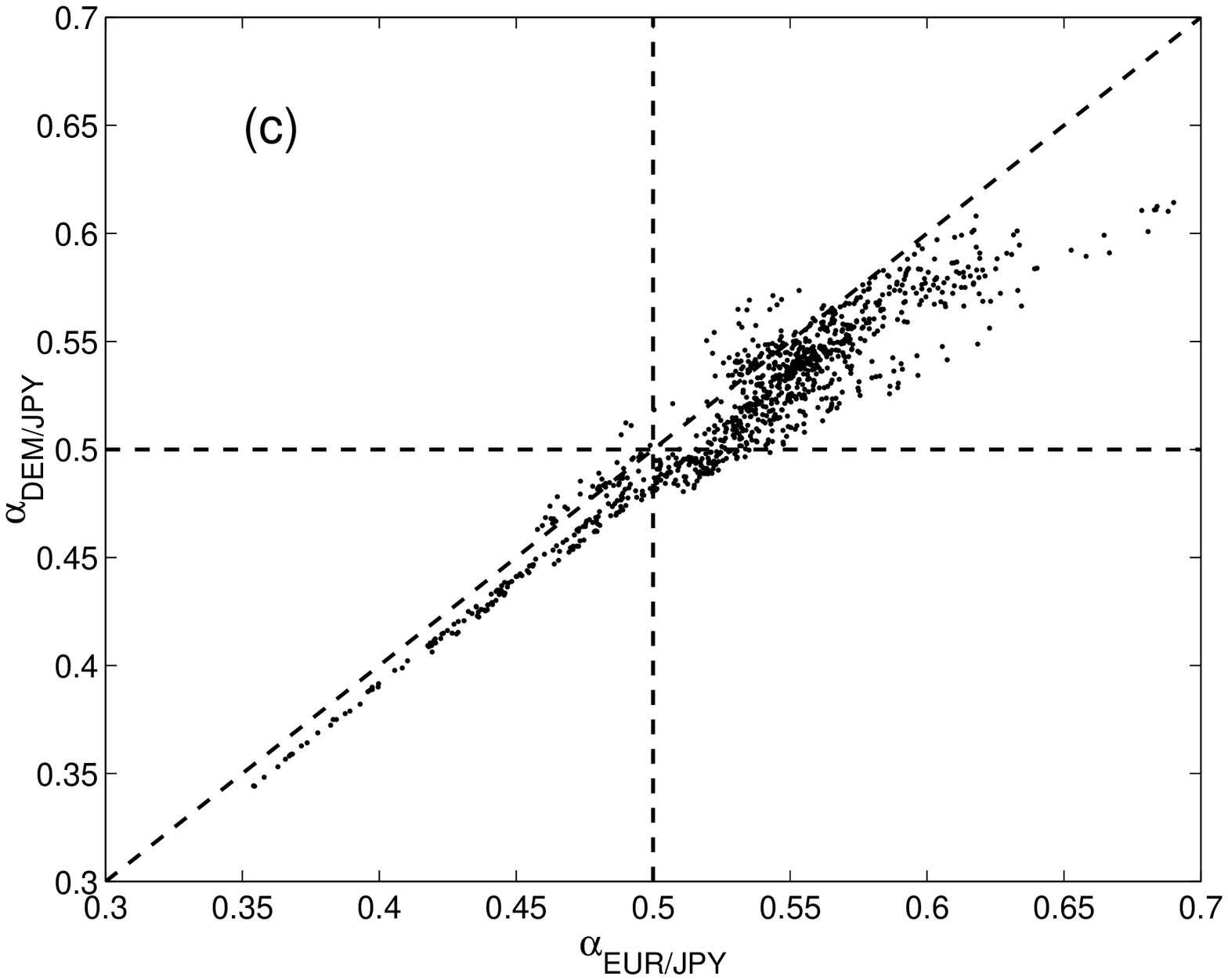}
\hfill
\leavevmode
\epsfysize=6.5cm
\epsffile{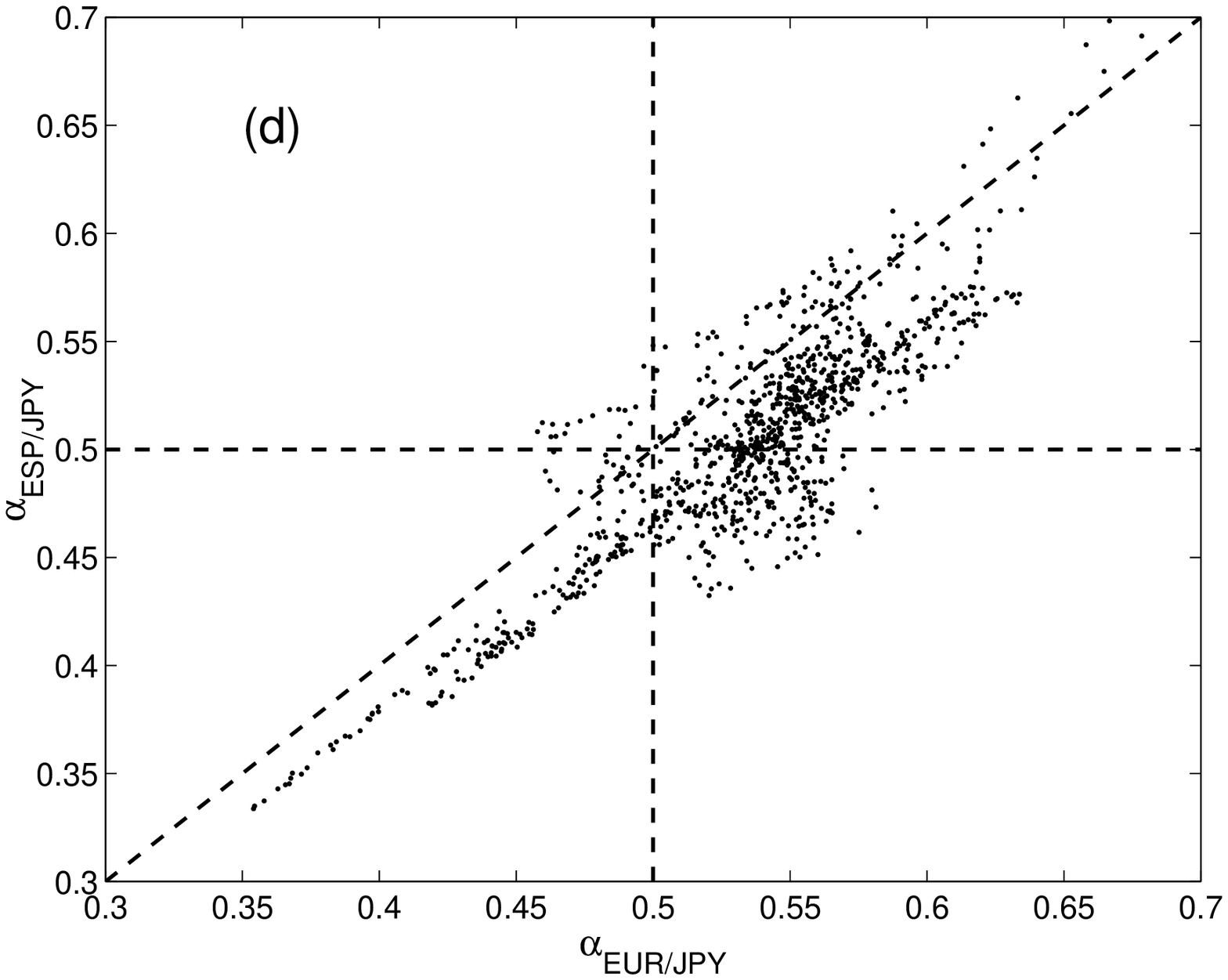}
\end{center}

\newpage
\begin{center}
\leavevmode
\epsfysize=6.5cm
\epsffile{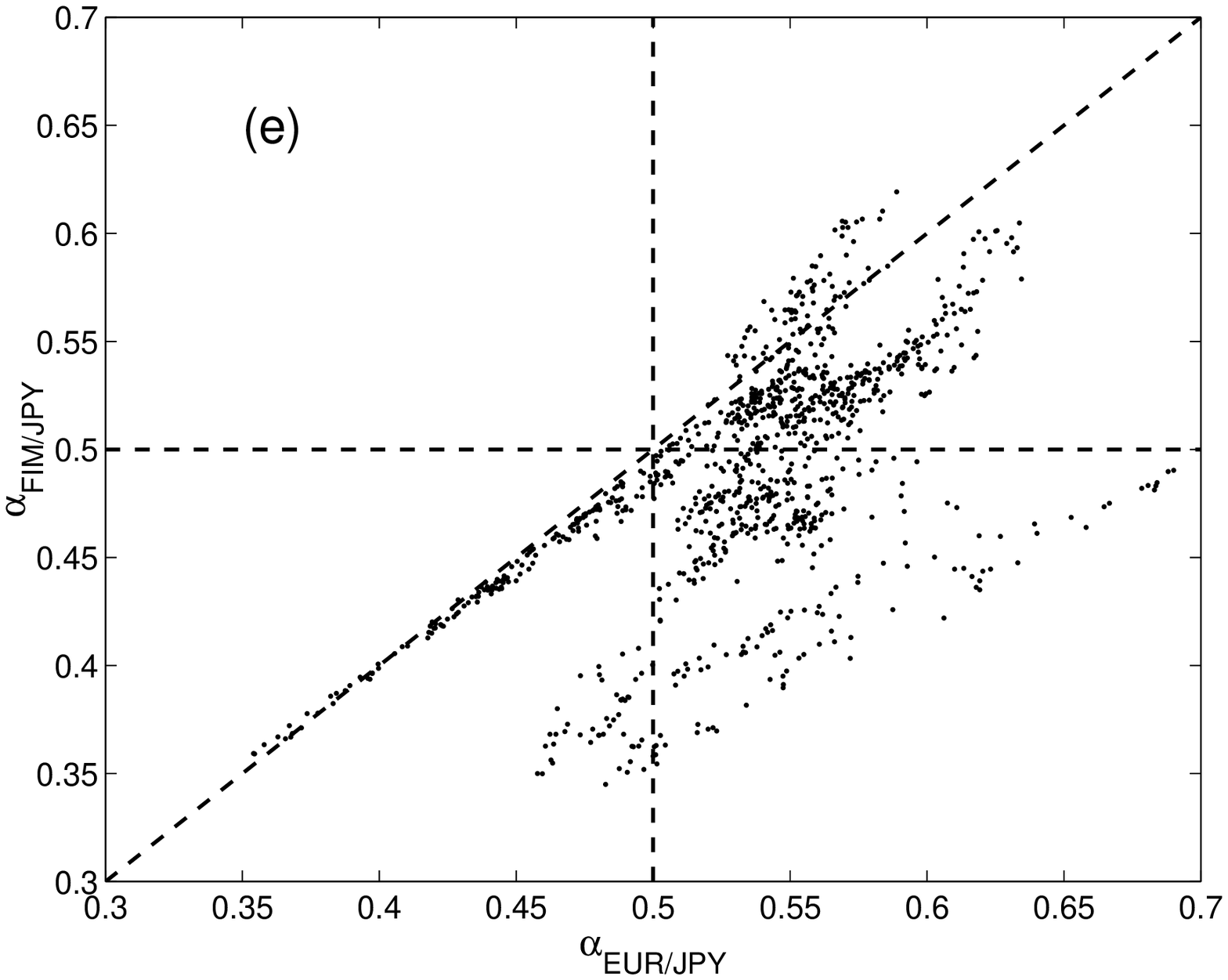}
\hfill
\leavevmode
\epsfysize=6.5cm
\epsffile{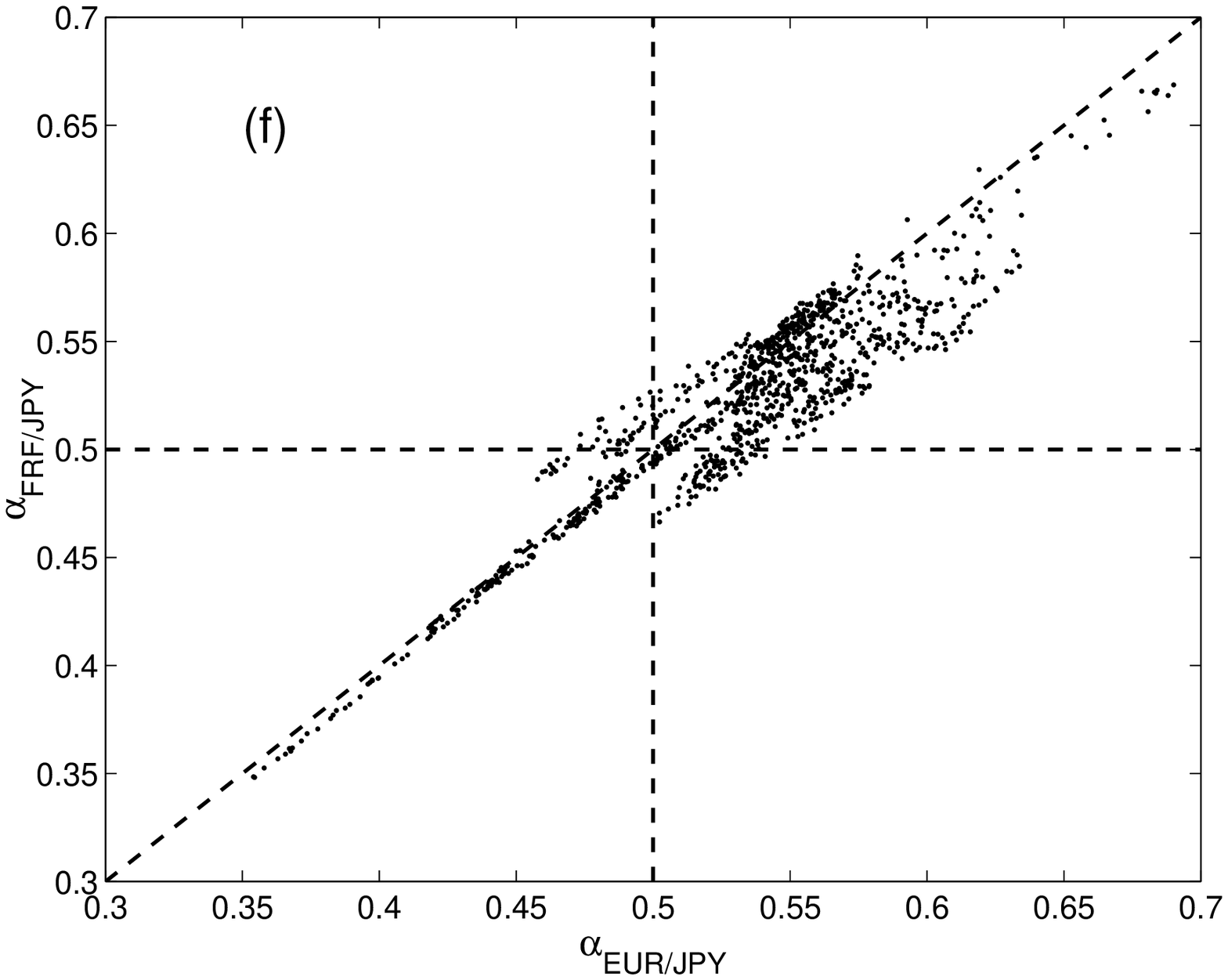}
\vfill
\leavevmode
\epsfysize=6.5cm
\epsffile{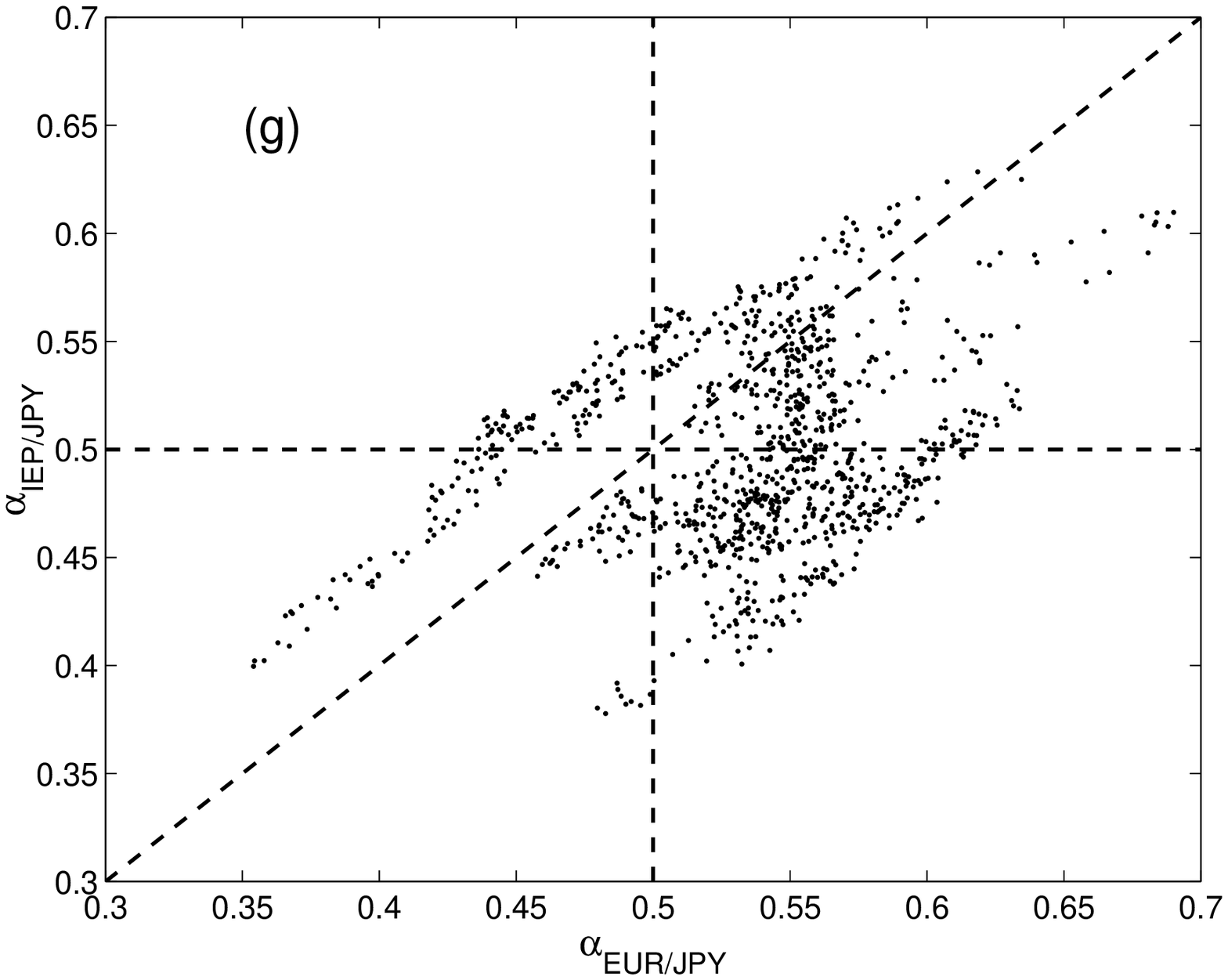}
\hfill
\leavevmode
\epsfysize=6.5cm
\epsffile{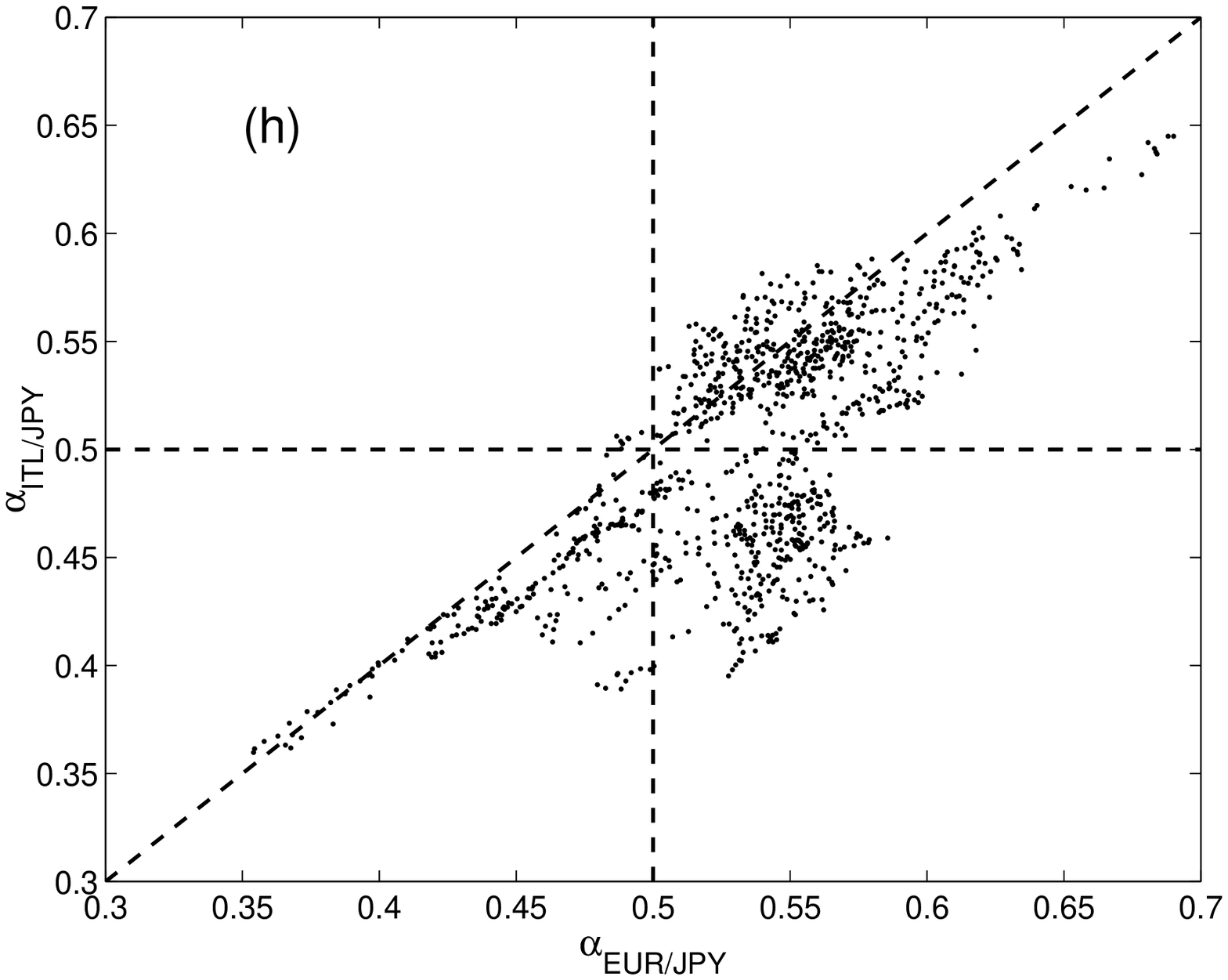}
\vfill
\leavevmode
\epsfysize=6.5cm
\epsffile{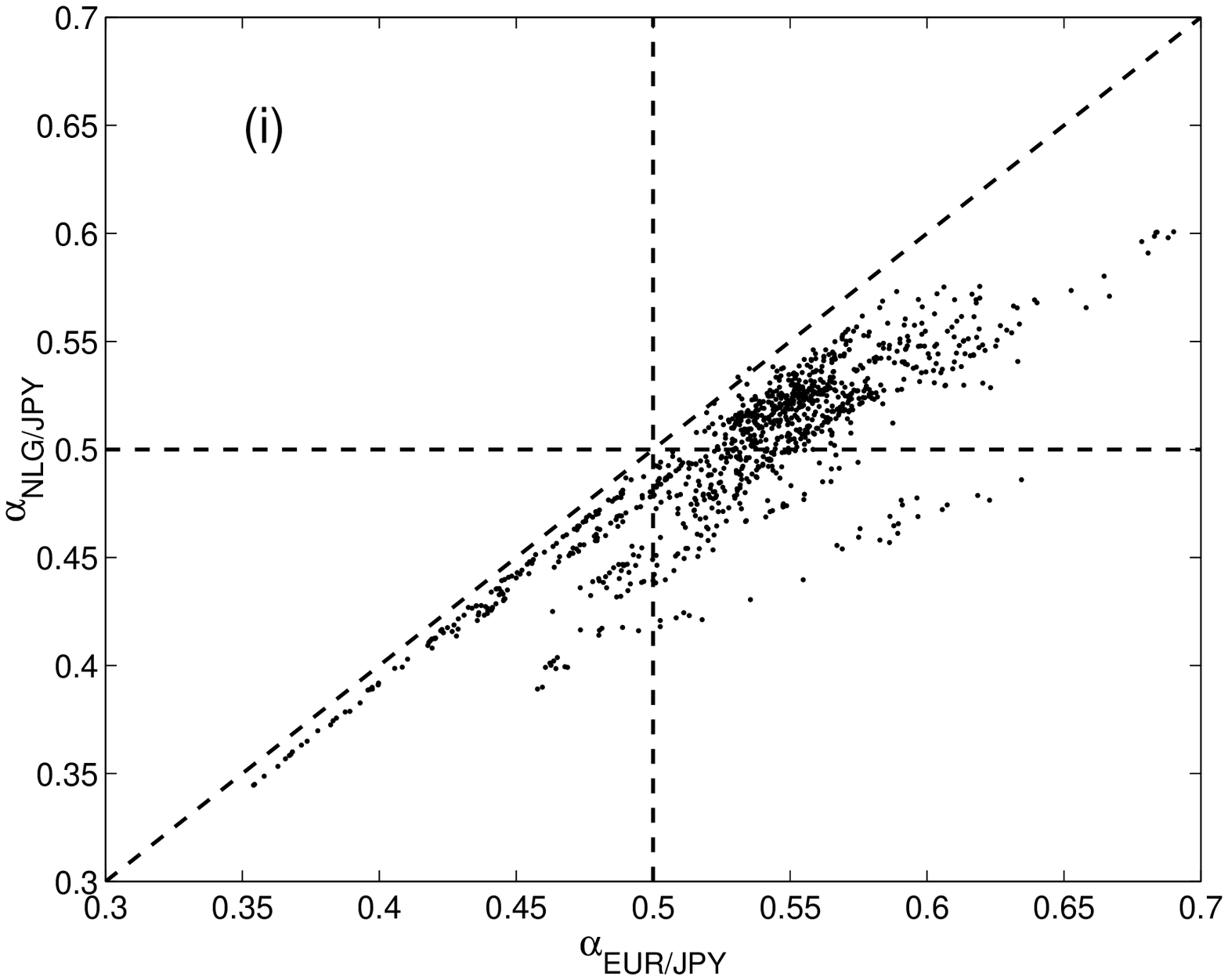}
\hfill
\leavevmode
\epsfysize=6.5cm
\epsffile{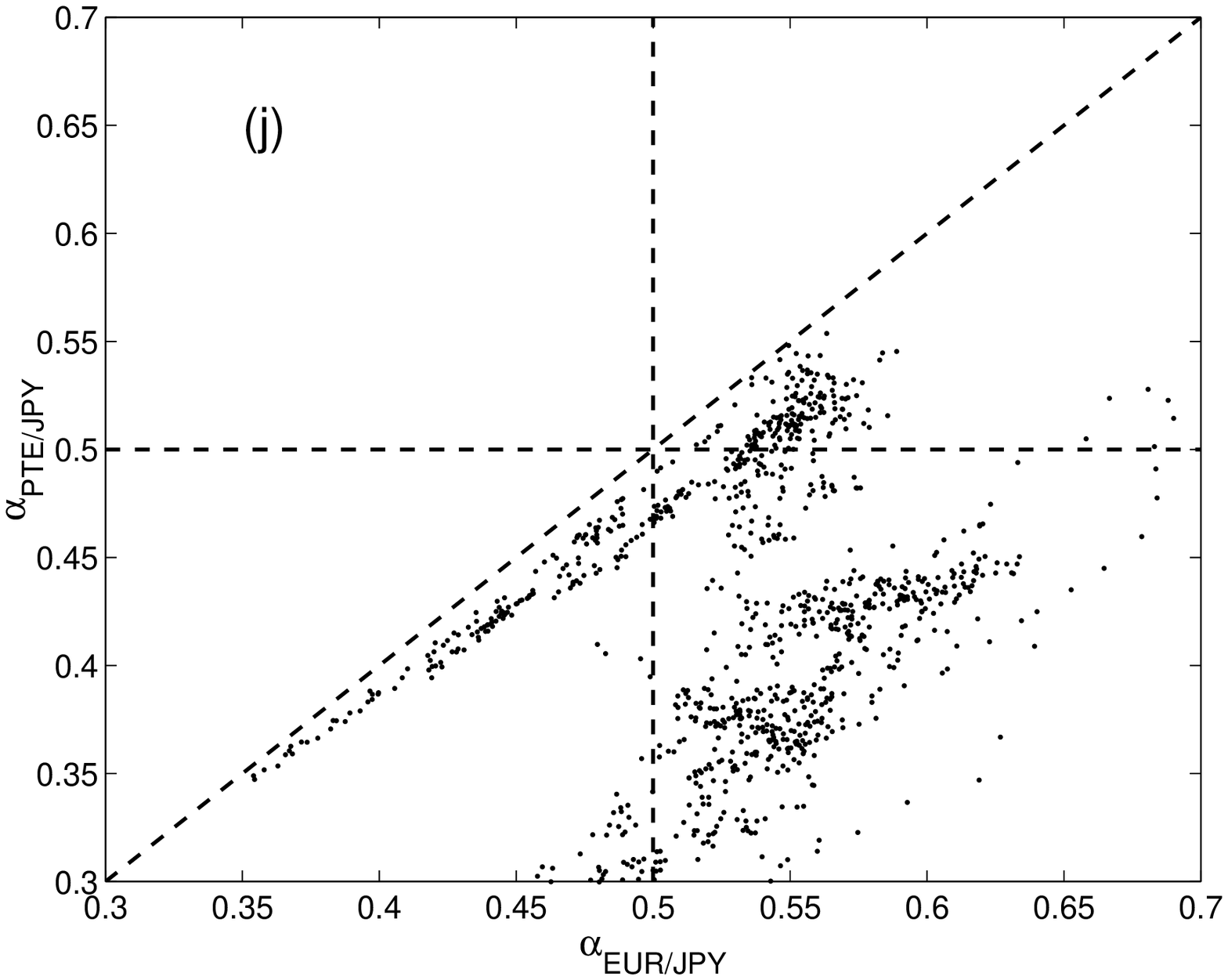}
\end{center}
\caption{Continue Fig 7d}
\label{fig7d}
\end{figure}

\begin{figure}[ht]
\begin{center}
\leavevmode
\epsfysize=6.5cm
\epsffile{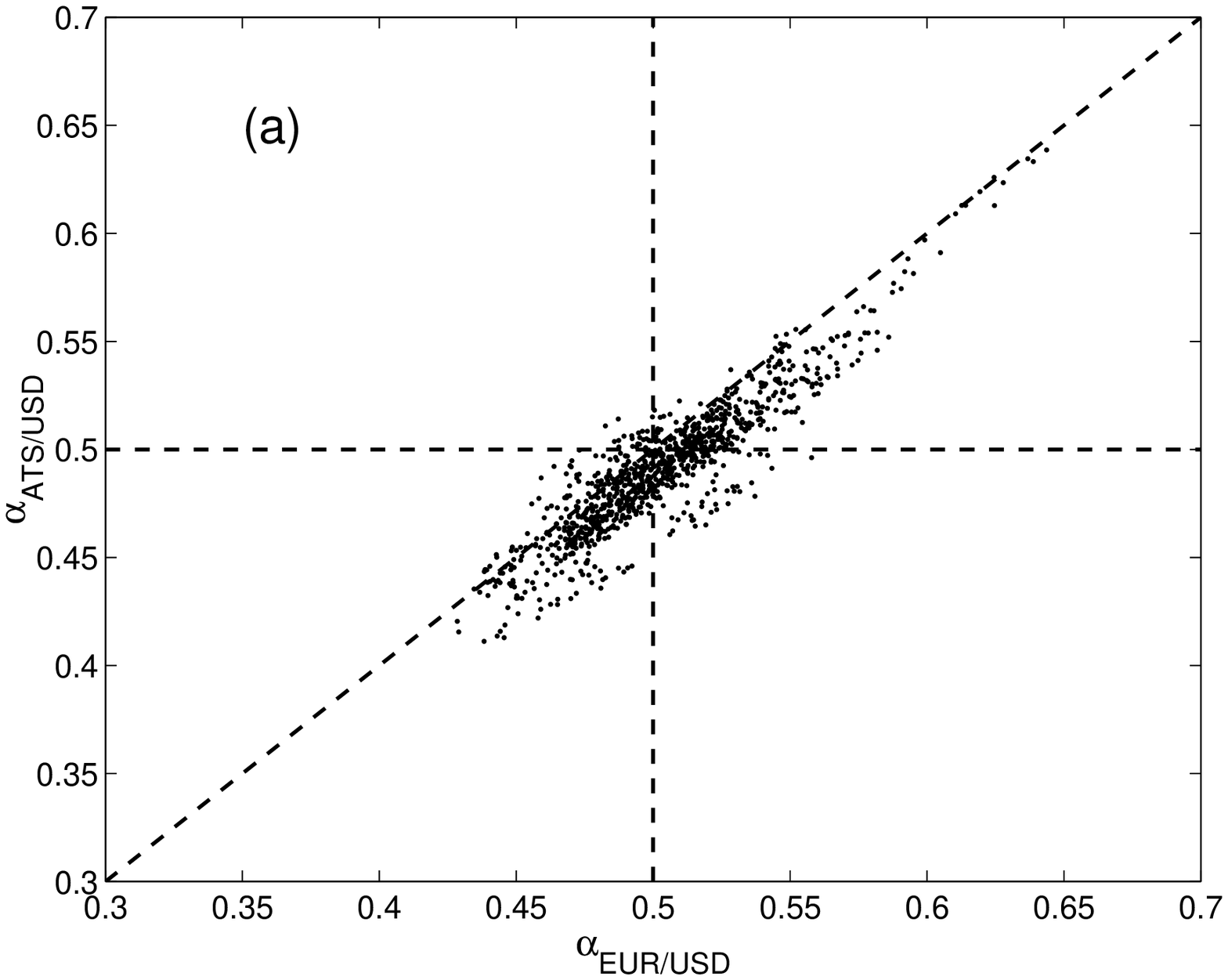}
\hfill
\leavevmode
\epsfysize=6.5cm
\epsffile{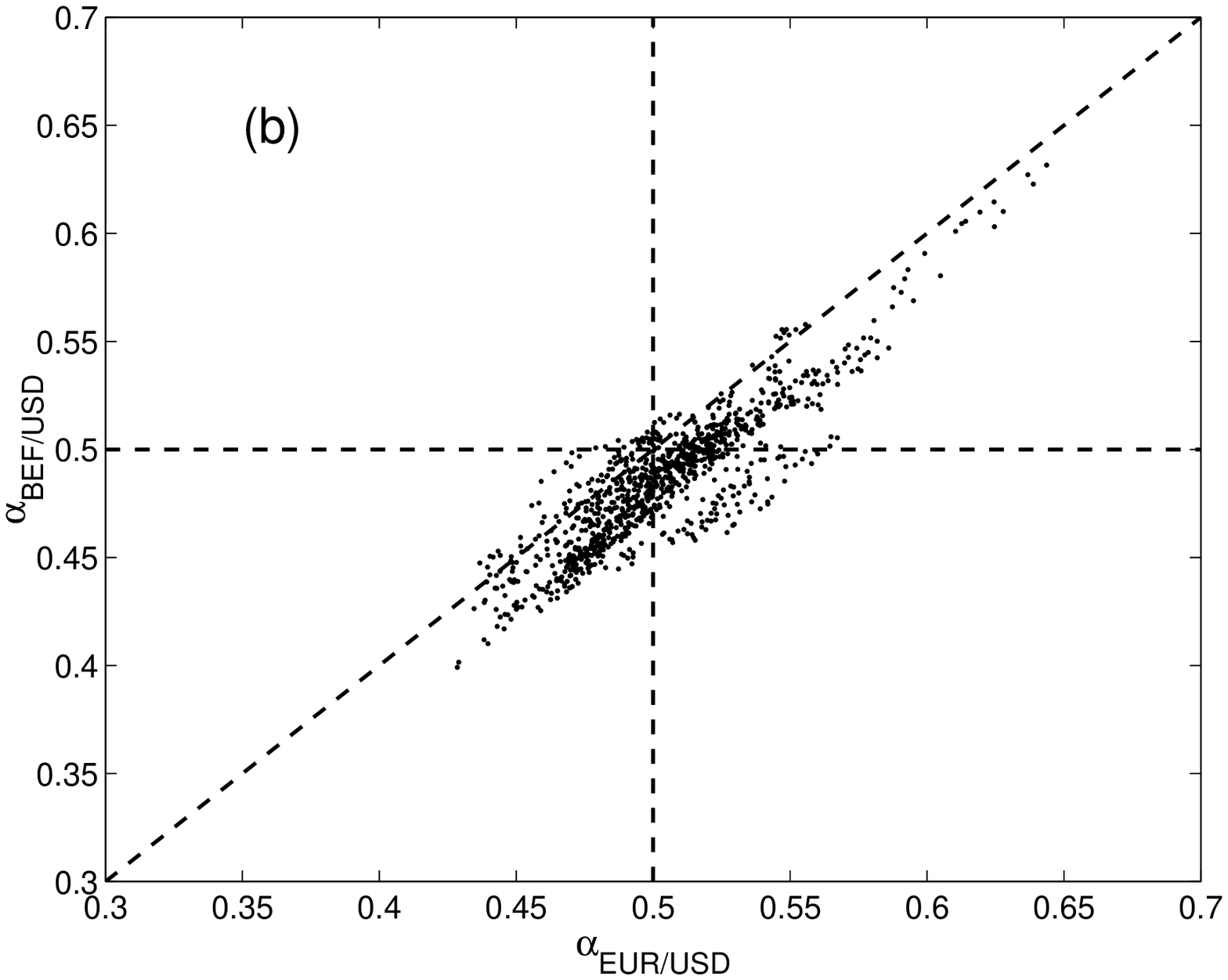}
\vfill
\leavevmode
\epsfysize=6.5cm
\epsffile{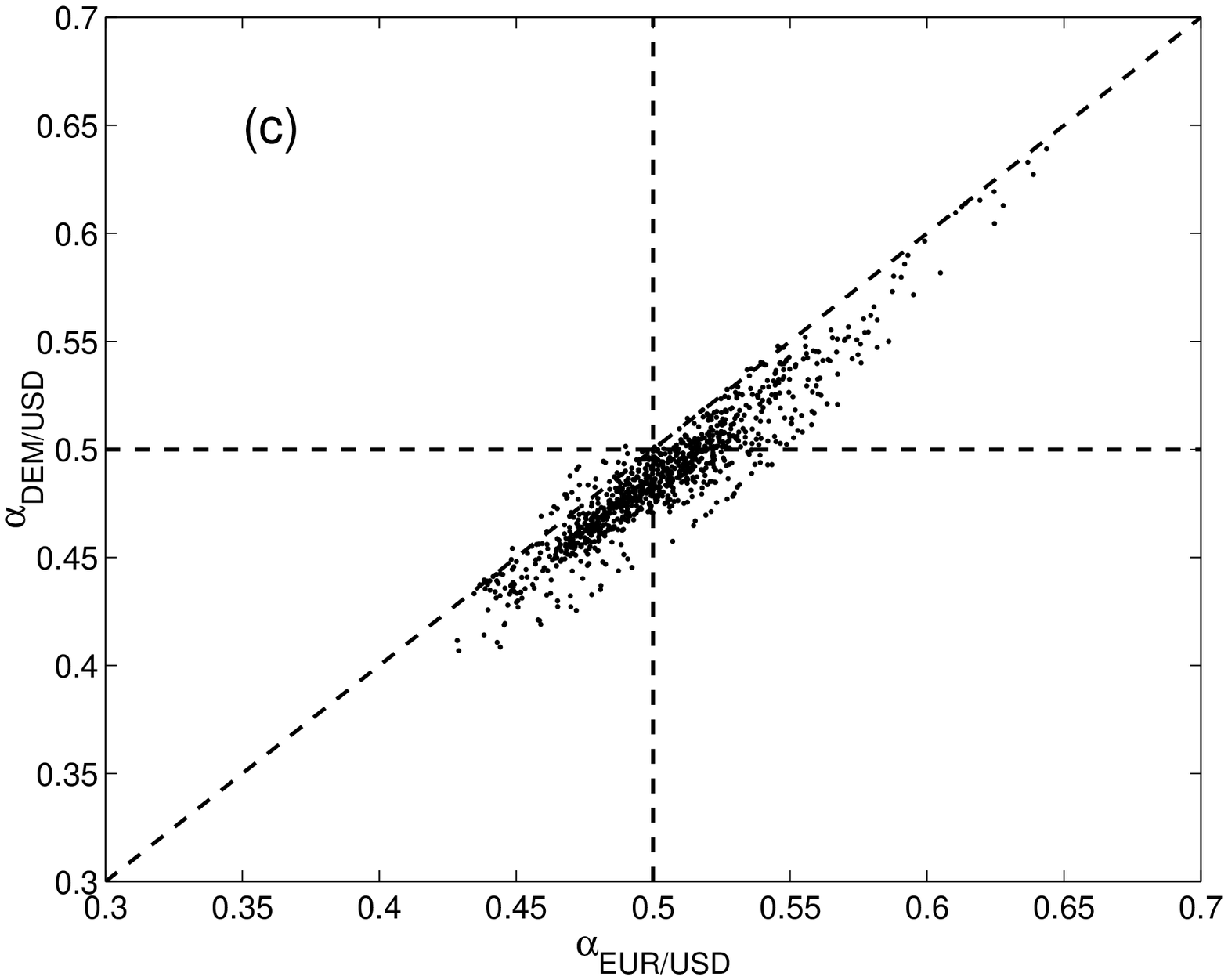}
\hfill
\leavevmode
\epsfysize=6.5cm
\epsffile{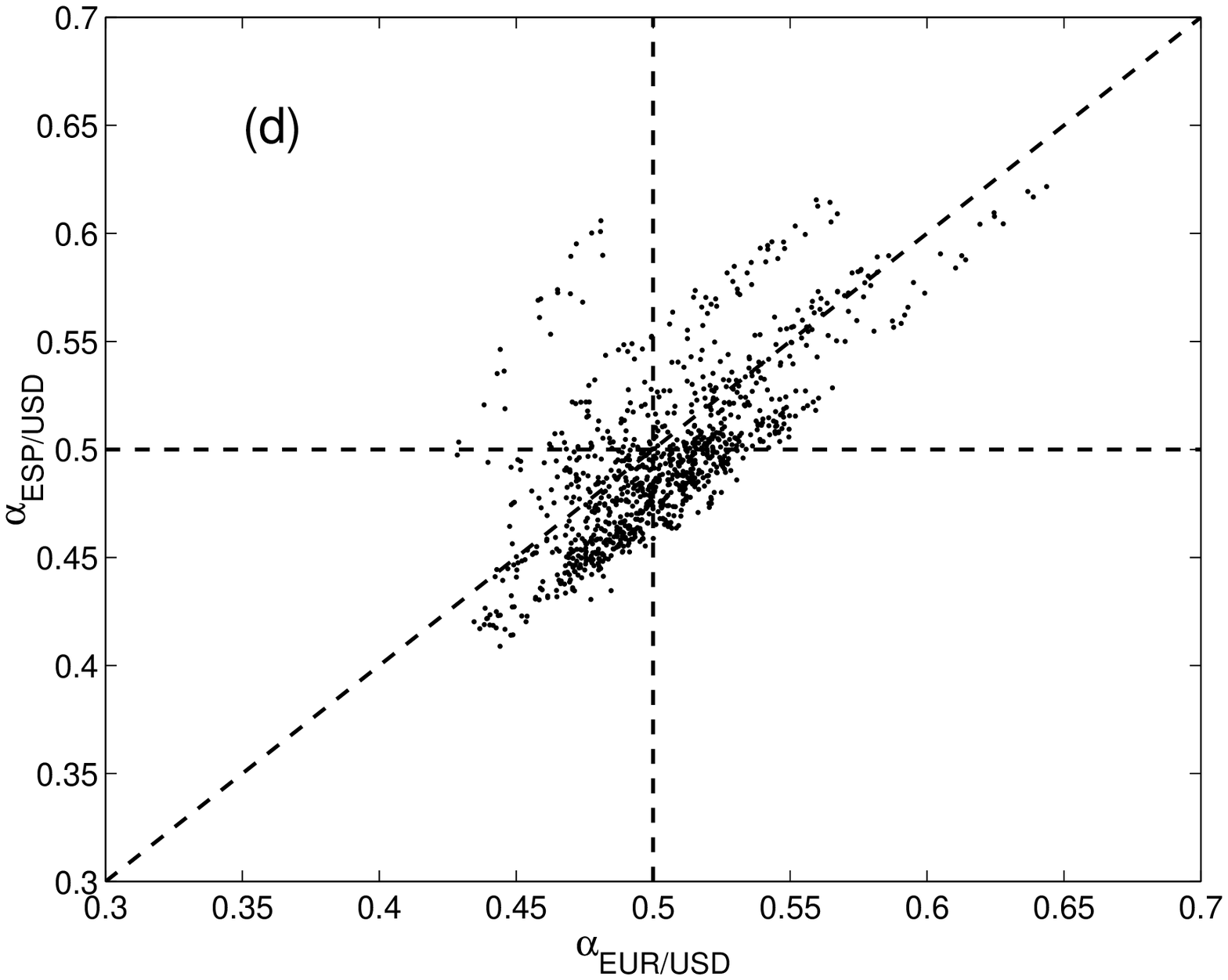}
\end{center}

\newpage

\begin{center}
\leavevmode
\epsfysize=6.5cm
\epsffile{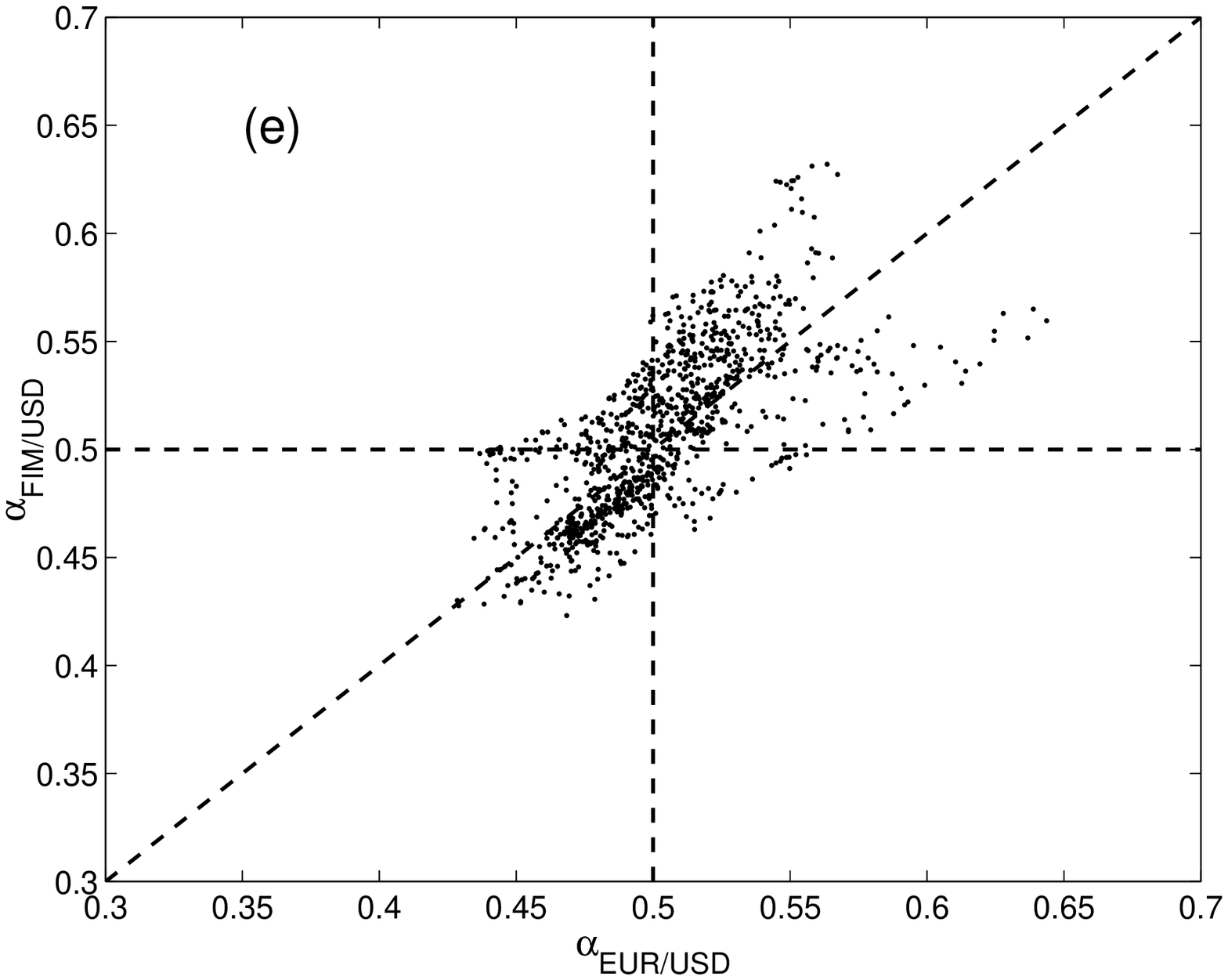}
\hfill
\leavevmode
\epsfysize=6.5cm
\epsffile{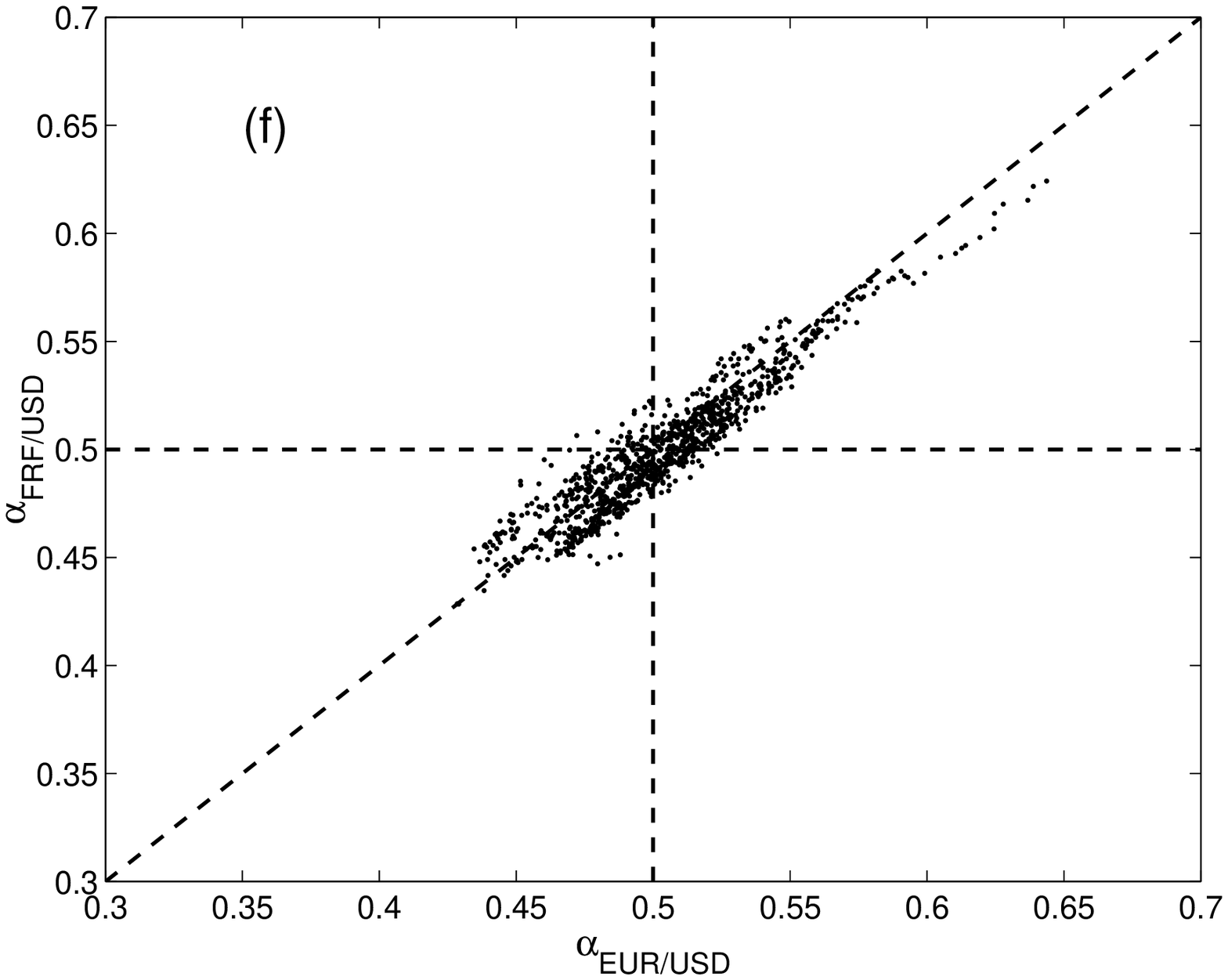}
\vfill
\leavevmode
\epsfysize=6.5cm
\epsffile{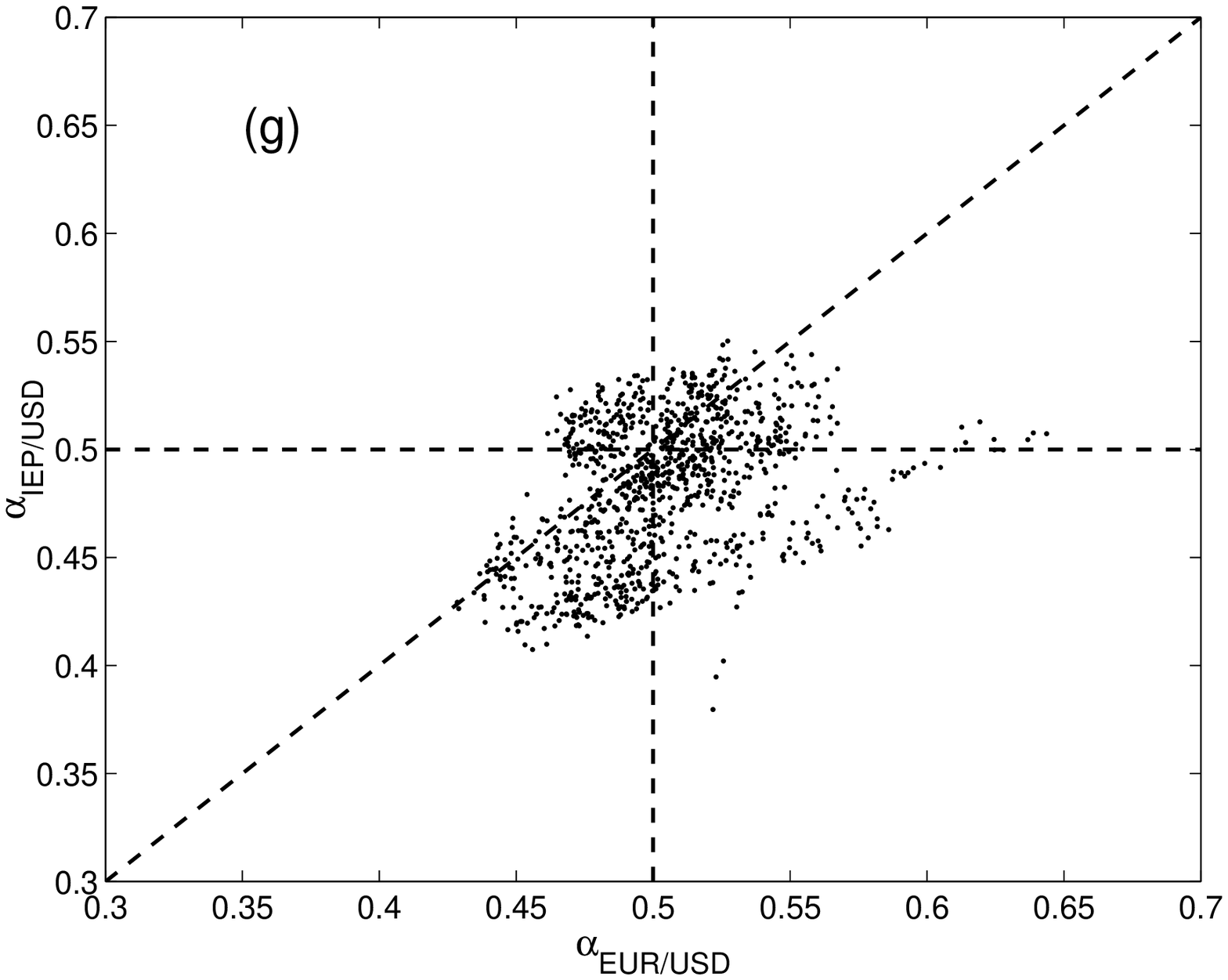}
\hfill
\leavevmode
\epsfysize=6.5cm
\epsffile{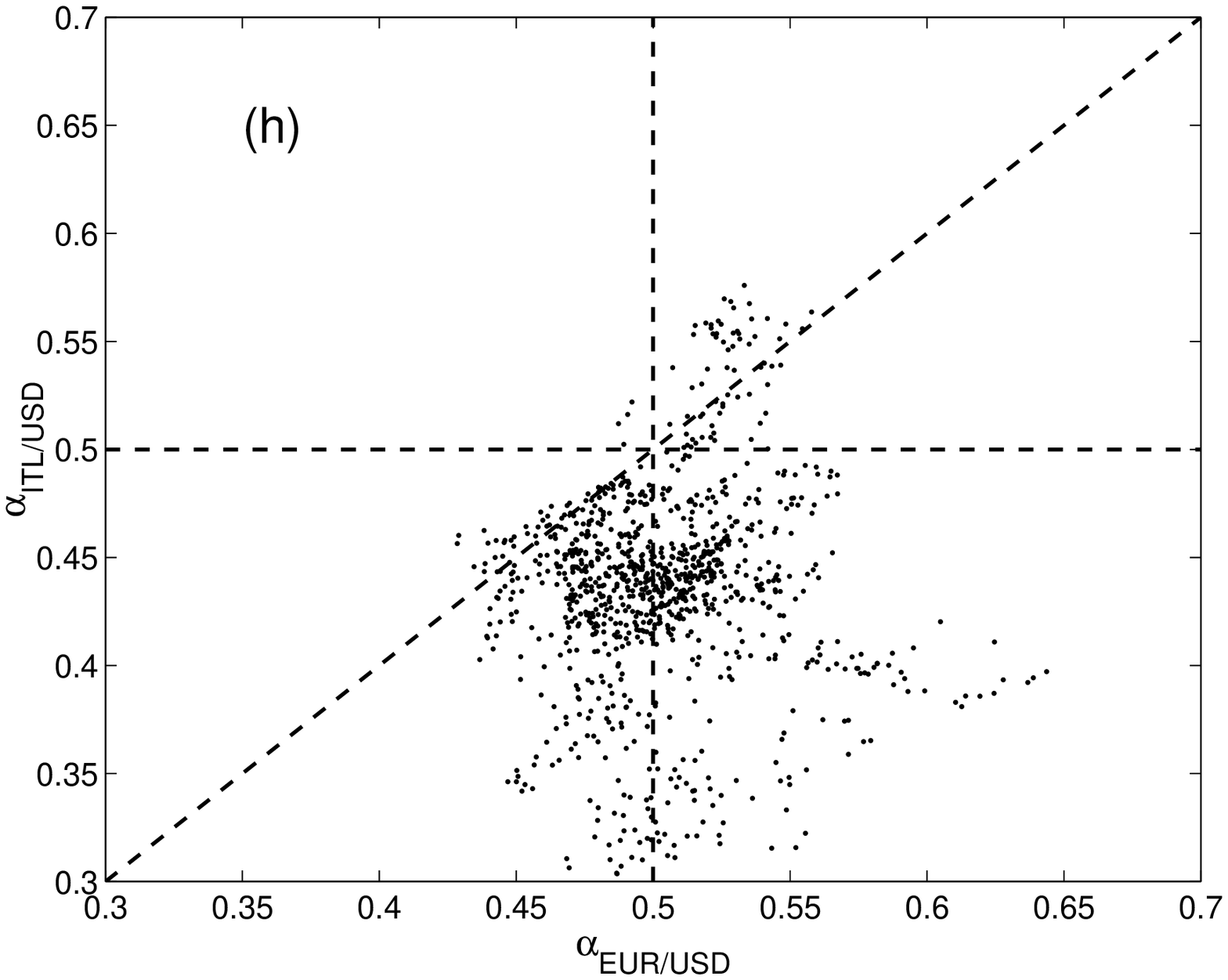}
\vfill
\leavevmode
\epsfysize=6.5cm
\epsffile{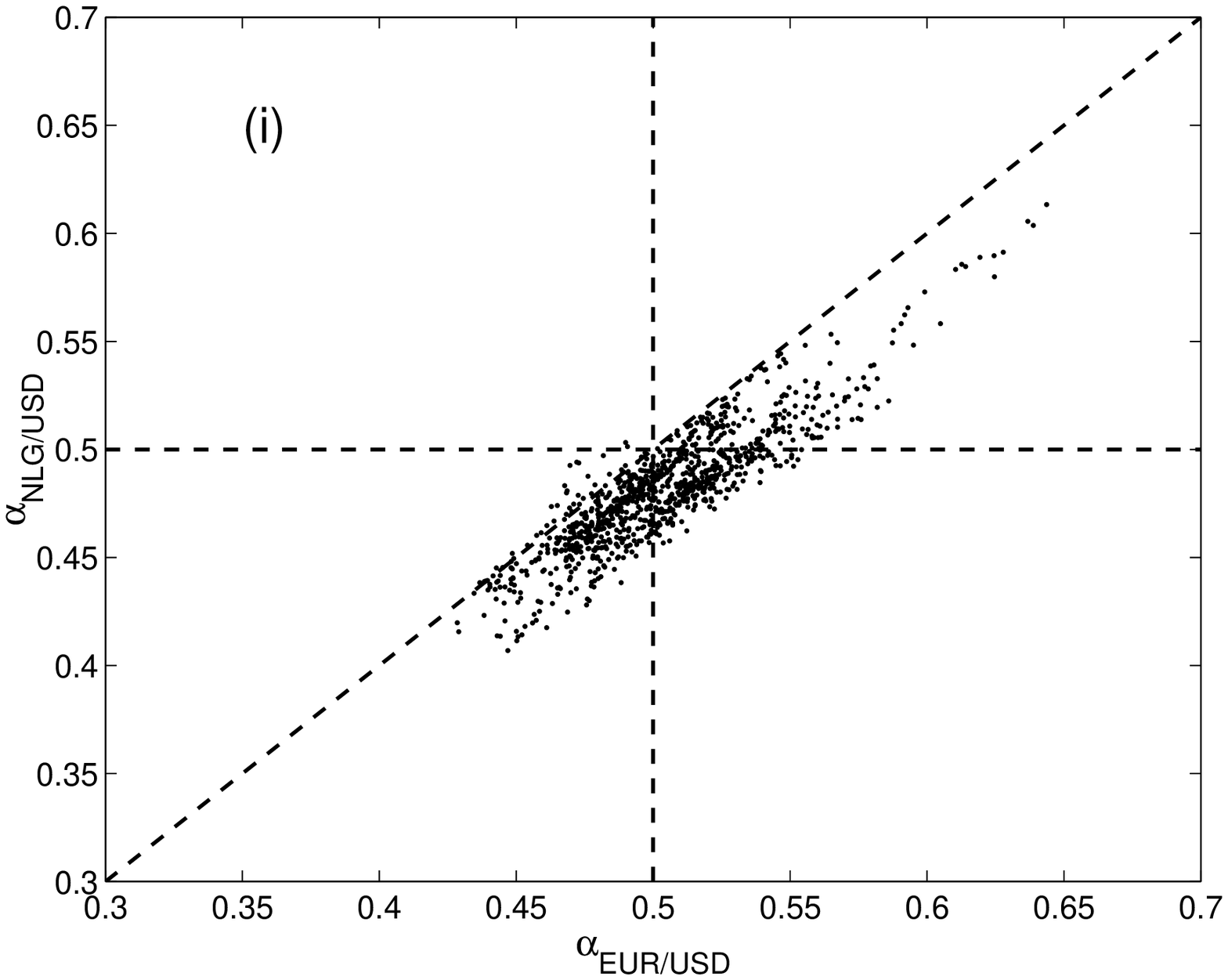}
\hfill
\leavevmode
\epsfysize=6.5cm
\epsffile{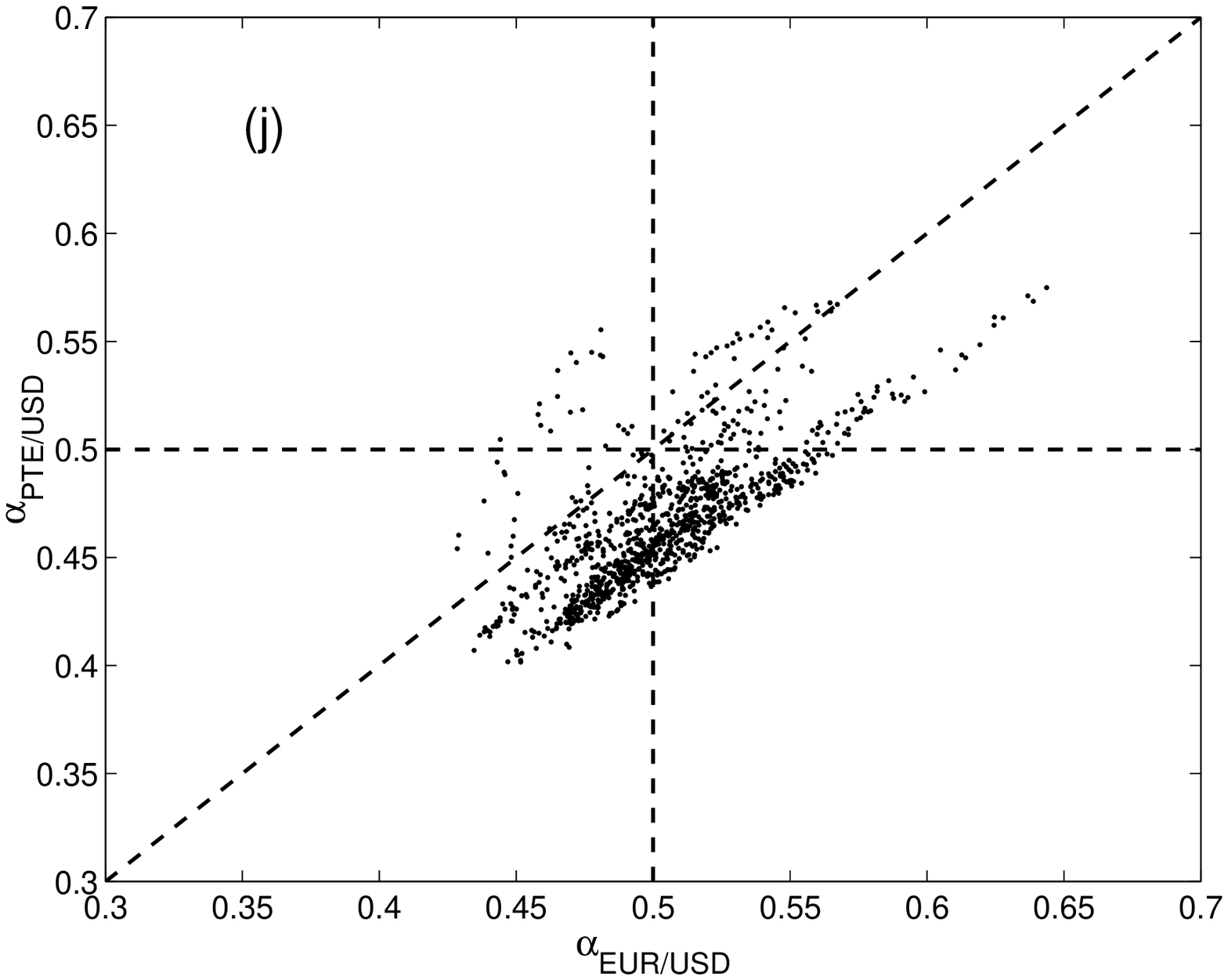}
\end{center}
\caption{Continue Fig 7e}
\label{fig7d}
\end{figure}

\begin{figure} 
\begin{center}
\leavevmode
\epsfysize=8cm
\epsffile{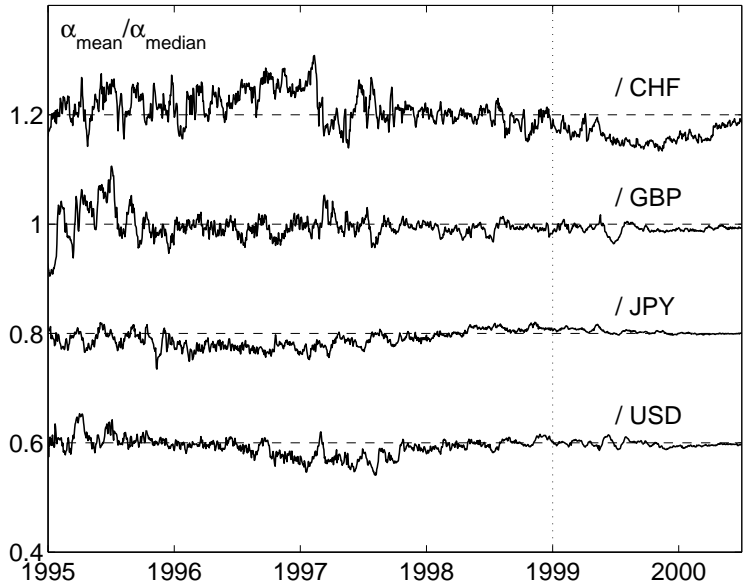}
\end{center}
\caption{Time evolution of the
$\alpha_{mean}/\alpha_{median}$ ratio extracted from the $\alpha$ exponents for
the currency exchange rates, currencies forming the $EUR$, with 
respect to $CHF$,
$GBP$, $JPY$ and $USD$. The curves are displaced for readability though the
y-axis scale is constant and the $GBP$ curve is not displaced. The horizontal
dashed lines correspond to $\alpha_{mean}/\alpha_{median}=1$.
} \label{fig8} \end{figure}

\end{document}